\documentclass[]{aa}  
\usepackage{lscape}
\usepackage{longtable,threeparttable}
\usepackage{float}
\usepackage{graphicx}
\usepackage{placeins}
\usepackage{txfonts}
\usepackage{caption}
\usepackage{subcaption}
\usepackage{comment}

\usepackage{hyperref}
\hypersetup{
	unicode=false,
	pdftoolbar=true,
	pdfmenubar=true,
	pdffitwindow=false,
	pdfstartview={FitH},
	pdftitle={My title},
	pdfauthor={Author},
	pdfsubject={Subject},
	pdfcreator={Creator},
	pdfproducer={Producer},
	pdfkeywords={keyword1} {key2} {key3},
	pdfnewwindow=true,
	colorlinks=true,
	linkcolor=red,
	citecolor=blue,
	filecolor=magenta,
	urlcolor=cyan
}

\begin{document} 
   \title{Hot subdwarfs in close binaries observed from space II:\\ Analysis of the light curves}

    \titlerunning{Close sdB binaries from TESS II}
   \author{V. Schaffenroth
          \inst{1}
          \and
          B.~N. Barlow\inst{2}
          \and I. Pelisoli\inst{3,1}
          \and S. Geier\inst{1}
          \and T. Kupfer\inst{4}
          }
\authorrunning{Schaffenroth et al.}
   \institute{Institute for Physics and Astronomy, University of Potsdam, Karl-Liebknecht-Str. 24/25, 14476 Potsdam, Germany\\
              \email{schaffenroth@astro.physik.uni-potsdam.de}
              \and
              Department of Physics and Astronomy, High Point University, High Point, NC 27268, USA
         \and
             Department of Physics, University of Warwick, Gibet Hill Road, Coventry CV4 7AL, UK
                 \and
        Department of Physics and Astronomy, Texas Tech University, PO Box 41051, Lubbock, TX 79409, USA
             }

   \date{Received 08 August 2022; accepted  24 February 2023}

 
  \abstract
   {Hot subdwarfs in close binaries with either M dwarf, brown dwarf or white dwarf companions show unique light variations. In hot subdwarf binaries with M dwarf or brown dwarf companions we can observe the so-called reflection effect and in hot subdwarfs with close white dwarf companions ellipsoidal modulation and/or Doppler beaming. }
   {The analysis of these light variations can be used to derive the mass and radius of the companion and hence determine its nature. Thereby we assume the most probable sdB mass and the radius of the sdB derived by the fit of the spectral energy distribution and the \textit{Gaia} parallax.}
   {In the high signal-to-noise space-based light curves from the Transiting Exoplanet Survey Satellite  and the \textit{K2} space mission, several reflection effect binaries and ellipsoidal modulation binaries have been observed with much better quality than possible for ground-based observations. The high quality of the light curves allowed us to analyse a large sample of sdB binaries with M dwarf or white dwarf companions using \textsc{lcurve}. }
   {For the first time we can constrain the absolute parameters of 19 companions of reflection effect systems covering periods from 2.5 to 19 hours and companion masses from the hydrogen burning limit to early M dwarfs. Moreover, we could determine the mass of eight white dwarf companion to hot subdwarf binaries showing ellipsoidal modulations, covering a so far unexplored period range from 7 to 19 hours. The derived masses of the white dwarf companions show that all but two of the white dwarf companions are most likely helium-core white dwarfs. Combining our results with previously measured rotation velocities allowed us to derive the rotation period of seven sdBs in short-period binaries. In four of those systems the rotation period of the sdB agrees with a tidally locked orbit, in the other three systems the sdB rotates significantly slower. }
   {}


   \maketitle
%

\section{Introduction}
Hot subdwarfs of spectral type O and B (sdO/B) are stars with temperatures from $\sim 25000-60000\,\rm K$ and luminosities placing them between main sequence stars and white dwarfs. Most of the sdBs are found on the extreme horizontal branch (EHB). Their formation is still unclear, but most of the sdBs are believed to be core-He burning objects that lost most of their envelope on the tip of the red giant branch (RGB). The H-rich sdOs are believed to be the progeny of the sdBs after the helium in the core is exhausted showing He-shell burning. Their evolution is much faster and hence they are rarer than the sdB \citep[][]{heber:2009,heber:2016}.  

\citet[][]{pelisoli:2020} suggested that the formation of typical sdBs requires binary interaction. Indeed one third of the sdBs is found in sdB+F/G/K type main-sequence companions with periods of several hundred days \citep[][]{vos:2013,vos:2018}. Another third of the sdBs is found in close binaries with low-mass main sequence stars of spectral type M close to the hydrogen burning limit or even brown dwarf companions (dM/BD) or white dwarf (WD) \citep[][]{maxted:2002,kupfer:2015,erebos} with periods from under one hour to 27 days. Such short periods can only be explained by a previous common envelope phase \citet[][]{han:2002,han:2003}. The remaining sdBs are apparently single.

The nature of the companions in many of these close sdB binaries can easily be identified by their characteristic light variations using high signal-to-noise light curves. Close binaries with dM/BD companion show a significant quasi-sinusoidal variation over each orbit with an amplitude from a few percent up to $\sim$~20\% (see Fig. \ref{refl1}). The strength of this variation, called the reflection or irradiation effect, increases from blue to red wavelengths. It results from a large temperature difference between the sdB primary and the cooler companion, but a similar or even larger size of the secondary compared to the sdB. Due to the large irradiating flux from the sdB, one side of the companion is heated up from temperatures originally around $3000\,\rm K$ to temperatures from $10\,000-20\,000\,\rm K$ \citep[][]{aador,kiss00}. Consequently, the contribution of the companion to the total system flux significantly increases when the hot side is visible. As those systems have small separations from $\sim 0.5$ to a few solar radii \citep[e.g.][]{schaffenroth14,schaffenroth15,schaffenroth21}, a significant percentage of them also show eclipses. They are referred to as HW Vir binaries, named after the prototype system. Both the shape and strength of the reflection effect depend strongly on the orbital inclination \citep[][]{schaffenroth14a} and so light curves with sufficient signal--to--noise (S/N) can be used to constrain system parameters, even without eclipses.

Since white dwarfs are much smaller than M dwarfs or brown dwarfs, the reflection effect cannot be observed in sdB+WD systems. However, a close WD companion can cause an ellipsoidal deformation of the hot subdwarf, which leads to a quasi--sinusoidal variation with half the orbital period. The amplitude of this ellipsoidal modulation can be up to almost 10\% in the most extreme cases \citep[e.g.][]{maxted_kpd,bloemen11}. Due to gravity darkening, the depths of the two minima are usually different, and lower flux is observed when the side of the hot subdwarf facing the companion is visible. As the orbital velocities are quite high, Doppler beaming from the hot subdwarf is also observed, resulting in more flux when it approaching Earth than when it moves away \citep[e.g.][]{cd-30,telting14,kup17,kup17a,kupfer22,pelisoli21}. The amplitude strongly scales with the separation and the companion mass, and longer period systems ($>$ a few hours) have ellipsoidal modulation amplitudes below 0.5\% and can only be found in space-based light curves. This fact can also be used to distinguish between WD and dM/BD companions, when high signal-to-noise light curves are available, as the amplitude of the reflection effect is much higher and would be visible up to several days in the \textit{TESS} light curves \citep[see][hereafter paper I]{Schaffenroth2022}. Hence a dM/BD companion can be excluded, if no variations can be detected. 

In  paper I we used this method to determine the nature of the companion for 75\% of the known close hot subdwarf binaries. Moreover, we performed a search for more sdB binaries showing a reflection effect, ellipsoidal modulation or Doppler beaming using light curves provided by the \textit{TESS} (Transiting Exoplanet Survey Satellite) \citep{TESS} and \textit{K2}  \citep{k2} missions. In total we found 85 new reflection effect systems \citep[including also systems found by][]{tess_north,tess_south,barlow22}, 8 new ellipsoidal systems and 16 systems showing Doppler beaming in the light curve, in addition to the 17 reflection effect and 11 ellipsoidal systems already known.

In this paper we present the analysis of 19 sdB+dM/BD systems showing a reflection effect and 25 sdB+WD systems showing ellipsoidal modulation or Doppler beaming. 
In Sect. 2 we discuss the target selection and data sources. In Sect. 3 we discuss the analysis of the sdB binaries with cool, low mass companions. In Sect. 4 we discuss the analysis of the sdB with white dwarf companions, and in Sect. 5 we give a short summary and a discussion of the results.


\section{Target selection and data sources}
We selected all sdB binaries with radial velocity curves published in the literature, which were observed by \textit{TESS} or \textit{K2} and show light variations indicating a hot subdwarf binary. In the case of the sdB+dM systems we only concentrated on the non-eclipsing systems. Moreover, we also included three bright sdB binaries, for which we could obtain spectroscopic follow-up.

All light curves were downloaded, phase-folded to the orbital period determined by a periodogram around the orbital period known from radial velocity (RV) variations, and binned using the Python package \textsc{lightkurve} \citep[][]{lightkurve}\footnote{\url{https://docs.lightkurve.org}}.

\section{The cool, low mass companions to the sdB stars}
\label{reflection}

\subsection{Method}
The presence of a reflection effect indicates a cool, low-mass companion of similar size in close orbit with the hot subdwarf. Without eclipses it is difficult to determine the inclination of the system, as the amplitude is degenerate in inclination and size of the companion \citep[][]{schaffenroth14a}. In this paper the authors also showed that the shape of the reflection effect changes with inclination, suggesting that it might be possible for high-quality light curves to provide tight constraints on the inclination angle even without eclipses. The same was also shown by \citet{oestensen2013}. With the space-based light curves available from \textit{TESS} and \textit{K2}, this is now possible for the first time. With the light curves from the original Kepler mission it was not possible, as only very few hot subdwarfs were observed, amongst them only one reflection effect system also showing eclipses. We analysed all reflection effect systems with solved RV curve atmospheric parameters derived from spectroscopy and space-based light curve (19 systems in total). 

For the analysis of the light curves we used \textsc{lcurve} \citep[see][for more details]{copperwheat10} as described in \citet[][]{schaffenroth21}. As we do not see any eclipses, the mass ratio and the radii are not well constrained from the light curve alone. To get a good solution we therefore had to make some assumptions.

All studied reflection effect systems are single-lined binaries. Therefore, the mass ratio cannot be determined with time-resolved spectroscopy. The sdB mass was derived by the fitting of the spectral energy distribution (SED) and combining this with the \textit{Gaia} parallax in paper I. However, in this paper we have seen that the masses for the reflection effect systems have to be taken with caution, as the mass distribution of the HW Vir systems shows discrepancies with the distribution of the reflection effect systems. Therefore, we use the assumption of the canonical He-flash mass of $0.47\,\rm M_\odot$  for the sdB for this analysis and is the most likely mass of a sdB in a sdB+dM binary \citep[paper I,][]{han:2002,han:2003,fontaine12}.

With this assumption and the inclination determined from the light curve analysis (which is not dependent on the mass ratio but sensitive to the light curve shape, see also Barlow et al. in prep, hereafter paper III), it is possible to get the mass ratio together with the separation of the system from the mass function determined by the radial velocity curve (see Table \ref{param_rv} for the parameters of the analysed reflection effect systems). The effective temperature is fixed to the value derived from spectral fitting (see paper I for a summary of all atmospheric parameters). As the contribution of the dark side of the companion to the flux is negligible, the temperature of the companion cannot be constrained and is hence fixed to a typical value for an M dwarf of 3000 K. Changes in the temperature of the companion therefore have a negligible effect on the other derived parameters. The SED fitting in combination with the \textit{Gaia} parallax (see paper I) provides the radius of the sdB $R_1$, and with the derived separation $a$, we can fix the relative radius of the sdB ($r_1=R_1/a$), which is used as parameter in the light curve fitting.

For simplicity the absorb factor, which is the percentage of the flux of the sdB used to heat up the irradiated companion side using a blackbody approximation, was fixed to 1. The gravitational and limb darkening coefficients were fixed according to the tables of \citet[][]{claret_limb}. We adopted their values closest to the atmospheric parameters for the \textsc{TESS} filter. Only the inclination and the radius of the companion were varied; all other parameters were fixed as explained above.

We performed Markov Chain Monte Carlo (MCMC) computations using \textsc{emcee} \citep[][]{emcee} in order to derive the distribution of the inclination and radius of the companion and to determine the uncertainties of both parameters. We tried to vary also the radius of the sdB using a Gaussion prior. Due to the residuals in the light curves (see Fig. \ref{refl1}), which are discussed later, this did not work. So, we fixed the radius of the sdB, neglecting its uncertainty.  As a result, the uncertainties in $i$ and $r_2$ will be underestimated. We performed several tests to quantify this by also varying the sdB radius. The uncertainty in $r_2$ and $i$ doubled in our test. As the uncertainty of the separation is dominating the overall uncertainty, the increase in uncertainty of the companions radius can be neglected. However, doubling the uncertainty in the inclination results in 50\% larger error bars in the mass ratio and companion mass and radius in our test. These results also depend on the quality of the light curve and the inclination. In the future this uncertainty will be included, when the mass of the sdB is constrained as well by the SED fit, as soon as reliable atmospheric parameters are available.  

\begin{figure}
\includegraphics[width=\linewidth]{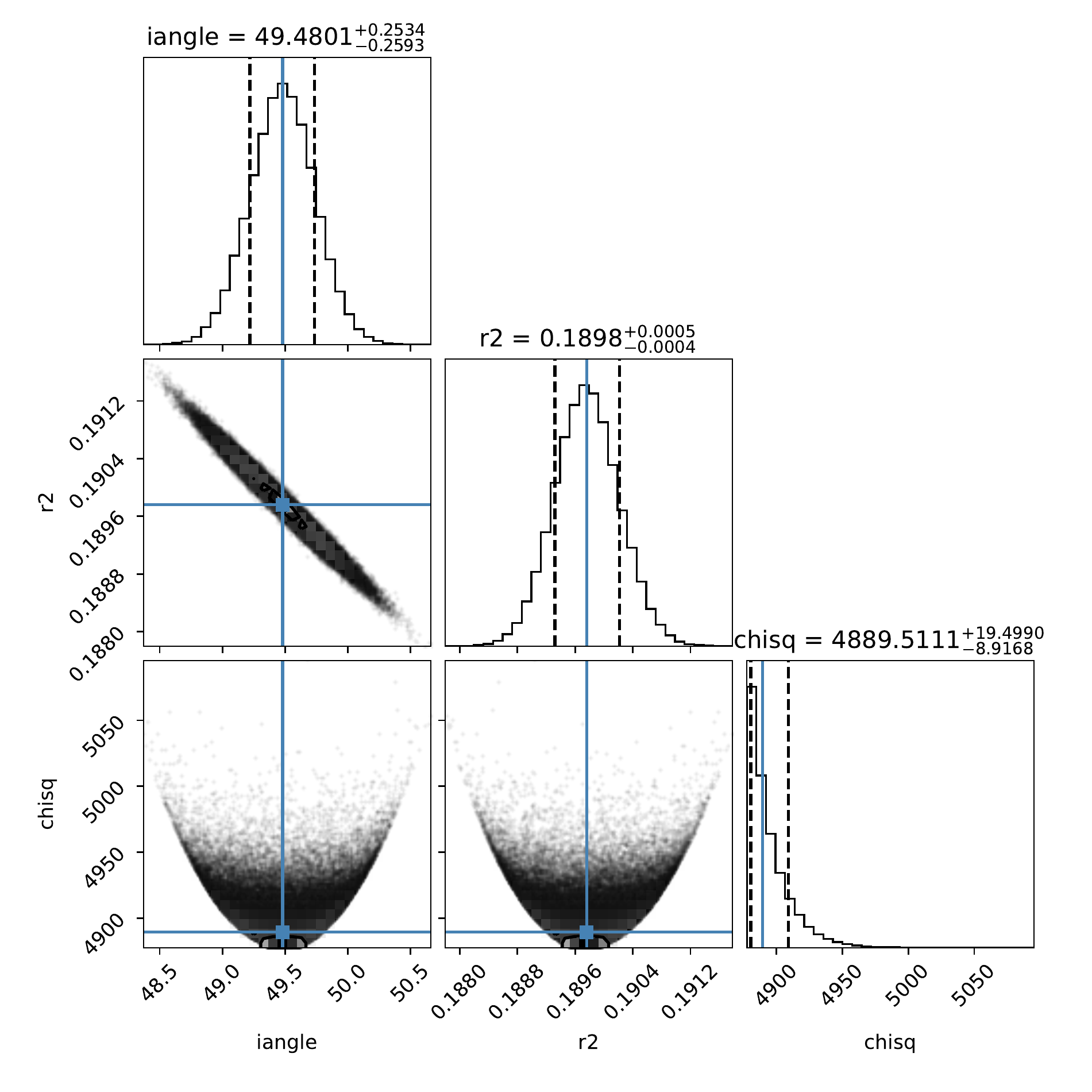}
\caption{Corner plot of the MCMC calculations of EC 01578-1743 showing the distribution of the orbital inclination (iangle), the radius of the companion (r2) and the $\chi^2$ (chisq).}
\label{mcmc}
\end{figure}

 In Fig. \ref{mcmc} an example of the MCMC results is shown.
There is some degeneracy between the orbital inclination and the radius of the companion visible, but the $\chi^2$ distribution is symmetric around the minimum representing the best solution. 
Some sdBs show short-period pulsations on the order of minutes (for sdO/B with $T_{\rm eff}>30000\,\rm K$) and long-period pulsations of low-amplitude (for sdO/B with $T_{\rm eff}<30000\,\rm K$) on the order of hours \citep[see][for a summary]{lynasgray21,kupfer2019}. In some sdB binaries we see hence a superposition of the pulsation and the binary signal, which complicates the analysis.
The phasing and binning of the light curves smoothed out the pulsations present in some of the systems, however the noise is still increased compared to the non-pulsating systems, leading to larger errors for the parameters compared to other systems with similar magnitudes. The pre-whitening of the pulsation frequencies could improve this, especially for longer period pulsations.  However, characterising the pulsations is out of the scope of this paper.

Due to the large pixel size, the \textit{TESS} light curves must be treated with care, especially if bright, unresolved stars fell on the same pixel. The PDCSAP flux provided in the  Science Processing Operations Center (SPOC) light curves tries to correct for this additional flux. However, this correction is not always perfect and can lead to the amplitude of the variation being over- or underestimated. For some targets a different amplitude in different sectors could be observed. The comparison with published light curves in other filters was used to choose the light curve with the correct amplitude. As we have discussed, the shape of the light curve is determined by the inclination. The amplitude on the other side is determined by the temperature of the primary and the orbital separation derived by time-resolved spectroscopy, as well as the radius of the companion. An overestimated amplitude would therefore result in an overestimated radius for the companion. We checked the field-of-view of \textit{TESS} and the CROWDSAP parameter for all our targets. With the exception of HS2333+3927 and TYC5977-517-1, the CROWDSAP parameter was close to 1, showing that no stars are blending into the target pixel, and hence no correction of the amplitude of the variation by the \textit{TESS} team was necessary.

\subsection{Results of the light curve analysis}
\begin{table*}

\caption{Orbital parameters of the solved reflection effect systems with space-based light curves from time-resolved spectroscopy and the analysis of the lightcurve. The objects are ordered following their inclination.}
\label{param_rv}
\begin{tabular}{llllllll}
\hline\hline
target & $P_{\rm RV}$ & $\gamma$ & $K_1$ & $P^*$$_{\rm lc,\it TESS/Kepler}$ \\ 
 & [d] & [$\rm km\,s^{-1}$]  & [$\rm km\,s^{-1}$] & [d] \\\hline 
BPSCS22169-0001$^d$  &  0.214     &  -  &  16.2  $\pm$  0.5  &  0.216895 \\
PHL457$^d$  &  0.3128  $\pm$  0.0007  &  -  &  12.8  $\pm$  0.08  &  0.313012 \\
KBS13$^d$  &  0.2923  $\pm$  0.0004  &  7.53  $\pm$  0.08  &  22.82  $\pm$  0.23  &  0.292365 \\
Feige48$^e$  &  0.343608  $\pm$  0.0000005  &  -47.9  $\pm$  0.1  &  28.0  $\pm$  0.2  &  0.343608 \\
GALEXJ2205-3141$^c$  &  0.341543  $\pm$  0.000008  &  -19.4  $\pm$  1.7  &  47.8  $\pm$  2.2  &  0.341552 \\
GALEXJ09348-2512$^a$  &  0.1429032  $\pm$  0.0000011  &  50.6  $\pm$  2.1  &  37  $\pm$  4  &  0.142903 \\
EQ Psc$^b$  &  0.800880  $\pm$  0.000097  &  25.9  $\pm$  1.3  &  34.9  $\pm$  1.6  &  0.800970 \\
PG1329+159$^d$  &  0.249699  $\pm$  0.0000002  &  -22.0  $\pm$  1.2  &  40.2  $\pm$  1.1  &  0.249696 \\
CPD-64481$^d$  &  0.277263  $\pm$  0.000005  &  94.1  $\pm$  0.3  &  23.9  $\pm$  0.05  &  0.277264 \\
JL82$^d$  &  0.73710  $\pm$  0.00005  &  -1.6  $\pm$  0.8  &  34.6  $\pm$  1.0  &  0.733799 \\
TYC5977-517-1$^a$  &  0.14387147  $\pm$  0.0000025  & -  &  87  $\pm$  2  &  0.143871 \\
GALEXJ0321+4727$^c$  &  0.265856  $\pm$  0.000003  &  69.6  $\pm$  2.2  &  60.8  $\pm$  4.5  &  0.265857 \\
SDSSJ012022+395059$^d$  &  0.252013  $\pm$  0.000013  &  -47.3  $\pm$  1.3  &  37.3  $\pm$  2.8  &  0.251975\\ 
UVEX0328+5035$^d$  &  0.11017  $\pm$  0.00011  &  44.9  $\pm$  0.7  &  64.0  $\pm$  1.5  &  0.110163 \\
HS2333+3927$^d$  &  0.1718023000  $\pm$  0.0000009  &  -31.40  $\pm$  2.1  &  89.60  $\pm$  3.2  &  0.171801 \\
V1405Ori$^d$  &  0.398    &  –33.6  $\pm$  5.5  &  85.1  $\pm$  8.6  &  0.398005 \\
HE1318-2111$^d$  &  0.487502  $\pm$  0.0000001  &  48.9  $\pm$  0.7  &  48.5  $\pm$  1.2  &  0.487424 \\
EC01578-1743$^a$  &  0.2581015  $\pm$  0.0000025  &  -23.19  $\pm$  0.4  &  86.5  $\pm$  0.5  &  0.258104 \\
HE0230-4323$^d$  &  0.45152  $\pm$  0.00002  &  16.6  $\pm$  1.0  &  62.4  $\pm$  1.6  &  0.450029 \\

\hline
\end{tabular}

$^a$ this paper \qquad $^b$ \citet{baran19} \qquad $^c$ \citet{nemeth:2012,kawka:2015} \qquad $^d$ \citet[][and references therein]{kupfer:2015} \qquad $^e$ \citet[][]{latour14}\qquad $^*$ typical error 0.0001 d
\end{table*}

\begin{table*}
\caption{Inclination, separation, mass ratio of the analysed reflection effect systems together with the calculated mass and radius of the companion.}
\label{param_lc}
\begin{tabular}{lllllll}
\hline\hline
target & $i$ & $q$ & $a$ & $M_2$ & $R_2$ & $P_{\rm rot}$\\ 
 & [$^\circ$] &  & $\rm [R_\odot]$ &$\rm [M_\odot]$ & $\rm [R_\odot]$ &[d]\\\hline
BPSCS22169-1  &  7.7  $\pm$  1.1  &  0.492  $\pm$  0.132  &  1.35  $\pm$  0.30  &  0.231  $\pm$  0.062  &  0.309  $\pm$  0.072 & $0.16 \pm0.04$\\
PHL457  &  9.3  $\pm$  1.6  &  0.416  $\pm$  0.085  &  1.69  $\pm$  0.37  &  0.196  $\pm$  0.040  &  0.157  $\pm$  0.035 &-\\
KBS13  &  10.1  $\pm$  0.4  &  0.552  $\pm$  0.038  &  1.040  $\pm$  0.063  &  0.260  $\pm$  0.018  &  0.284  $\pm$  0.020 &-\\
Feige48  &  16.3  $\pm$  1.4  &  0.495  $\pm$  0.089  &  1.83  $\pm$  0.30  &  0.232  $\pm$  0.042  &  0.266  $\pm$  0.044 &$0.36 \pm0.07$\\
GALEXJ2205-3141  &  17.3  $\pm$  2.6  &  1.12  $\pm$  0.21  &  2.05  $\pm$  0.36  &  0.527  $\pm$  0.100  &  0.419  $\pm$  0.075 &-\\
GALEXJ09348-2512  &  24.0  $\pm$  3.0  &  0.351  $\pm$  0.099  &  0.99  $\pm$  0.26  &  0.165  $\pm$  0.046  &  0.175  $\pm$  0.048 &-\\
EQPsc  &  25.4  $\pm$  0.5  &  0.724  $\pm$  0.073  &  3.07  $\pm$  0.24  &  0.222  $\pm$  0.019  &  0.181  $\pm$  0.014 &-\\
PG1329+159  &  31.8  $\pm$  2.1  &  0.356  $\pm$  0.037  &  1.44  $\pm$  0.15  &  0.167  $\pm$  0.018  &  0.199  $\pm$  0.021 & $0.64 \pm0.07$\\
CPD-64481  &  34.3  $\pm$  2.2  &  0.187  $\pm$  0.012  &  1.473  $\pm$  0.12  &  0.088  $\pm$  0.006  &  0.118  $\pm$  0.010 & $1.22 \pm0.30$\\
JL82  &  34.6  $\pm$  1.1  &  0.511  $\pm$  0.043  &  3.06  $\pm$  0.22  &  0.240  $\pm$  0.020  &  0.249  $\pm$  0.018 &$0.61 \pm0.07$\\
TYC5977-517-1  &  35.0  $\pm$  0.25  &  0.678  $\pm$  0.026  &  1.07  $\pm$  0.034  &  0.319  $\pm$  0.012  &  0.380  $\pm$  0.013 &-\\
GALEXJ0321+4727  &  38.6  $\pm$  0.9  &  0.495  $\pm$  0.060  &  1.55  $\pm$  0.17  &  0.233  $\pm$  0.028  &  0.298  $\pm$  0.033 &-\\
SDSSJ012022+395059  &  39.9  $\pm$  7.0  &  0.343  $\pm$  0.070  &  1.44  $\pm$  0.36  &  0.16  $\pm$  0.033  &  0.242  $\pm$  0.063 &-\\
UVEX0328+5035  &  41.4  $\pm$  0.5  &  0.341  $\pm$  0.012  &  0.83  $\pm$  0.03  &  0.160  $\pm$  0.006  &  0.250  $\pm$  0.010 &-\\
HS2333+3927  &  42.8  $\pm$  0.5  &  0.609  $\pm$  0.017  &  1.18  $\pm$  0.05  &  0.388  $\pm$  0.017  &  0.401  $\pm$  0.017 &-\\
V1405Ori  &  43.0  $\pm$  0.9  &  0.829  $\pm$  0.141  &  2.17  $\pm$  0.30  &  0.390  $\pm$  0.066  &  0.341  $\pm$  0.047 &-\\
HE1318-2111  &  48.5  $\pm$  1.7  &  0.335  $\pm$  0.018  &  2.23  $\pm$  0.11  &  0.158  $\pm$  0.008  &  0.277  $\pm$  0.014&- \\
EC01578-1743  &  49.5  $\pm$  0.25  &  0.591  $\pm$  0.009  &  1.548  $\pm$  0.013  &  0.278  $\pm$  0.004  &  0.294  $\pm$  0.003 &-\\
HE0230-4323  &  52.6  $\pm$  1.5  &  0.470  $\pm$  0.027  &  2.18  $\pm$  0.11  &  0.221  $\pm$  0.013  &  0.309  $\pm$  0.016 &-\\
\hline
\end{tabular}
\end{table*}

To investigate the blending effect on the results further, we had a closer look at the \textit{TESS} light curve of HS2333+3927.
\textit{TESS} observed this system in two different sectors (sector 16/17; CROWDSAP=0.78/0.75). The amplitude in the first sector is about 10\% larger. We choose the light curve from the second sector, as the amplitude of the reflection effect of the first sector (50\%) is much larger than the amplitude expected for the \textit{TESS} filter compared to the observations by \citet[][]{heber:reflection} in B (20\%),V (25\%) and R (30\%). This indicates that the light curve of the first sector seems to be over-corrected.
The analysis of the \textit{TESS} light curve from sector 17 (see Fig. \ref{lc_hs2223}) can confirm the results of \citet[][]{heber:reflection}, showing that the correction can be trusted in this case, as we suspected.

To check the influence of an amplitude too large, we also fitted the light curve from sector 16. We get the same inclination but a radius of $0.47\pm0.02\,\rm R_\odot$, 17\% larger than the companion radius determined by the light curve of the other sector. As expected, the higher amplitude does not affect the determination of the inclination but will lead to a higher companion radius. Fortunately this affects only two of our targets, as discussed before. A more detailed discussion on the other object is given in Sect. \ref{tyc}.

When investigating the residuals of our highest S/N light curve fits (see Fig. \ref{refl1}), we can see a recurring pattern that grows in strength with increasing inclination. For example, the residuals for GALEX J0321+4727 and EC 01578-1743 show that the fit overestimates the flux right at the moment the reflection effect peaks, but immediately underestimates the flux on either side of the peak. The inability of the LCURVE models to fit the reflection effect shape precisely in this region reveals the limitations of the reflection effect model and the way it handles irradiation. The models improve with smaller sdB radii, suggesting that the illumination of the side of the companion facing the hot subdwarf is not homogeneous. This effect is very small (on the order of 0.25-0.5\%) compared to the amplitude of the reflection effect (5-20\%). Hence, we do not expect that this will have a large impact on the results.

In Table \ref{param_rv} and \ref{param_lc} a summary of all derived parameters can be found. The period derived from the RV curve and the light curve agree very well in most cases. As the \textit{TESS} is covering at least 27~d continuously, the period from the light curve is more trustworthy and the error on the period from the RV curve taken from the literature might be underestimated in some cases. All companions are likely M dwarf companions with masses from 0.088 to $0.5\,\rm M_\odot$. For five systems with published rotational velocities, we could also derive the rotational period by combining the velocity with the inclination and radius derived in paper I ($P_{\rm rot}=\frac{2\pi R \sin i}{v_{\rm rot}\sin i}$). We find that the sdB is rotating significantly (with more than three sigma) slower than the orbital period in three systems.
More details on the individual systems can be found in the appendix in Sect. \ref{refl_indiv}.
In the subsections that follow, we introduce and discuss the newly discovered reflection effect systems.

\subsection{Newly discovered reflection effect systems}

\subsubsection{TYC5977-517-1}\label{tyc}
\begin{figure}
\includegraphics[width=\linewidth]{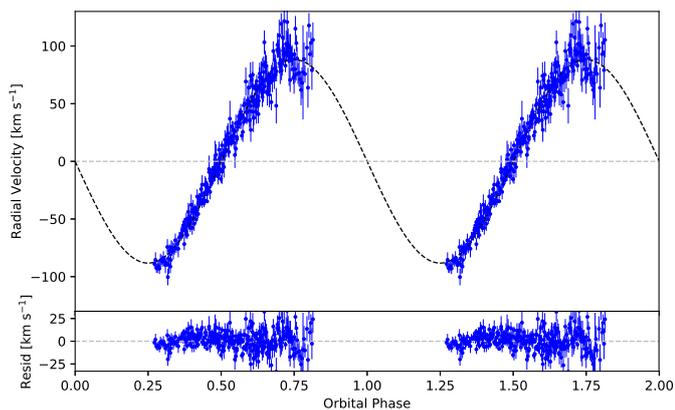}
\caption{Radial velocity curve of TYC5977-517-1 phased to the orbital period. The black line is the best fit model, the blue dots are the data including uncertainties.} 
\label{rv_tyc}
\end{figure}
\begin{figure}
\includegraphics[width=\linewidth]{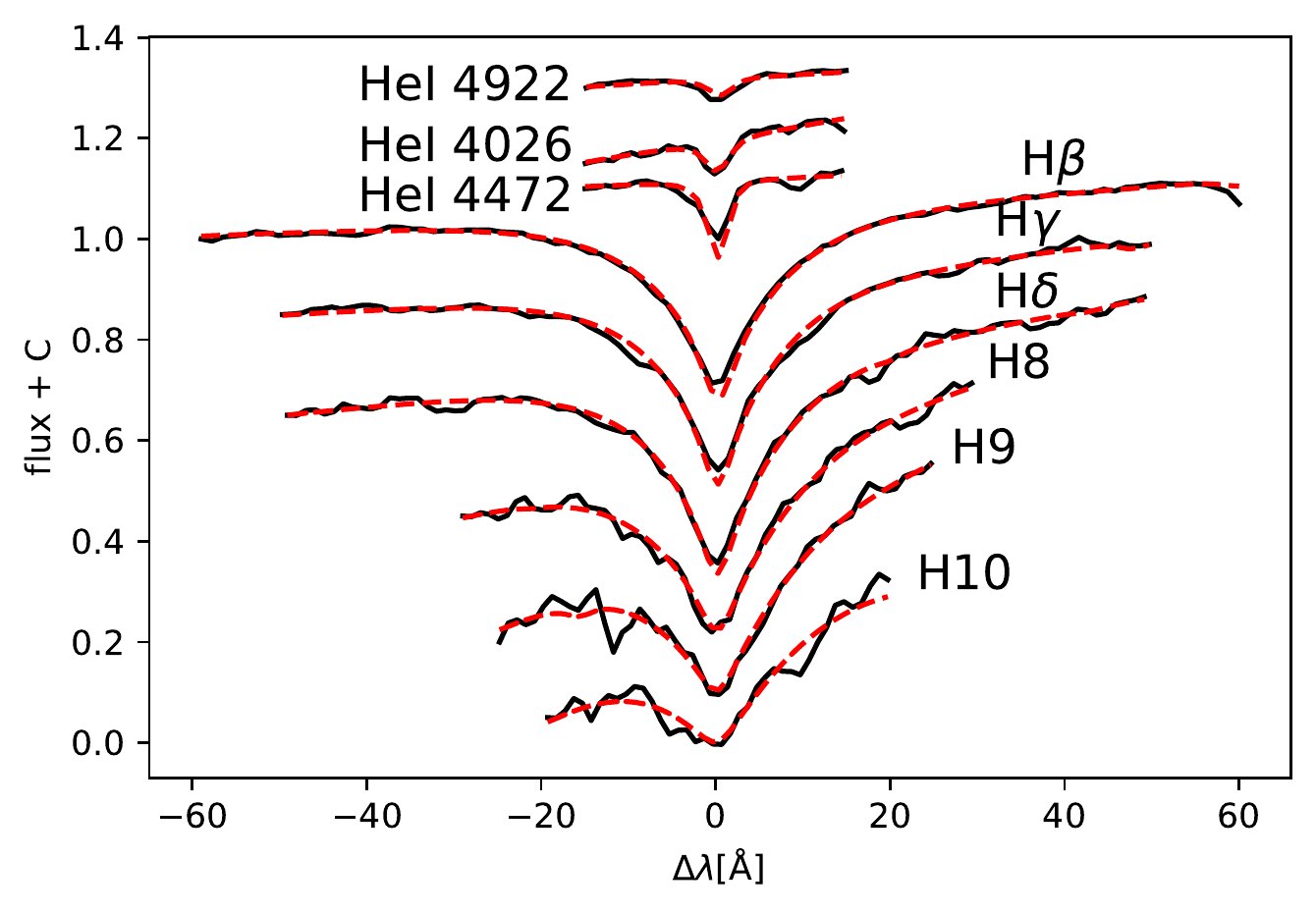}
\caption{Spectral line fit of hydrogen and He lines of the SOAR spectrum of TYC5977-517-1. The best fit is shown in the dashed red line, the black line shows the data. } 
\label{line_tyc}
\end{figure}

\begin{figure}
\includegraphics[width=\linewidth]{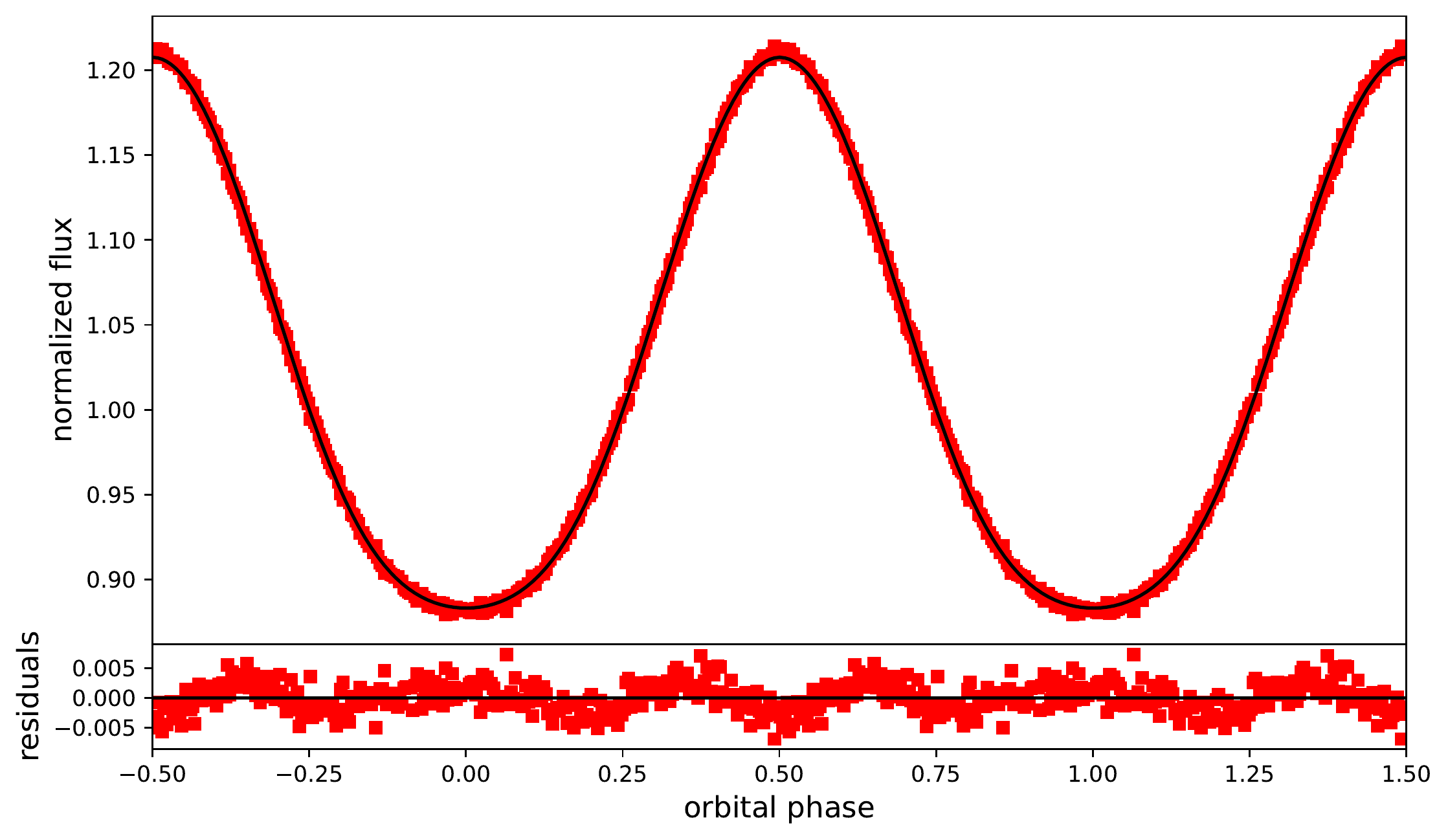}
\caption{Phased \textit{TESS} light curve of TYC5977-517-1 (given by the red squares) together with the best-fit model given by the black line. The lower panel shows the residuals.} 
\label{lc_tyc}
\end{figure}
TYC5977-517-1 was identified as a hot subdwarf candidate in the \textit{Gaia} catalogue of hot subluminous star candidates \citep[][]{geier_gaia_catalog}. It was also reported to be an eclipsing contact binary candidate in the ATLAS survey \citep[][]{heinze18}. By inspecting the \textit{TESS} light curve (see Fig. \ref{lc_tyc}) we discovered it not to be an eclipsing binary but instead a reflection effect binary with a period of 0.14387147 d.  We obtained time--series spectroscopy with the Goodman spectrograph on the SOAR telescope over three consecutive nights in June 2019, taking 67, 120, and 113 spectra each night. Each spectrum had an integration time of 30 s, and each series had a cycle time of $\approx$42 s. Since the target was setting for the season and only visible for the first 1.25 hr, we could only follow it for one-third of its orbit each night. Unfortunately, its nearly integer-value orbital frequency (6.95 $d^{-1}$) meant our starting observing phase did not drift much night to night, and in total our spectra only cover just over half of the orbit. Nonetheless, we successfully derived the RV curve ($K_1=87\pm2\rm\, km\,s^{-1}$, see Fig. \ref{rv_tyc} and Table \ref{RV_tyc}). This was possible due to fact that the orbital period could be determined by photomnetry. To derive the atmospheric parameters we fitted the hydrogen and helium lines of this spectrum with synthetic spectra calculated by a hybrid local thermodynamic equilibrium (LTE)/non-LTE (NLTE) model atmospheric approach as described in \citet[][]{schaffenroth21} using \textsc{spas} \citep[][]{hirsch}. We obtained $T_{\rm eff}=35200\pm500\,\rm K$, $\log g=5.69\pm0.05$, $\log{y}=-2.02\pm0.05$  (the spectral line fit is shown in Fig. \ref{line_tyc}).

\textit{TESS} observed the system in sectors 7, 33 and 34 (CROWDSAP=0.49/0.37/0.34). A significantly higher reflection effect amplitude was observed in the first sector compared to the other two sectors, which were similar, so we excluded the light curve observed in the first sector, as it is probably overcorrected, as discussed in the previous section.
The best fit results in a companion mass of $0.319\pm0.012\,\rm M_\odot$ and a radius of $0.380\pm0.010\,\rm R_\odot$.

\subsubsection{EC01578-1743}
EC01578-1743 was found to be a sdB by the Edinburgh-Cape Blue Object (EC) Survey \citep{Kilkenny2016}. This system was identified as a reflection effect system in the Evryscope survey \citep[][]{Evr03}. Inspecting the \textit{TESS} light curves, we also found this system to have a strong reflection effect with an amplitude of 20\% and period of 0.258104 d. Its shape indicated a higher inclination angle, but no eclipses are visible in the data.

\begin{figure}
\includegraphics[width=\linewidth]{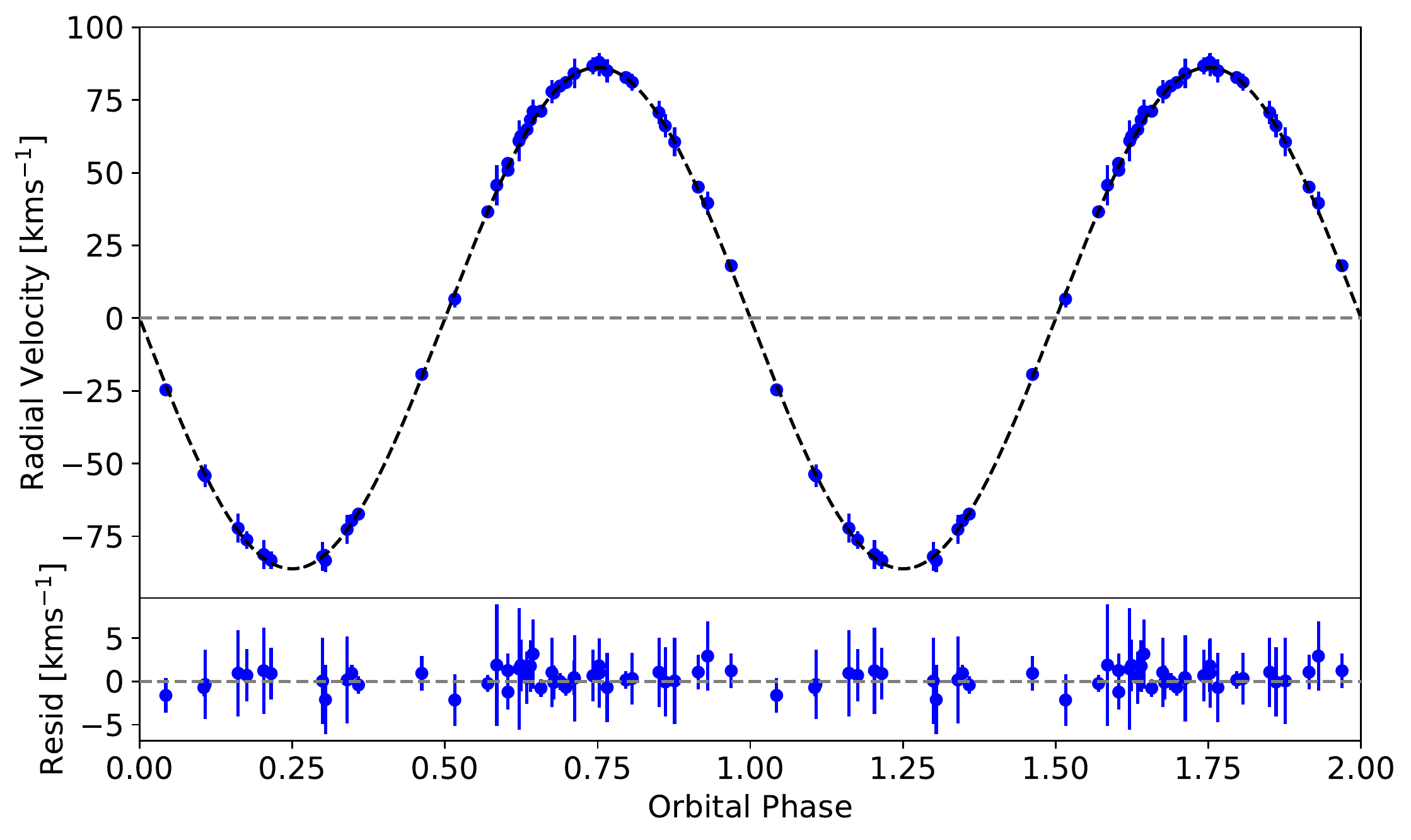}
\caption{Radial velocity curve of EC01578-1743 phased to the orbital period. The black line is the best fit model, the blue dots are the data including uncertainties.} 
\label{rv_ec}
\end{figure}

\begin{figure}
\includegraphics[width=\linewidth]{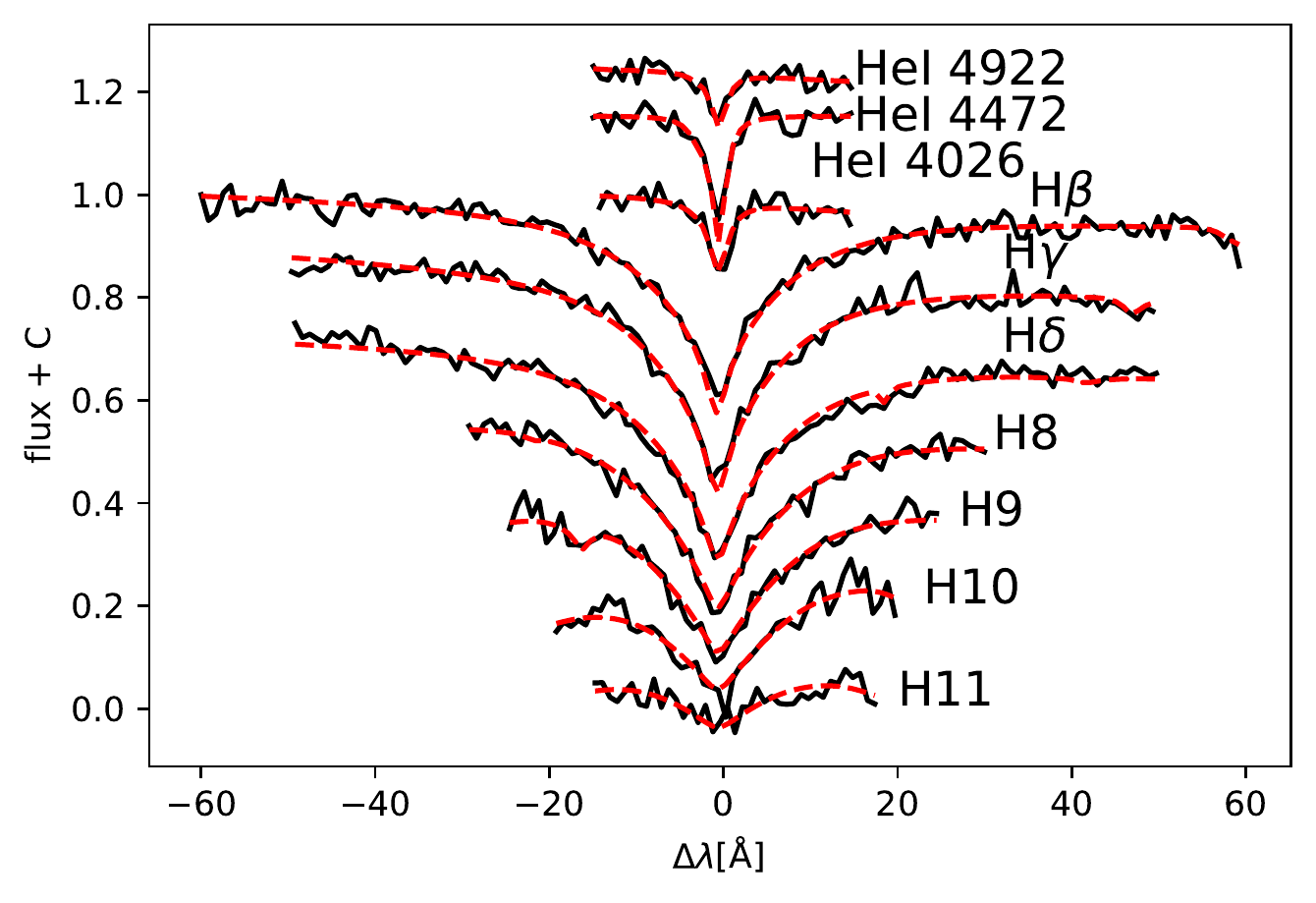}
\caption{Spectral line fit of hydrogen and He lines of the SOAR spectrum of EC01578-1743. The best fit is shown in the dashed red line, the black line shows the data. } 
\label{line_ec}
\end{figure}

In order to derive the radial velocity curve, we obtained high--resolution spectra using the CHIRON echelle spectrometer on the CTIO 1.5--m telescope \citep{tokovinin13}. Observations were taken at sporadic intervals from December 2017 to September 2018 (in total 39 single spectra) and cover the full range of orbital phases (see Fig. \ref{rv_ec} and Table \ref{RV_ec}). A 2.7\arcsec\ fiber was used to cover the wavelength range 4400--8800 \AA\ with a spectral resolution of R$\approx$28000. Extracted and wavelength-calibrated spectra were delivered by a pipeline running at Georgia State University \citep{brewer14}. In order to measure the radial velocities we used cross-correlation with the \textsc{iraf} task \textsc{fxcor}. To fit the radial velocity curve we used the Python package \textsc{radvel} \citep[][]{radvel}\footnote{\url{https://radvel.readthedocs.io/}} getting a semi-amplitude of the radial velocity curve of $K_1=86.5  \pm  0.5 \,\rm km\,s^{-1}$ (see Fig. \ref{rv_ec}). To derive the atmospheric parameters we additionally took one spectrum with SOAR/Goodman. The analysis was done the same way as for TYC5977-517-1 and resulted in $T_{\rm eff}=32000\pm500\,\rm K$, $\log g=5.75\pm0.06$, $\log{y}=-2.0\pm0.1$ (the spectral line fit is shown in Fig. \ref{line_ec}).


\begin{figure}
\includegraphics[width=\linewidth]{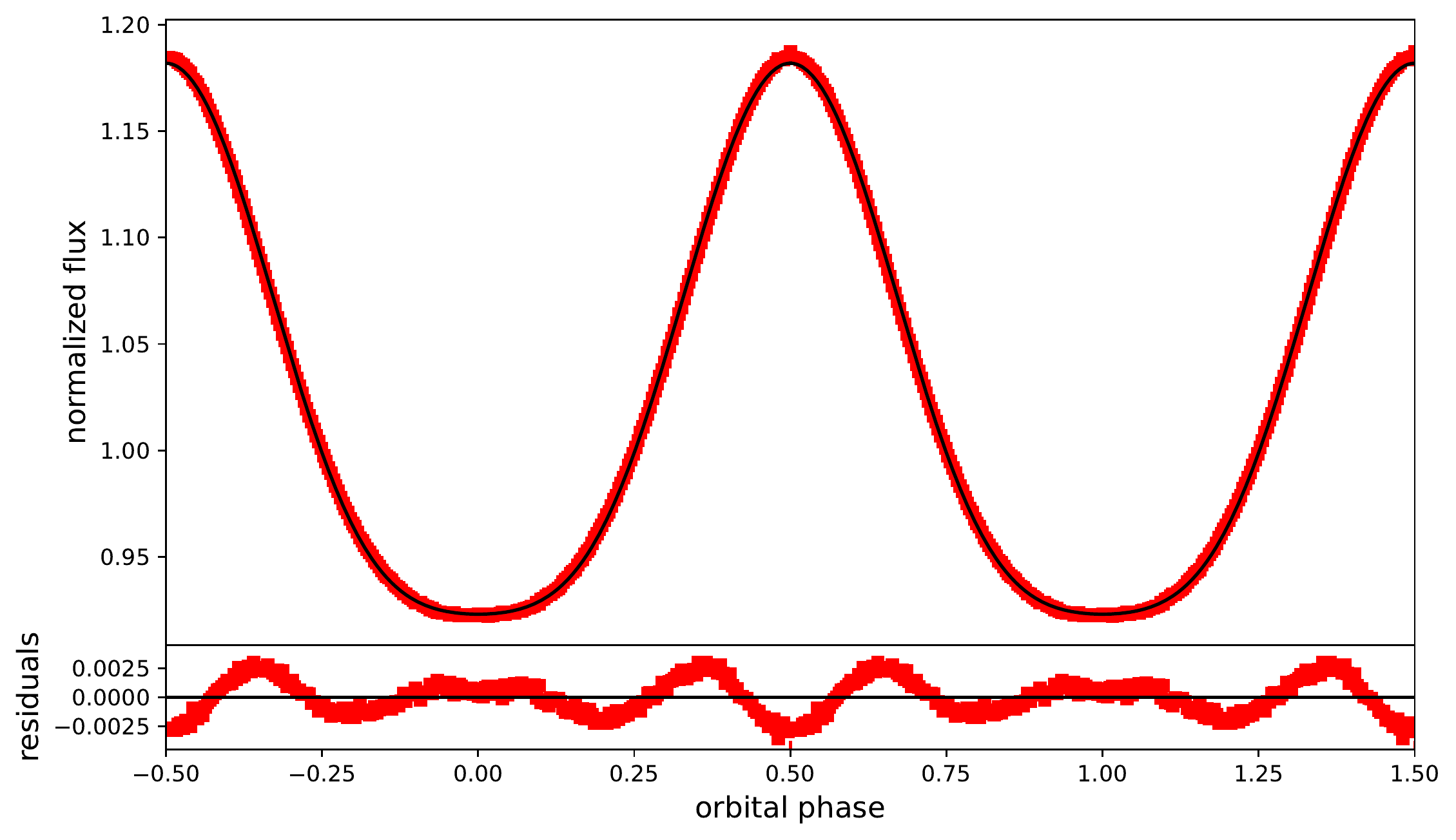}
\caption{Phased \textit{TESS} light curve of EC 01578-1743 (given by the red squares) together with the best-fit model given by the black line. The lower panel shows the residuals.} 
\label{lc_ec01678}
\end{figure}

The best fit of the light curve (see Fig. \ref{lc_ec01678}) was found for an inclination of $49.5^\circ\pm0.25^\circ$. From this we can constrain the mass and radius of the companion to $0.278\pm0.004\,\rm M_\odot$ and $0.294\pm0.0025\,\rm R_\odot$.


\subsubsection{KPD2215+5037}
\begin{figure}
    \centering
\includegraphics[width=1.05\linewidth]{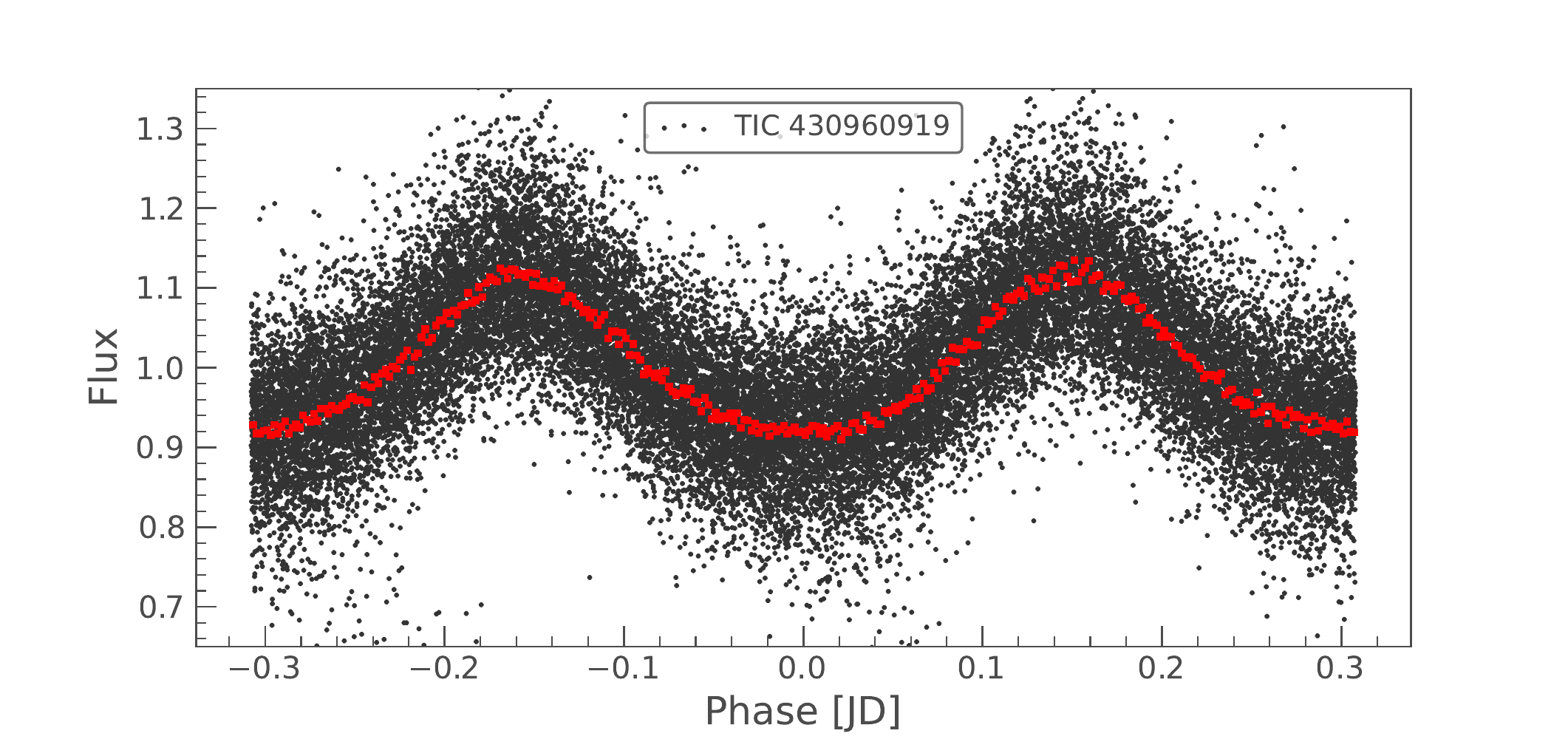}
    \caption{TESS light curve of KPD2215+5037 phase-folded to the dominant peak in the periodogram determined by the light curve.}
    \label{lc_kpd}
\end{figure}

KPD2215+5037 was identified as a subdwarf by \citet[][]{kpd} in the Kitt Peak-Downes Survey for Galactic Plane Ultraviolet-Excess Objects. A survey for RV variable hot subdwarfs by \citet[][]{Copperwheat11} found it to be varying with a period of 0.809146 d. The \textit{TESS} light curve shows a variation at 0.3078784 d with the typical shape of an reflection effect (Fig. \ref{lc_kpd}). An additional sinusoidal variation at 6.5 d is visible. However, this signal probably originates from a known red, long-period variable 25 arcsec away. To confirm that the 0.3 d signal is coming from our target, we also extracted the light curve from the \textit{TESS} fullframe images (FFI) from the single pixel, which should not be influenced by the brighter target. We confirm the 0.3 d signal most likely comes from our target, and that the longer-period variation does not. This is also confirmed using the Python package \textsc{tess-localize}\footnote{\url{https://github.com/Higgins00/TESS-Localize}} \citep[][]{localize}. To confirm the light curve period, additional time-resolved spectroscopy and photometry should be taken in the future.

\hfill

\subsubsection{GALEX J1753-5007 -- A triple system?}
\begin{figure}
    \centering
    \includegraphics[width=\linewidth]{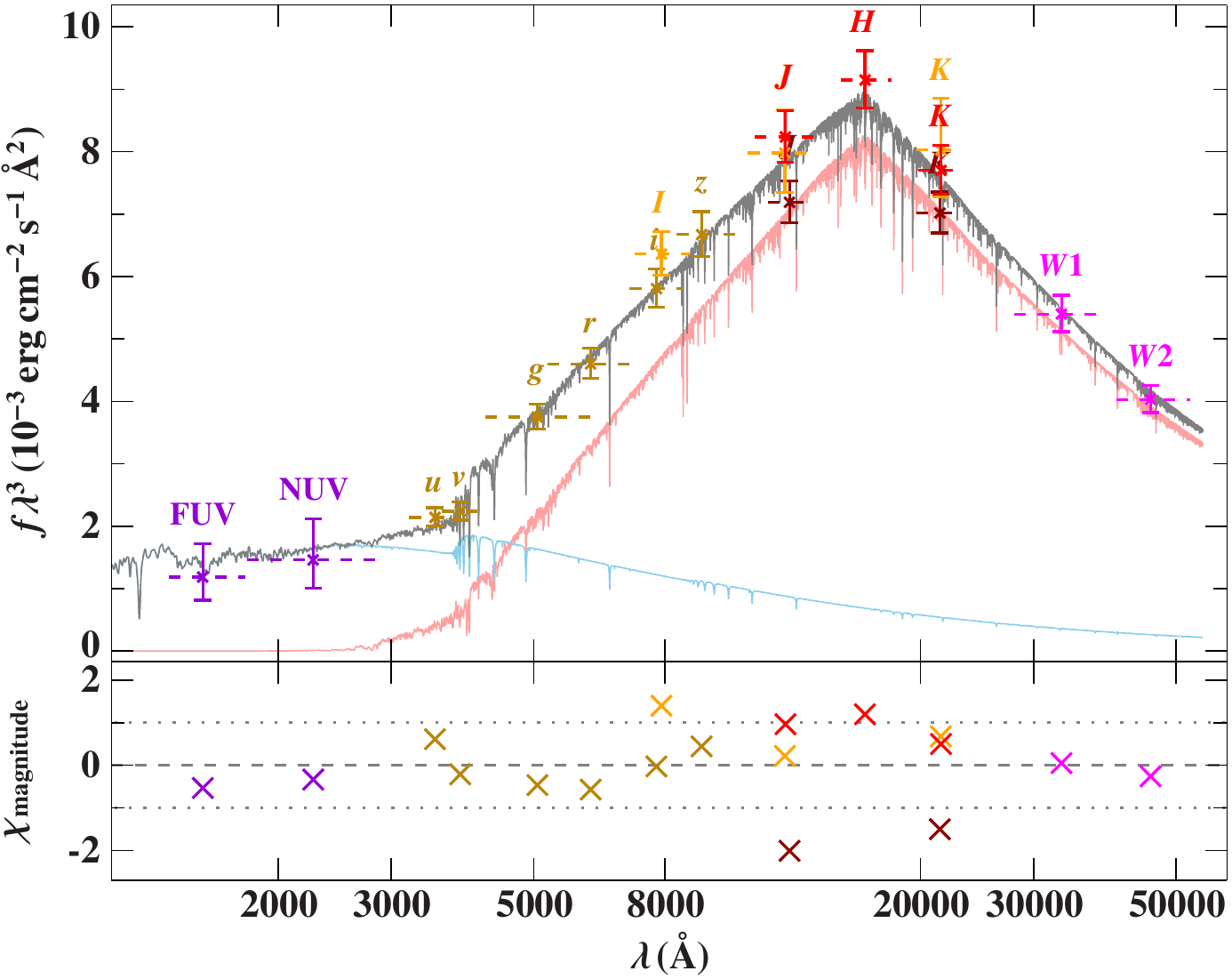}
    \caption{Spectral energy distribution of GALEX J1753-5007 showing a sdB with a F7V companion.}
    \label{sed_galex}
\end{figure}

GALEX J1753-5007 (GALEX J175340.5-500741) was discovered in the GALEX survey and classified as a sdB with a F7V companion due to an infrared SED excess by \citet[][]{nemeth:2012}. \citet[][]{kawka:2015} took spectroscopic follow-up of this target confirming it to be RV variable and hence in a close binary system. As they could not find any variation with an upper limit of 20 mmag in the ASAS light curve, they suggested that the companion is a WD. We can confirm from fitting the SED the same way as described in \citet[][]{heber:2018,sed_andreas} and paper I (see Fig. \ref{sed_galex}) that it is a sdB with a F type companion ($T_{\rm eff,2}=6000^{+400}_{-250}\,\rm K$).

\begin{figure}
    \centering
\includegraphics[width=1.05\linewidth]{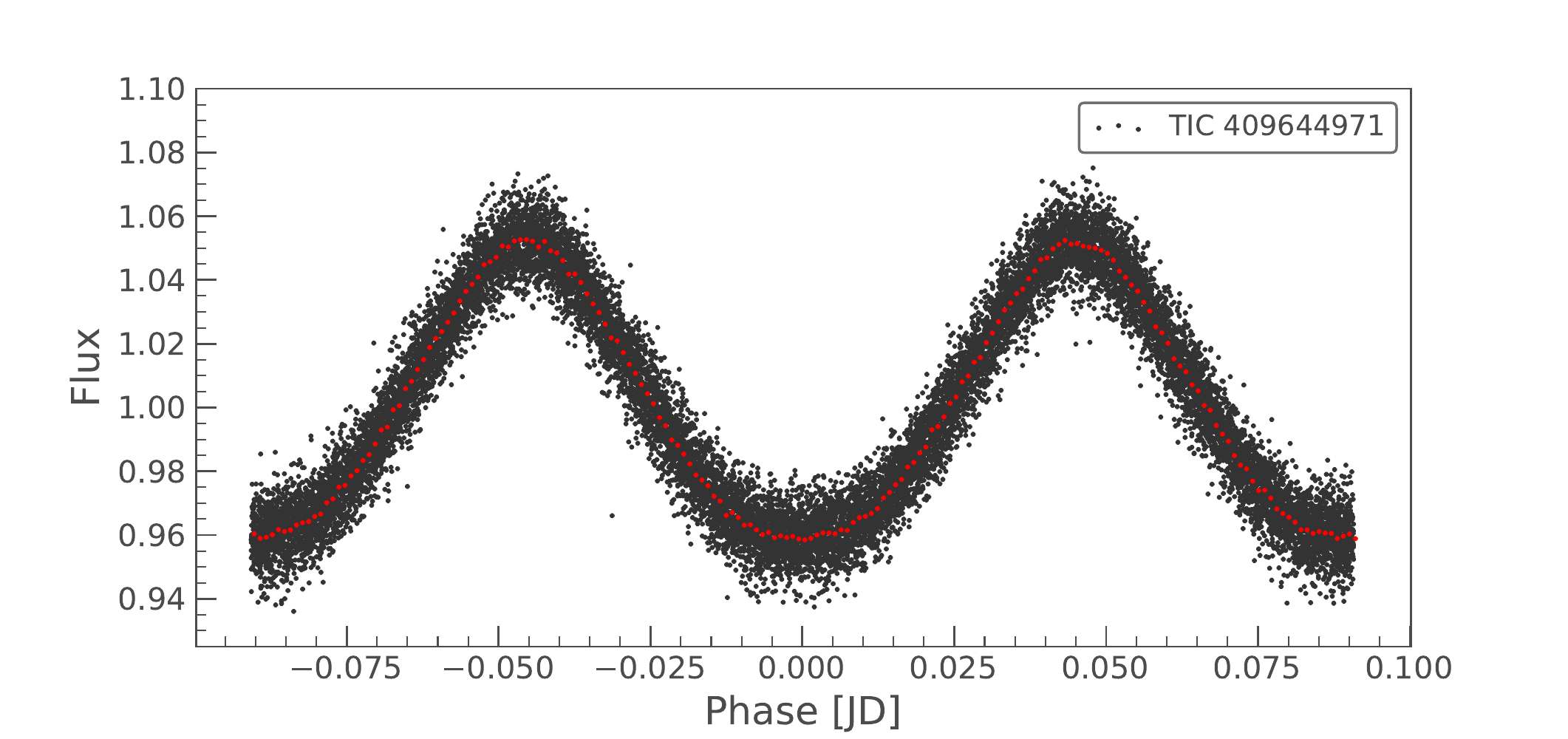}
    \caption{TESS light curve of GALEX J1753-5007 phase-folded to the dominant peak in the periodogram determined by the light curve.}
    \label{lc_galex1753}
\end{figure}\hfill


The \textit{TESS} light curve (Fig. \ref{lc_galex1753}) shows clearly a reflection effect with an amplitude of about 6\% with a period of 0.0907405 d. As the \textit{TESS} filter is much redder than the ASAS filter, the amplitude in \textit{TESS} is expected to be significantly higher. That could explain why the ASAS light curve did not show any variation. For an sdB + FV star, such a short period is not possible as the F star would be larger than the orbital separation. So it is most likely that an inner binary with a cool, low-mass companion is being orbited by a F star in a wide orbit and this is actually a triple system. Only very few confirmed sdBs in triple systems are known \citep[see e.g.][]{pelisoli:2020}. The astrometric orbits, which will be released by \textit{Gaia} eventually, could confirm this. 
The large RUWE of 31.846 already indicates that a single-star solution is not a good fit, suggesting non-negligible astrometric wobble that could be caused by a longer period companion.

\section{Hot subdwarfs with white dwarf companions showing ellipsoidal deformation and/or beaming in their light curves}\label{wd}

\subsection{Method}
While searching for light variations of the hot subdwarf binaries with known orbital periods, we detected several targets showing small variations with half of the orbital period and uneven minima/maxima. The most probable explanation for these variations is ellipsoidal deformation of the hot subdwarf due to a nearby white dwarf companion. We also found several systems showing tiny variations ($\sim 0.01\%-0.1\%$) with the orbital period that are most likely due to Doppler beaming of the hot subdwarf, also indicative of a nearby white dwarf companion. No ellipsoidal deformation is observed in these cases because the separation of the components is too large (more details in the next subsection). Most of these variations would not have been found without previous knowledge of the orbital period since their periodogram peaks are barely visible above the noise. 

To confirm the ellipsoidal deformation and/or beaming in the light curves, we fitted them using \textsc{lcurve} as we did for the reflection effect systems. We assumed the mass and radius of the sdB as determined by the SED fitting and \textit{Gaia} parallax (see paper I), and used the RV semi-amplitude to derive the orbital separation from the mass function, and the atmospheric parameters from the spectral analysis to constrain as many parameters as possible. For the limb-darkening, gravity darkening and beaming coefficients, we used the values closest to the atmospheric parameters of the hot subdwarf from the tables of \citet[][]{claret_beaming,claret_limb} for the \textit{TESS} and \textit{Kepler} filters. 

Similar to the analysis of the reflection effect systems as shown in Sect. \ref{reflection}, we performed an MCMC varying the inclination as well as the mass ratio and the radius of the sdB using a prior to include the uncertainties of both parameters.

\subsection{Results of the light curve analysis}
\begin{table*}[h]
\setlength{\tabcolsep}{2pt}
    \caption{Period, RV curve parameters, inclination, mass ratio, separation, and companion mass of the analysed ellipsoidal systems together with the minimum companion mass. The objects are ordered following their orbital period.}
    \label{ell_result}
    \begin{tabular}{llllllllll}
\hline\hline
target&$P_{\rm RV}$&$\gamma$&$K_1$&$i$&$q$&$a$&$M_2$&$M^*_{2,\rm min}$&$P_{\rm rot}$\\
&[d]&[km/s]&[km/s]&[$^\circ$]&&[$\rm R_\odot$]&[$\rm M_\odot$]&[$\rm M_\odot$]&[d]\\
\hline
PG1043+760$^a$ & 0.1201506 & 24.80 & 63.60 & $15\pm0.6$ & $1.65\pm0.11$ & $0.94\pm0.04$ & $0.48\pm0.08$ & 0.09 &-\\
GALEXJ075147.0+092526$^b$ & 0.178319 & 15.50 & 147.70 & $74\pm10$ & $0.85^{+0.09}_{-0.04}$ & $1.19\pm0.08$ & $0.31^{+0.07}_{-0.03}$ & 0.31 &-\\
HS1741+213$^a$ & 0.2 & - & 157 & $47\pm11$ & $1.45^{0.65}_{0.3}$ & $1.4\pm0.3$& $0.58_{-0.15}^{+0.3}$ & 0.36 &-\\
PG1136-003$^a$ & 0.207536 & 23.30 & 162.00 & $75\pm11$ & $0.90^{+0.10}_{-0.04}$ & $1.4\pm0.1$ & $0.45^{+0.08}_{-0.04}$ & 0.38 &-\\
GD687$^a$ & 0.37765$^a$ & 32.30 & 118.30 & $58\pm8$ & $1.23^{+0.24}_{-0.14}$ & $1.9\pm0.2$ & $0.35^{+0.09}_{-0.06}$ & 0.32&$0.39 \pm 0.05$ \\
GALEXJ234947.7+384440$^a$ & 0.462516 & 2.00 & 87.90 & $70\pm10$ & $0.64^{+0.08}_{-0.04}$ & $2.2\pm0.2$ & $0.26_{-0.04}^{+0.04}$ & 0.24 &-\\
PG0101+039$^a$ & 0.569899 & 7.30 & 104.70 & $89.4\pm0.6$ & $0.8174^{+0.0001}_{-0.0001}$ & $2.53\pm0.01$ & $0.34^{+0.04}_{-0.04}$ & 0.33 & $0.85 \pm 0.09$\\
EC13332-1424$^a$ & 0.82794 & -53.20 & 104.10 & $82\pm2$ & $1.0^{+0.1}_{-0.1}$ & $3.4\pm0.2$ & $0.40^{+0.06}_{-0.06}$ & 0.39 &-\\
PG1232-136$^a$ & 0.363 & 4.10 & 129.60 & - & - & - & - & 0.36 &-\\
PG1743+477$^a$ & 0.515561 & -65.80 & 121.40 & - & - & - & - & 0.39 &-\\
PG1519+640$^a$ & 0.54029143 & 0.10 & 42.70 & - & - & - & - & 0.10 &-\\
GALEXJ025023.8-040611$^b$ & 0.6641 & 0.00 & 93.90 & - & - & - & - & 0.30 &-\\
PG1648+536$^a$ & 0.6109107 & -69.90 & 109.00 & - & - & - & - & 0.36 &-\\
EC22202-1834$^a$ & 0.70471 & -5.50 & 118.60 & - & - & - & - & 0.44 &-\\
EC02200-2338$^a$ & 0.8022 & 20.70 & 96.4 & - & - & - & - & 0.35 &-\\
TONS183$^a$ & 0.8277 & 50.50 & 84.80 & - & - & - & - & 0.29 &-\\
EC21556-5552$^a$ & 0.834 & 31.40 & 65.00 & - & - & - & - & 0.21 &-\\
PG1000+408$^a$ & 1.049343 & 56.60 & 63.50 & - & - & - & - & 0.22 &-\\
GALEXJ225444.1-551505$^b$ & 1.22702& 4.20 & 79.70 & - & - & - & - & 0.32 &-\\
PG0133+114$^a$ & 1.23787 & -0.30 & 82.00 & - & - & - & - & 0.34 &-\\
PG1512+244$^a$ & 1.26978 & -2.90 & 92.70 & - & - & - & - & 0.41 &-\\
UVO1735+22$^a$ & 1.278 & 20.60 & 103.00 & - & - & - & - & 0.48 &-\\
PG0934+186$^a$ & 4.051 & 7.70 & 60.30 & - & - & - & - & 0.38 &-\\
CD-24731$^a$ & 5.85 & 20.00 & 63.00 & - & - & - & - & 0.50 &-\\
\hline
    \end{tabular}
    
    $^a$ \citet[][and references therein]{kupfer:2015}\qquad $^b$ \citet[][]{kawka:2015}  \qquad $^*$ under the assumption: $M_{\rm sdB}=0.4\,\rm M_\odot$
\end{table*}

 All light curves can be found in Fig. \ref{ell1} along with their best-fitting models, which agree well with the data. For the systems showing ellipsoidal modulation and Doppler beaming we were able to derive inclinations and hence also the mass of the companion. The results are summarized in Table \ref{ell_result}. All but two companions are more likely He WDs rather than CO WDs. We also could derive the rotational velocity of the sdB in two systems the same way as for the sdB+dM systems. In one system the sdB seems to rotate a bit slower than the orbital rotations, which has implications on the light curve. More details can be found in Sect. \ref{pg0101}.

The systems showing only Doppler boosting are not sensitive to the mass or inclination, as shown in Sect. \ref{pg1232-136}, so we are only overplotting the synthetic light curve calculated using the sdB radius and mass, as well as the mass ratio and orbital separation derived by the RV semi-amplitude and orbital period from the light curve to show that the variation can indeed be explained by beaming (Fig. \ref{ell1} i-x). 
In total we could detect Doppler beaming in 16 sdB binaries with periods from 9 hours to 5. days. 
In the next sections we discuss two newly confirmed sdB+WD systems, and in appendix \ref{ell_indiv}, we provide more details on some of the individual sdB+WD systems.

\subsection{Newly confirmed sdB+WD systems}
\subsubsection{PG 1232-136}
\label{pg1232-136}

PG 1232-136 was found to be a sdB star in the Palomar-Green survey \citep[][]{green86}. Spectroscopic follow-up by \citet[][]{edelmann:2005} revealed that it is in a close binary system with 0.3630 d period and a quite large RV amplitude ($129.60\pm0.4\,\rm km\,s^{-1}$). \citet[][]{geier:2010} constrained the rotational velocity of the sdB to $v_{\rm rot}\sin i < 5\,\rm km\,s^{-1}$. Assuming synchronization they derived the minimum mass of the companion to be $6\,\rm M_\odot$. As the companion is not visible in the spectrum, they claimed it to be a black hole candidate.

\begin{figure}
\includegraphics[width=\linewidth]{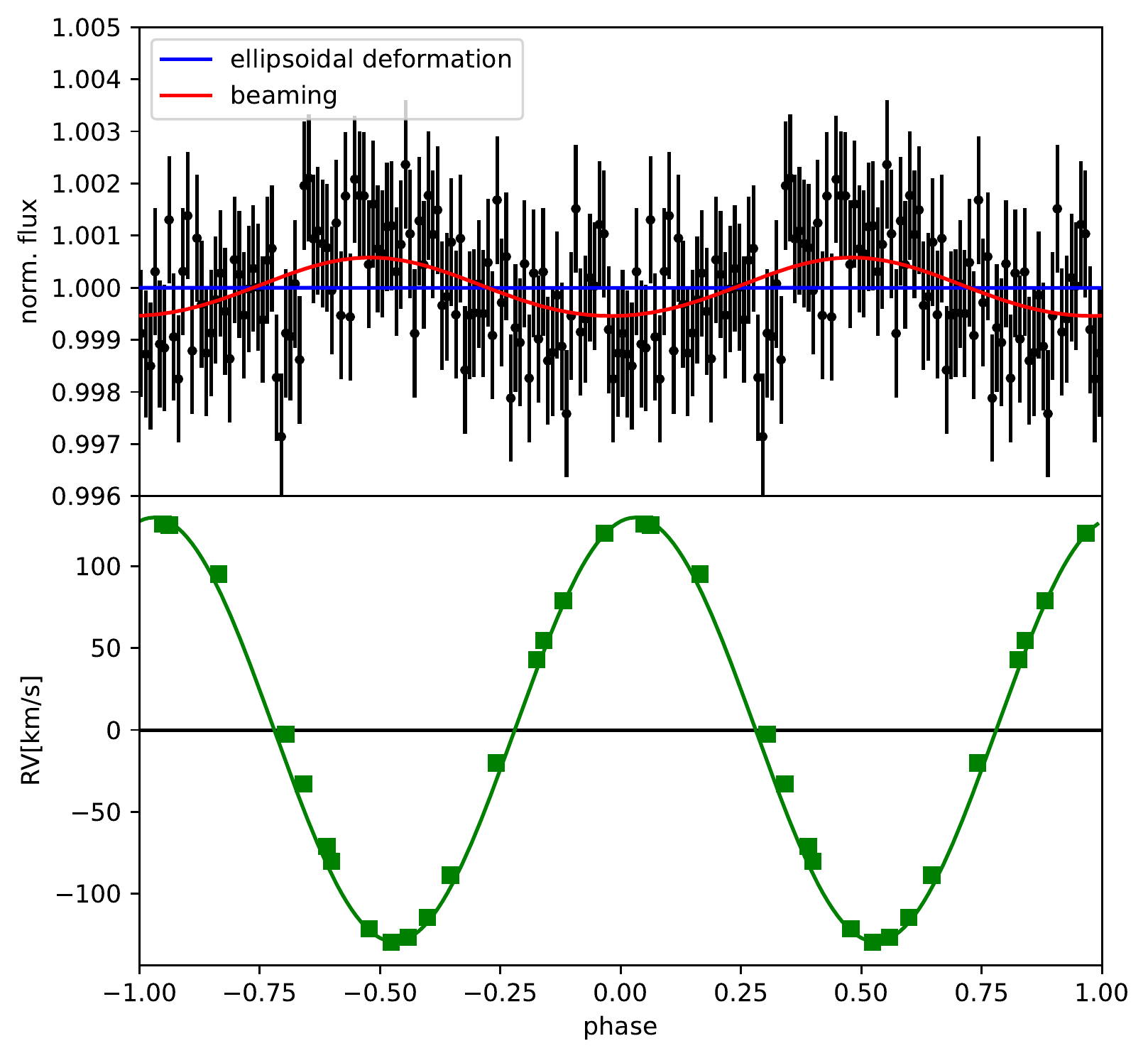}
\caption{Phased \textit{TESS} light curve and RV curve of PG 1232-136. The upper panel shows the \textit{TESS} light curve phased with the period determined from the \textit{TESS} light curve. The lower panel shows the RV curve measured by \citet[][]{edelmann:2005} phased with the same period.}
\label{rv_pg1232}
\end{figure}

The \textit{TESS} light curve showed a tiny variation with an amplitude of only about 0.2\% amplitude with the orbital period derived be the RV curve suggesting a reflection effect. The light curve analysis showed that the variation could be explained by a reflection effect with the size and mass of a He-WD companion. Phasing the RV curve and the light curve to the same ephemeris however showed that the variation is not a reflection effect but most likely Doppler beaming resulting from the high velocity of the sdB, as the light variation is aligned with the RV variation and we observe the highest flux when the sdB is moving towards us (see Fig. \ref{rv_pg1232}). To check if this can help us constrain the nature of the companion, we calculated several light curve models with different inclinations. Unfortunately, even at lower inclinations (i.e., higher companion masses), one would not expect to detect the ellipsoidal deformation of the sdB (see Fig. \ref{rv_pg1232}). Hence, we observe only the beaming of the sdB, which varies with the radial velocity curve, and thus the light curve does not give us any additional information. If we detect the same signal in an unknown sdB binary, we could predict the amplitude of the RV curve however. This also means that it is not possible to constrain the mass of the companion, as we cannot constrain the inclination.

\subsubsection{KPD 0629-0016}
\begin{figure}
    \centering
    \includegraphics[width=\linewidth]{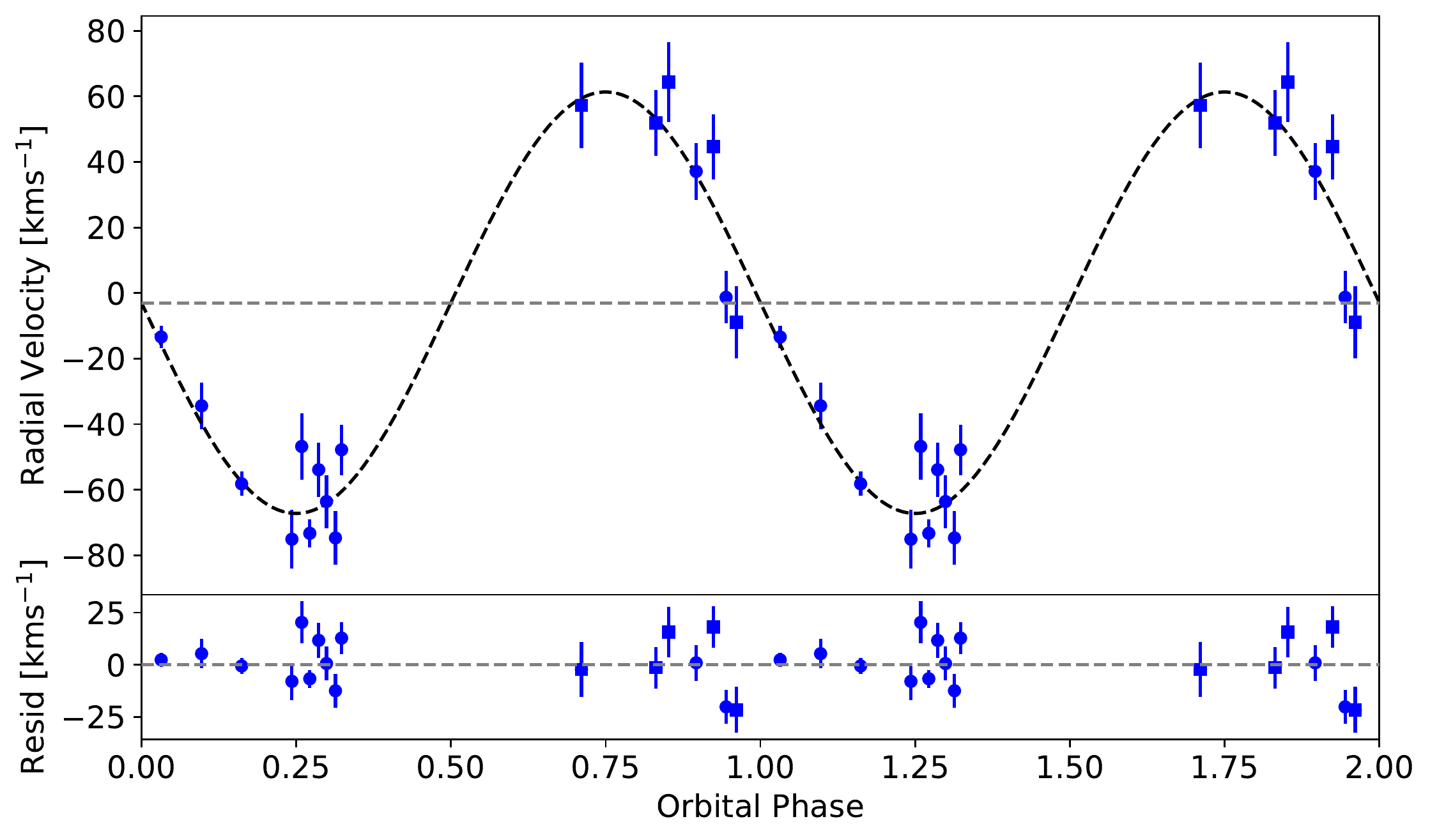}
    \caption{RV curve of KPD 0629-0016 phased to the most probable orbital period with the best-fitting RV model curve shown with the black sinusoidal curve. The data observed with the EMMI spectrograph is shown with the blue circles, the EFOSC2 data with the blue squares. The lower panel shows the residuals. }
    \label{rv_kpd}
\end{figure}

KPD 0629-0016 was first discovered to be a slowly pulsating sdB star by \citet[][]{koen:2007}. The observation of the sdB by the CoRoT (COnvection, ROtation, and planetary Transits) satellite \citep[][]{corot} opened a  new era in sdB asteroseismology leading to the detection of a large number of g-mode pulsations \citep[][]{charpinet:2010}. This rich spectrum could be used to derive the structural and core parameters of the sdB \citep[][]{vangrootel:2010}. An additional binary signal could not be found in the CoRoT data, however as a lot of binary systems have orbital periods in the same range as the g-mode pulsations, so it is not easy to find them in the light curve. We took spectroscopic follow-up of this sdB to search for RV variations in three runs with the EMMI and EFOSC2 spectrograph mounted at the ESO/NTT telescope in Chile (080.D-0685(A), 082.D-0649(A), 092.D-0040(A), PI: S. Geier). More details to the observations and the RV determination can be found in \citet[][]{geier:2014}. The RV curve phased to the most probable orbital period ($0.8754\pm0.0001$ d) can be found in Fig. \ref{rv_kpd} and is resulting in a semi-amplitude of $K_1=64.4\pm3.4\,\rm km/s$, from which a minimum mass of $0.22\,\rm M_\odot$ can be derived for the companion. As in the CoRoT light curve no period near the orbital period could be detected, a M dwarf companion can most likely be excluded and the companion has to be a WD. The system was also observed by \textit{TESS} in sector 6 and 33. The analysis of the light curve also shows no detectable period close to the orbital period with an upper limit of 1.0\% confirming that the companion is most likely a WD, but the quality of the light curve is not high enough to detect light variations.

\section{Discussion and summary}

\begin{figure}
    \centering
    \includegraphics[width=\linewidth]{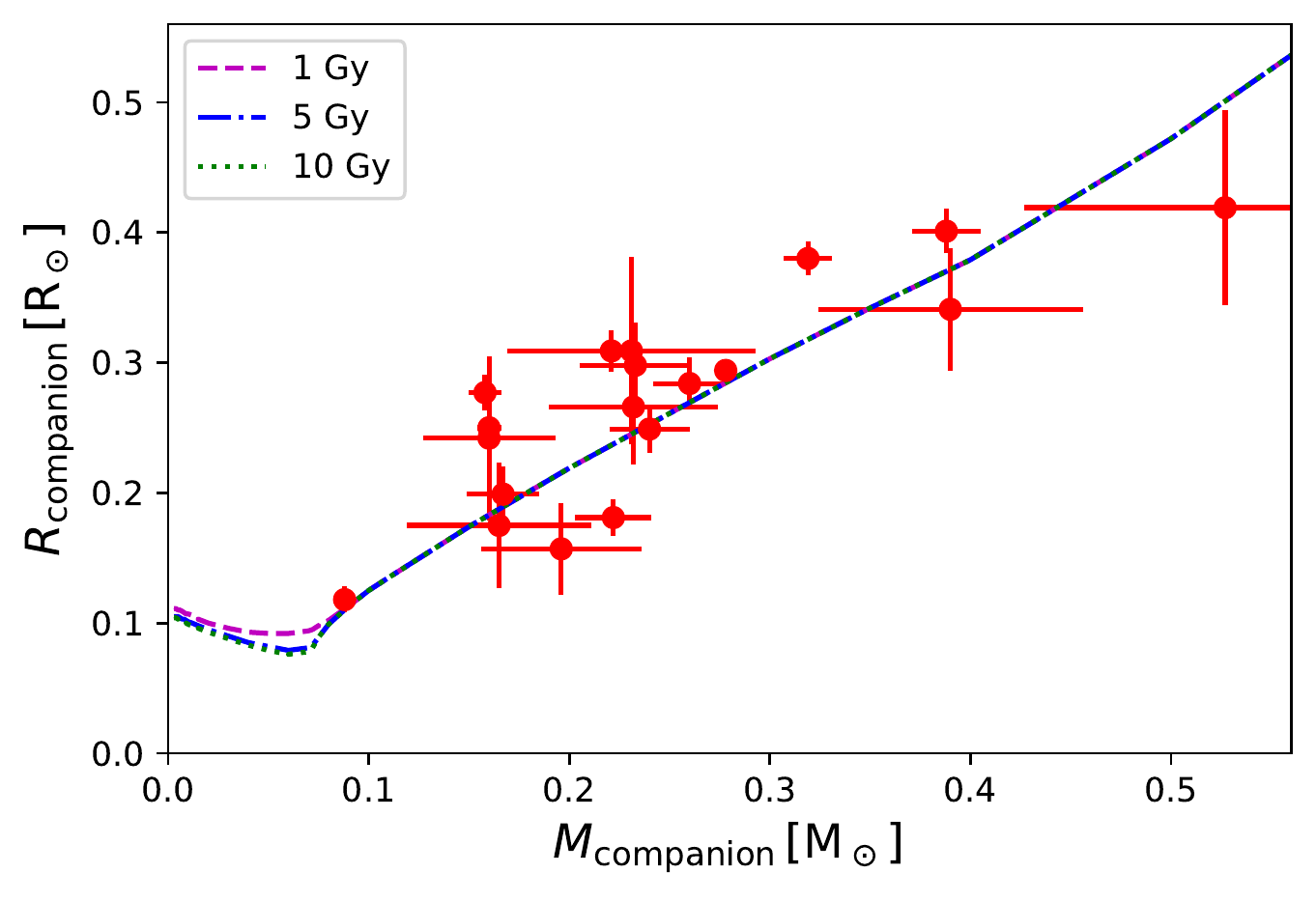}
    \caption{Mass-radius relation of the companions in the analyzed reflection effect systems compared to theoretical calculations for an age from 1 to 10 Gy by \citet[][]{baraffe:15}.}
    \label{m-r}
\end{figure}
\begin{figure}
    \centering
    \includegraphics[width=\linewidth]{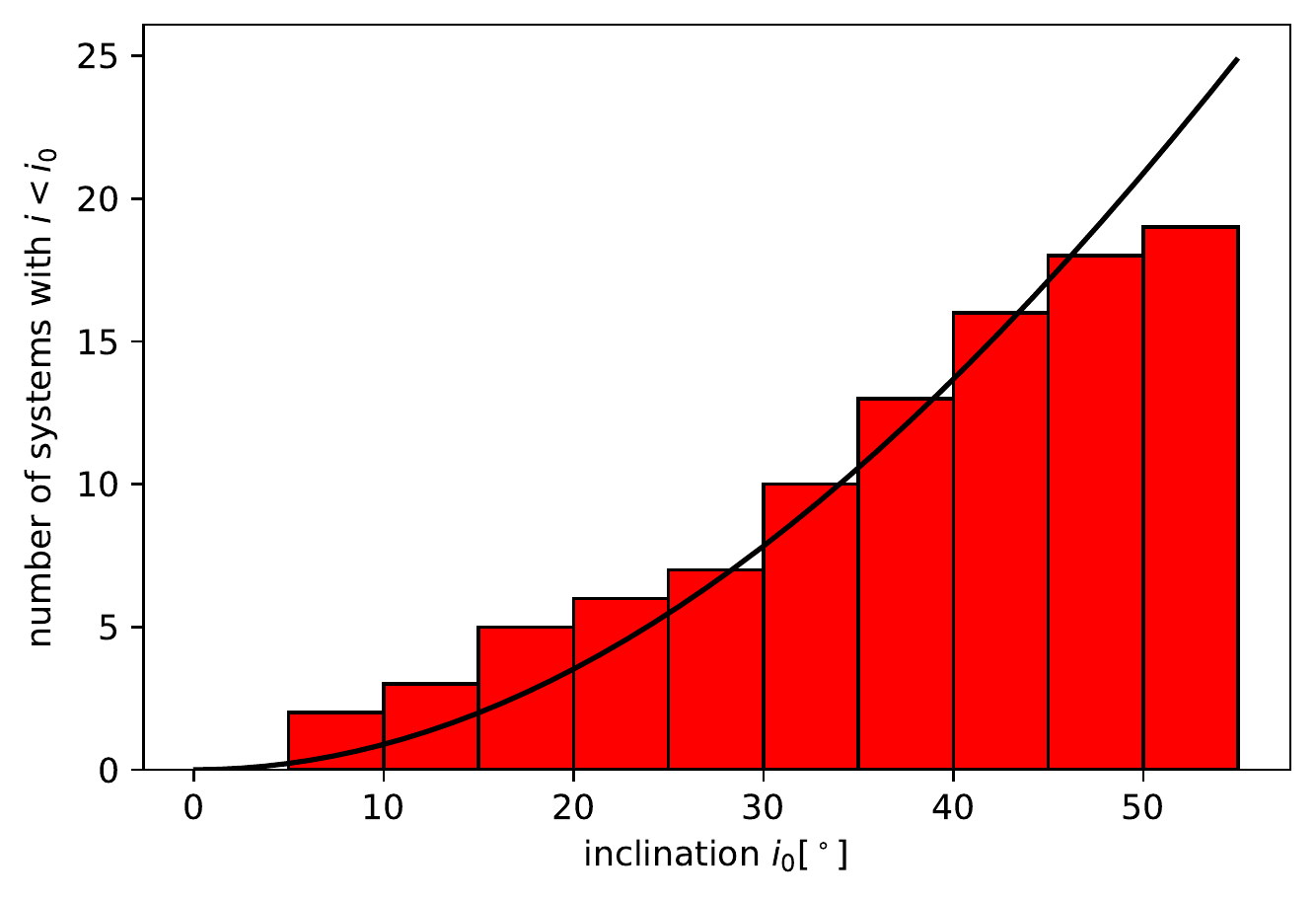}
    \caption{Inclination distribution of the analyzed reflection effect systems. The black line shows the number of systems we expect, when we assume that the orientation of a sdB binary is uniformly distributed. Due to the projection effect it is much more
likely to find binary systems at high rather than low inclinations.}
    \label{incl}
\end{figure}

\begin{figure}
    \centering
    \includegraphics[width=\linewidth]{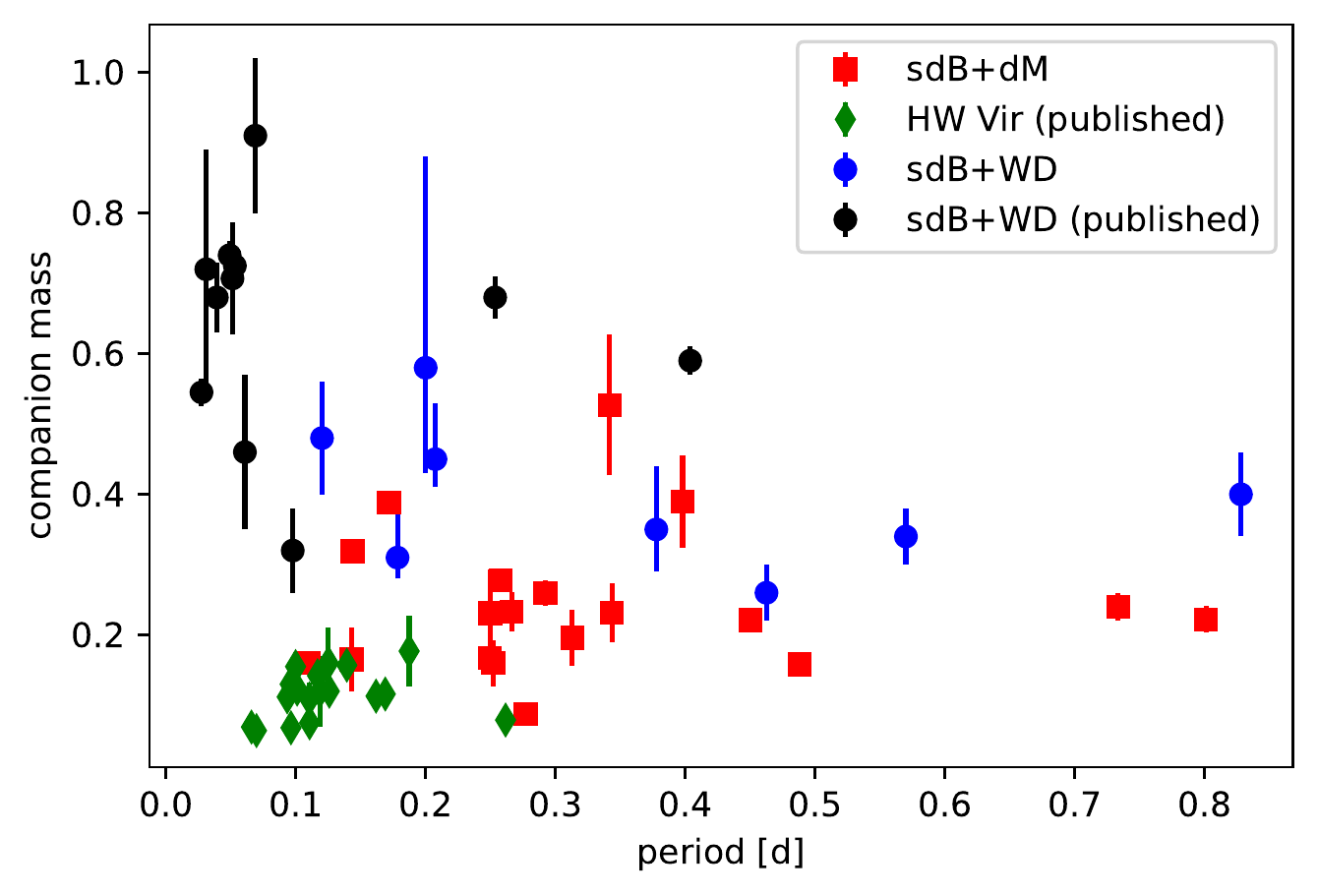}
    \caption{Period vs companion mass for the reflection effect systems (sdB+dM, with red squares) and the ellipsoidal modulation systems (sdB+WD, with blue circles). In comparison we show the parameters of the published reflection effect systems \citep[see][and references therein]{schaffenroth18} as well as the published sdB+WD systems: KPD1946+4340 \citep[][]{bloemen11}; CD-3011223 \citep[][]{cd-30}; PTF1J0823+0819 \citep[][]{kup17a}; OWJ074106.0–294811.0 \citep[][]{kup17}; EVR-CB-001 \citep[][]{evr01}; EVR-CB-004 \citep[][]{evr04}; HD265435 \citep[][]{pelisoli21}; ZTFJ2130+4420 \citep[][]{kupfer20}; ZTFJ2055+4651 \citep[][]{kupfer20b}; PTF1J223857.11+743015.1 \citep[][]{kupfer22}; OWJ081530.8-342123.5 \citep[][]{ramsay22}}
    \label{mass_comp}
\end{figure}

For the first time we analyze a larger sample of reflection effect systems and derive the masses and radii of the companion under the assumption of a canonical mass sdB, which is the most likely mass as shown by the mass distribution of the sdBs in HW Vir systems (see paper I), as well as the radius of the sdB derived by the fit of the SED and the Gaia parallax (paper I). 

To check the validity of our method we also compared the mass and radius we derive for the companion to theoretical mass-radius relations by \citet[][]{baraffe:15}. This is shown in Fig. \ref{m-r}. There is some scatter, which is also found by other investigations \citep[e.g.][]{parsons18}, but most of the companions agree well with the theoretical predictions. This shows that our assumption of the canonical mass is not so far off and our method works quite well. A change in the assumed mass of the sdB will lead to a systematic shift of the companion mass. In the future the sdB mass should be constrained using the SED fit and \textit{Gaia} parallax, after a careful determination of the atmospheric parameters at phase 0. This will allow the determination of reliable companion masses and radii as well as reliable error bars, as the companion is much fainter and the contribution of the dark side to the spectrum is negligible.

The orientations of sdB binaries in space should be uniformly distributed. Thus, higher inclinations should be more likely than low inclinations due to the projection effect. The probability of a system to have an inclination lower than a certain value $i_0$ can easily be calculated by $P_{i<i_0}=1-\cos i_0^\circ$ \citep[][]{gray}. We can use this to estimate how many systems we expect to find below a certain inclination and compare this to the inclination distribution we measure. This comparison is shown in Fig. \ref{incl}. As we are only including non-eclipsing systems, we do not find any systems with inclinations higher than $\sim 55^\circ$. Moreover, most of the analyzed systems are systems which have been found before in different ways and are not homogeneously selected. 
At the highest inclinations the measured distribution starts to deviate from expectation. It looks like we are finding too few systems at high inclinations, maybe because we are starting to see (tiny) eclipses at this inclinations.
Nevertheless, despite the inhomogeneous target selection we still get a good agreement showing that systematic effects seem to play a minor role and we can indeed derive inclinations from the reflection effect systems.

Our sample of reflection effect systems includes sdBs with companions covering the entire mass range of dM stars from the hydrogen burning limit to early M dwarfs with masses around $0.4\,\rm M_\odot$, and also a large part of the period range from 0.1 to 0.8 d. For the previously found reflection effect systems with brown dwarf candidates (KBS13, BPS CS 22169-1, PHL457, CPD-64 481) we have shown that most of them just have low inclinations, and so the companions are M dwarfs instead. This means so far still no BDs around sdBs in longer periods have been confirmed, and the BDs in sdB binaries are still preferentially found at close periods below 0.1 d \citep{erebos}.

We also found 24 sdB+WD systems showing tiny variations with amplitudes below $\sim 0.1\%$ due to Doppler beaming or ellipsoidal deformation. Fitting of the light curves allowed us to derive the masses of the WD companion for 8 systems. We derive masses for the companions from 0.25 to 0$.6\,\rm M_\odot$ with orbital periods of these systems from 0.132 to 0.83 d. 

16 longer period systems show only Doppler beaming, and so no additional information can be derived from the light curve in this case, as the semi-amplitude of the RV curve $K_1$ was derived before. We could show that the variation could indeed be explained by Doppler beaming by overplotting a model calculated using the sdB radius and the $K_1$. Finding more Doppler beaming in systems without solved orbits, will allow us to derive the period and the $K_1$ without spectroscopy.

To compare our sample to the sample of published sdB+dM/BD and sdB+WD systems we also plotted the period-companion mass diagram (see Fig. \ref{mass_comp}). It is evident that the sample known so far only covers a very small parameter range. For sdB+dM/BD systems, only those with short orbital periods and low companion masses have been studied. The same has been true for the sdB+WD systems, for which those with the shortest periods and highest-mass companions have been studied preferentially. Our new sample covers a much larger orbital period and companion mass range than before.

For the sdB+WD systems we can see that the highest companion masses are found at the shorter periods below 0.2 d, where we have two WD companions which are more likely to be CO WDs. The rest of the companions have masses below $0.45\,\rm M_\odot$ and are most likely He WDs. If they evolved from higher mass stars with $2-3\,\rm M_\odot$, WDs with masses $>0.33\,\rm M_\odot$ could also be CO WDs, in principle, but such objects are expected to be much more rare. 
This sample of studied post-common envelope systems over a large parameter range is ideal to constrain the common envelope phase as done e.g. in \citet[][]{ge22}. A large sample of post-common envelope binaries with known masses of both primary star and companion, as well as orbital separations and orbital periods is necessary for such studies. This is beyond the scope of this paper, but we are already preparing a paper to use this sample to constrain sdB formation by a common envelope phase (Vos et al. in prep.). Moreover, such a sample is important to be compared to the parameters predicted by hydrodynamical simulations, as done in \citet{kramer20}, who simulated a red giant of $1\,M_\sun$ being stripped by a substellar companion in a common envelope phase evolving to an sdB.

Using the previously measured projected rotational velocity of some the sdBs and the radius of the sdB derived by the SED together with the \textit{Gaia} parallax as well as the determined inclination we also could measure the rotation period for sdBs with dM as well as WD companions. We find that in three systems (with orbital periods from 0.25 to 0.56 d) out of seven the sdB is rotating significantly slower than the orbital period. On the other hand, systems with even longer periods of 0.7 d seem to be (almost) synchronized. This agrees well with the findings of \citet[][]{silvotti22} and \citet[][]{schaffenroth21} that both synchronized and non-synchronized systems are found on the EHB suggesting that synchronization is happening on the EHB. 
Theoretical synchronization theories \citep{preece:18} expect that none of the sdBs in close binary systems should be synchronized and cannot explain the observations yet.

The high S/N of the \textit{TESS} light curves allowed us to almost double the sample of studied sdB+dM/BD and sdB+WD systems. Additional sectors of \textit{TESS} data are already available, and the future photometric surveys as e.g. the survey that will be obtained by the Vera Rubin observatory or \textit{PLATO} as the succesor of \textit{TESS} will allow us to obtain a statistically significant sample of post-common envelope systems with hot subdwarf primaries.


\begin{acknowledgements}
     This research made use of Lightkurve, a Python package for Kepler and \textit{TESS} data analysis \citep[][]{lightkurve}.
     
     Based on observations collected at the European Organisation for Astronomical Research in the Southern Hemisphere under ESO programme(s) 080.D-0685(A), 082.D-0649(A), 092.D-0040(A).
     
     This paper includes data collected by the \textit{TESS} mission, which are publicly available from the Mikulski Archive for Space Telescopes (MAST). Funding for the \textit{TESS} mission is provided by NASA's Science Mission directorate.

     VS and SG acknowledge funding from the German Academic Exchange Service (DAAD PPP USA 57444366) for this
     project and would like to thank the host institution Texas Tech University for the hospitality. VS was funded by the Deutsche Forschungsgemeinschaft under grant GE2506/9-1. IP was partially funded by the Deutsche Forschungsgemeinschaft under grant GE2506/12-1 and by the UK’s Science and Technology Facilities Council (STFC), grant ST/T000406/1.

     IP acknowledges support from the UK's Science and Technology Facilities Council (STFC), grant ST/T000406/1. 
     BNB was supported by the National Science Foundation grant AST-1812874 and by NASA through grant 80NSSC21K0364.
     
     TK acknowledges support from the National Science Foundation through grant AST \#2107982, from NASA through grant 80NSSC22K0338 and from STScI through grant HST-GO-16659.002-A.
     
     We thank Uli Heber for comments on the manuscript and Lars Möller for sharing his radial velocity measurements with us. We thank Andreas Irrgang for the development of the SED fitting tool and making it available to us. We thank Alfred Tillich for observing some of the spectra used in this paper.
     We thank Stephen Walser for helping with some of the SOAR and Chiron observations used in this paper. 
\end{acknowledgements}


\bibliography{biblio}
\bibliographystyle{aa}
\begin{appendix}

 \section{The known reflection effect systems}\label{refl_indiv}
 \subsection{BPS CS 22169-1}
BPS CS 22169-0001 was discovered to be a sdB binary with very small RV amplitude by \citet{edelmann:2005}. \citet{bpsc} reported a tiny reflection effect with a period of 0.214\,d. The minimum mass of the companion is only $0.026\,\rm M_\odot$. \citet{geier:2010} also derived the rotational velocity of the sdB and calculated a companion mass of $0.19\,\rm M_\odot$ assuming tidal synchronization. However, in recent years this assumption became questionable \citep[][]{schaffenroth21,preece:18} and this requires a low inclination, which is quite unlikely.

The analysis of the \textit{TESS} light curve (Fig. \ref{lc_bpsc}) results in an inclination of $7.6^\circ\pm1.0^\circ$ resulting in a mass and radius of the companion of $0.23 \pm 0.03\,\rm M_\odot$ and  $0.31 \pm 0.05\,\rm R_\odot$ demonstrating that it is in deed a M dwarf companion rather than a brown dwarf. With the radius of the sdB from the SED fit together with the determined inclination and the projected rotational velocity measured by \citet[][]{geier:2010} it is possible to derive the rotational velocity ($P_{\rm rot}=0.16\pm 0.04\rm \, d$). This shows that the sdB rotation is almost tidally locked to the orbit.

\subsection{PHL 457}
PHL\,457 was also discovered to be a close sdB binary with small RV amplitude by \citet{edelmann:2005}. Light variations caused by long-period pulsations were found by \citet[][]{blanchette08}. \citet{schaffenroth14a} observed a small reflection effect in PHL\,457 with a period of 0.3128\,d and confirmed a small RV amplitude of only $K=12.8\pm0.08\,\rm km\,s^{-1}$. This results in a minimum mass of $0.027\,\rm M_\odot$ for the companion, making it a brown dwarf if the inclination exceeds $21^\circ$. The likelihood of this being the case is 94\%. 

PHL\,457 was observed in \textit{K2}  and TESS. \citet[][]{baran19} analysed the pulsations of PHL\,457 and detected short- and long-period pulsations from 4.5 min to 1.8 hours. Our analysis of the \textit{K2}  light curve (Fig. \ref{lc_phl}) results in an inclination of $9.3^\circ\pm1.6^\circ$, which translates to a companion mass and radius of $0.19 \pm 0.04\,\rm M_\odot$ and  $0.16 \pm 0.04\,\rm R_\odot$. Hence, the companion is a low-mass M dwarf instead of a brown dwarf.


\subsection{KBS 13}
This sdB binary systems was found to show a reflection effect with a period of 0.2923\,d  by \citet{kbs13}. They derived a semi-amplitude of the radial velocity curve of $K_1=22.82 \pm 0.23\,\rm km\,s^{-1}$. Using the mass function, a canonical mass for the sdB and the period from the \textit{K2}  and \textit{TESS} light curves ($P=0.292365\,\rm d$), we derive a minimum mass of only 0.045 $\rm M_\odot$ --- well below the limit for hydrogen burning.

From analysis of the \textit{TESS} light curve (Fig. \ref{lc_kbs13}), we could derive an inclination of the system of $10.1^\circ\pm0.4^\circ$, which gives us a mass for the companion of $0.260 \pm 0.008\,\rm M_\odot$ and a radius of  $0.284 \pm 0.015\,\rm R_\odot$. Thus the companion is an M dwarf and not a brown dwarf.




\subsection{Feige 48}
Feige\,48 was identified to be a sdB star by \citet[][]{green86}. \citet[][]{koen98} observed this target and found that it is one of the coolest sdBs showing short-period pulsations. \citet[][]{otoole04} analyzed UV spectra of Feige 48 proving that it is a close binary with a period of 0.376 d and a RV semi-amplitude of $K_1=28\pm0.2\,\rm km\,s^{-1}$. Assuming that the sdB rotation is tidally locked to the orbit, they derived a mass of $0.46\,\rm M_\odot$ for the companion and claimed it is most likely a white dwarf as they did not detect a reflection effect.
\citet[][]{van_grootel08} corroborated this by performing an asteroseismic analysis with the best model being for an object having a solid-body rotation with the orbital period. \citet[][]{geier:2010} re-measured the rotational velocity and derived a slightly higher $v\sin i=8.5\pm1.5\rm km/s$. A re-analysis of this system with time-resolved spectroscopy and photometry by \citet[][]{latour14} found a shorter orbital period of only 0.3438~d  and also an reflection effect with the same period and therefore they claimed that the companion is an M dwarf instead of a white dwarf.

The \textit{TESS} light curve confirms the reflection effect (see Fig. \ref{lc_feige48}). We were able to fit it and could so derive the inclination ($i=16.3^\circ\pm1.4^\circ$) and the mass and radius of the companion ($M_{\rm comp}=0.232\pm0.020\,\rm M_\odot$, $R_{\rm comp}=0.266\pm0.033\,\rm R_\odot$) confirming the M dwarf nature of the companion. Using the inclination, the radius of the sdB and projected rotational velocity by \citet[][]{geier:2010} we can calculate the rotational period of the sdB ($P_{\rm rot}=0.36\pm0.07\,\rm d$) showing that the rotation is most likely synchronized to the orbital period.


\subsection{GALEX J2205-3141}

\citet[][]{nemeth:2012} identified  GALEX J2205-3141 (GALEX J220551.8-314105) to be a sdB star from spectroscopic follow-up from the hot subdwarf candidates identified in the Galaxy Evolution Explorer (GALEX)/Guide star catalogue (GSC) survey by an UV excess. Photometric and spectroscopic follow-up done by \citet[][]{kawka:2015} showed that the sdB is in a close binary system with a M dwarf companion with a period of 0.341543 d, as it shows a RV variation of $K_1=47.8\pm2.2\,\rm km/s$ and a 4\% amplitude reflection effect.

The \textit{TESS} light curve also shows this reflection effect (see Fig. \ref{lc_galexj22}). From the best fit we could derive an inclination of $17.3^\circ\pm2.6^\circ$, giving a companion mass and radius of $0.53 \pm 0.10\,\rm M_\odot$ and $0.42 \pm 0.08\,\rm R_\odot$, which means the companion is an early M dwarf. This would be the highest mass companion found so far. At a mass this high it should be possible to detect spectral line contamination from the companion in the sdB spectrum. The radius of the companion is a bit smaller than expected for such an object, so the mass might be overestimated. The SED fitting (paper I) indicates that the mass of the sdB is not canonical but higher. More spectroscopic follow-up is necessary to determine the $\log{g}$ around phase 0, when the contribution of the companion is smallest, to constrain the companion better.

\subsection{GALEX J09348-2512}
GALEX\,J09348-2512 (GALEX\,J093448.2-251248) was found to be an sdB star by \citet[][]{nemeth:2012}. When searching for short-period variables in the ATLAS survey, \citet{koen19} discovered light variations with a period of 0.143\,d and an amplitude of 0.05 mag, indicating the presence of reflected light from the companion.  The analysis suggested a companion mass close to $0.1\,M_\odot$ but was lacking spectroscopic confirmation.

\citet[][]{Moeller} analysed archival spectra of this system and  
confirmed it to be a sdB binary ($T_{\rm eff}=40800\pm1000\,\rm K$, $\log g=5.55\pm0.10$) with RV semi-amplitude of $K_1=37 \pm 4 \,\rm km\,s^{-1}$. The minimum companion mass can be calculated to $0.06\,\rm M_\odot$, which is below the hydrogen burning limit.

This system was also observed by \textit{TESS}, which confirmed it to be a reflection effect system with a period of 0.142903 d. 
The analysis of the \textit{TESS} light curve (Fig. \ref{lc_galexj09}) resulted in an inclination $i=24.0^\circ\pm3.0^\circ$ and a mass and radius of the companion of $0.165 \pm 0.022\,M_\odot$ and $0.175 \pm 0.035\,\rm R_\odot$, showing that it is a low-mass M dwarf companion. The mass of the sdB derived by the SED method and \textit{Gaia} parallax (paper I) is $0.737^{+0.176}_{-0.143}\rm M_\odot$, higher than the canonical sdB mass. However, the radius from the SED agrees with the radius derived by the light curve assuming a canonical mass for the sdB, while no consistent solution could be found using the mass derived by SED and parallax. This suggests that the determination of the mass with this method relying mainly on the $\log g$ determination from the co-added spectrum is not reliable because of contamination by light from the companion. This further demonstrates why we prefer the assumption of the canonical mass for the sdB for now.


\subsection{EQ Psc}


EQ Psc (PB 5450) was identified as a sdB star by \citet[][]{berger80}. \citet[][]{green:2003} discovered long-period pulsations. From the \textit{K2}  light curve \citet[][]{jeffrey:2014} found that it not only shows several pulsation periods, but also a reflection effect with a period of 0.801 d. \citet[][]{baran19} re-analyzed the photometry after combining it with additional time-resolved data. They found RV variations ($34.9\pm1.6\,\rm km\,s^{-1}$) with the same period and confirmed the primary to be a sdB star.

The best fit of \textit{TESS} the light curve (see Fig. \ref{lc_eqpsc}) was found for an orbital inclination $i=25.4^\circ\pm1.5^\circ$, giving us a mass and radius of the companion of $0.222\pm0.019\,\rm M_\odot$ and  $0.181\pm0.014\,\rm R_\odot$. The radius of the companion is significantly smaller as expected by theoretical calculations (see Fig. \ref{m-r}). The SED and together with the parallax preferred a sdB mass of $0.35\,\rm M_\odot$, which results in a mass and radius of the companion of $0.253\pm0.012\,\rm M_\odot$ and  $0.179\pm0.014\,\rm R_\odot$, which agrees better. Its mass is below the minimum mass for core helium burning, which could indicate that the hot subdwarf originates from a intermediate-mass progenitor or is a pre-He WD.

\subsection{PG 1329+159} 
PG 1329+159 was discovered to be a sdB star by the Palomar Green (PG) survey \citep[][]{green86}. In a survey to search for close sdB binaries, \citet[][]{morales-rueda03} found it to be RV variable with a period of 0.249699 d. Using follow-up photometry \citet[][]{maxted:2004} found this system to also show a reflection effect indicating it to be a sdB+dM system.

The analysis of the \textit{TESS} light curve (Fig. \ref{lc_pg1329}) gave an inclination of $i=37.8^\circ\pm2.1^\circ$, resulting in a mass and radius of the companion of $0.167\pm0.008\,\rm M_\odot$ and $0.199 \pm 0.016\,\rm R_\odot$. \citet[][]{geier:2010} also measured the rotational velocity of the sdB and combining this with sdB radius and inclination we derive a rotational period of $P_{\rm rot}=0.64\pm0.07\,\rm d$. This means that rotation period is significantly higher than the orbital period.


\subsection{CPD-64 481}
CPD-64\,481 is another close sdB binary with small RV amplitude discovered by \citet{edelmann:2005}. \citet{schaffenroth14a} found a small reflection effect in its light curve with a period of 0.277263\,d. In this case the minimum companion mass was found to be $0.048\,\rm M_\odot$ making it another brown dwarf candidate. 

CPD-64\,481 was observed by \textit{TESS} in 26 different sectors. From the analysis of the light curve (see Fig. \ref{lc_cpd}) we could derive an inclination of $i=34.3^\circ\pm 2.2^\circ$ resulting in a mass and radius for the companion of $0.088 \pm 0.006\,\rm M_\odot$ and  $0.118\pm 0.035\,\rm R_\odot$, showing that the companion is probably a low-mass M dwarf very close to the hydrogen-burning limit. Using the projected rotational velocity by \citet[][]{geier:2010} the rotation period is derived to be $P_{\rm rot}=1.22\pm0.30\,\rm d$. In this case hence the sdB is rotating significantly slower than synchronized.


\subsection{JL 82}
JL\,82 was identified as a sdB star by the EC survey \citep[][]{kilkenny95}. RV measurements from \citet{edelmann:2005} confirmed it to be a close sdB binary with a period of 0.737 d and a minimum companion mass of $0.1\,\rm M_\odot$.
\citet{koen_2009} showed that this star shows a reflection effect as well as long-period pulsations with periods between 1 and 4 hours. 

The \textit{TESS} light curve (Fig. \ref{lc_jl82}) also shows this reflection effect. The best fit results in an inclination of the system to $29.1^\circ\pm1.1^\circ$, which constrains the companion mass and radius to $0.240 \pm 0.009\,\rm M_\odot$ and $0.249 \pm 0.013\,\rm R_\odot$. With the radius of the sdB from the SED fit together with the inclination and the projected rotational velocity measured by \citet[][]{geier:2010} we can derive the rotational velocity ($P_{\rm rot}=0.61\pm 0.07\rm \, d$). This shows that the sdB rotation is probably tidally locked to the orbit.


\subsection{GALEX J0321+4727} 


\citet[][]{kawka:2010} found GALEX J0321+4727 (GALEX J032139.8+472716) to be a close sdB binary with a period of  0.26584 d  and $K_1=59.8 \pm 4.5\,\rm km\,s^{-1}$ showing a reflection effect of about 6\%. 

The \textit{TESS} light curve shows this reflection effect with a larger amplitude, as expected at longer wavelengths (Fig. \ref{lc_galexj03}). From the analysis of the light curve we derived an inclination of $38.6^\circ\pm0.9^\circ$, which gives a companion mass and radius of $0.233\pm0.013\,\rm M_\odot$ and $0.298\pm0.026\,\rm R_\odot$.

\subsection{SDSS J012022+395059}
SDSS J012022+395059 (FBS 0117+396) was found to be a sdB star by \citet[][]{geier:2011}. As this target showed quite a large radial velocity shift in a short time, it was flagged as a high-priority target for follow-up. \citet[][]{oestensen2013} showed that this star exhibits a reflection effect and short-period pulsations. Additionally they also obtained spectroscopy showing it is indeed a close binary system with a period of 0.252013 d.

The analysis of the \textit{TESS} light curve (Fig. \ref{lc_j0102}) resulted in an inclination of $i=40^\circ\pm7^\circ$, giving a companion mass and radius of $0.161\pm0.033\,\rm M_\odot$ and $0.241 \pm 0.062\,\rm R_\odot$.


\subsection{UVEX 0328+5035}
\citet[][]{verbeek12} identified UVEX 0328+5035 (UVEX J032855.25+503529.8) as a single-lined sdB in the UV-Excess Survey of the Northern Galactic Plane (UVEX) survey. \citet[][]{kupfer:2014} obtained spectroscopic and photometric follow-up observations and found it to be in a close binary with a period of 0.11017 d and a reflection effect amplitude of about $25\%$.

The \textit{TESS} light curve (Fig.~\ref{lc_uvex}) gives an inclination of $41.4^\circ\pm 0.6^\circ$ and a mass and radius of the companion of $0.160\pm0.003\,\rm M_\odot$ and $0.160\pm0.012\,\rm R_\odot$.


\subsection{HS 2333+3927}
This target was selected as a hot subdwarf candidate from the Hamburg-Schmidt (HS) Survey and confirmed as an sdB by \citet[][]{Edelmann2003}, who also identified it to be RV variable from two spectra.
\citet[][]{heber:reflection} carried out photometric and spectroscopic follow-up observations of this target, confirming it to be a close reflection effect binary with a period of 0.1718023 d and RV semi-amplitude of $K_1=89.6\,\rm km\,s^{-1}$. For the companion they derived a mass of $0.24-0.32 \rm\,M_\odot$.

In their analysis they find that the radius of the sdB is too small compared to the spectroscopic one. Even using a higher or lower mass for the sdB cannot solve this issue. It also does not agree with the radius determined by the SED (see paper I), which agrees well with the spectroscopic $\log g$, when assuming a canonical mass. No model could be found for the correct radius of the sdB, as the size of the companion would have to be larger than its Roche lobe to match the amplitude of the reflection effect. By increasing the absorb factor to absorb=2 we can solve this issue. This has also been found for other systems \citep[e.g.][]{schaffenroth14}. One possible explanation for this is that we are using a blackbody instead of a synthetic spectrum for modelling the reflection effect. This underestimates the flux of the sdB in the UV, and HS 2333+3927 is relatively hot.

We get an inclination of $42.8^\circ\pm0.5^\circ$ and a mass and radius of the companion of $0.286\pm0.008\,\rm M_\odot$ and $0.40\pm0.02\,\rm R_\odot$ with this assumption. The companion is hence an early M dwarf. 
To solve this issue the spectral analysis should be re-done with the newest generation of model spectra to see whether the problem can be solved, as a higher temperature of the sdB would also result in a higher reflection effect amplitude.


\subsection{V1405 Ori}
KUV 0442+1416, also called V1405 Ori, was identified as an sdO or sdB by \citet{wegener85}. 
\citet[][]{Koen1999} demonstrated an absence of He II lines, making it a sdB star. Additionally, they detected short-period pulsations and reddening in this target, which could either come from interstellar reddening or from a cool companion. \citet[][]{reed10} later discovered a reflection effect with an amplitude of about 20\%. They also found RV variations with a period of 0.398 d and a semi-amplitude of $K_1=85.1\pm8.6\,\rm km\,s^{-1}$, which results in a minimum companion mass of $0.25\,\rm M_\odot$, making it an early M type companion. 

The \textit{TESS} light curve confirms the observation of the reflection effect (Fig. \ref{lc_vori}). From the analysis we find an inclination of $43^\circ\pm0.9^\circ$ and a mass and radius of the companion of $0.390 \pm 0.031\,\rm M_\odot$ and $0.341 \pm 0.038\,\rm R_\odot$, making it an early type M dwarf, as expected.

\subsection{HE 0230-4323}
HE 0230-4323 was found to be a sdB star by \citet{lisker} and confirmed to be in a close binary system by \citet{edelmann:2005} with a period of 0.4515 d. \citet{koen:2007} found it to show a reflection effect as well as low-amplitude pulsations with periods between 24 to 45 min in the light curve.

The \textit{TESS} light curve (Fig. \ref{lc_he0230}) confirms this reflection effect. The best fit constrains the inclination of the system to $52.6^\circ\pm1.5^\circ$, which constrains the companion mass and radius to $0.209 \pm 0.006\,\rm M_\odot$ and $0.307 \pm 0.012\,\rm R_\odot$.


\subsection{HE 1318-2111}


HE 1318-2111 was discovered to be a sdB star in the EC survey \citep[][]{kilkenny:1997}. \citet[][]{christlieb} rediscovered it in the Hamburg/ESO objective-prism (HE) survey. In the ESO Supernovae type Ia Progenitor survey (SPY) \citet[][]{napiwotzki04b} took spectroscopic follow-up of several sdO/B stars and discovered that HE 1318-2111 was in a close binary system with a period of 0.487 d. In the \textit{TESS} light curves we discovered that it also shows a reflection effect. At the same time  \citet[][]{tess_south} also found the reflection effect in the light curves of sdB candidates from \citet[][]{geier_gaia_catalog} derived by the \textit{TESS} full-frame images.

From the light curve we derived an orbital inclination of $i=56.5^\circ\pm1.7^\circ$, giving us a mass and radius of the companion of $0.158\pm0.006\,\rm M_\odot$ and  $0.277\pm0.015\,\rm R_\odot$

\section{The sdB+WD systems}\label{ell_indiv}
\subsection{PG 1043+760}
PG 1043+760 was identified to be a hot subdwarf in the PG survey \citep[][]{green86}. \citet[][]{maxted:2001} discovered it to be in a close binary with a period of 0.12 d. They could not detect any light variations and claimed therefore that the companion must be a low-mass He WD. 

The light curve of PG 1043+760 (see Fig. \ref{lc_PG1043+760}) observed by \textit{TESS} shows ellipsoidal modulation and beaming with an amplitude of only 0.2\%. 
From modelling the light curve we derive a very low inclination of only $15^\circ\pm0.6^\circ$ and a mass ratio of $1.65\pm0.11$, which results in a companion mass of $0.48\pm0.08\,\rm M_\odot$ and hence the companion could be a CO WD or a He WD orbiting a sdB with a mass of  $0.289^{+0.038}_{-0.036}\,\rm M_\odot$.

\subsection{GALEX J0751+0925}
GALEX J0751+0925 (GALEX J075147.0+092526) was found to be a sdB by \citet[][]{nemeth:2012}. \citet[][]{kawka:2015} carried out spectroscopic follow-up of this star and discovered that it is in a close binary with a period of 0.178319 d. They also checked the light curve of the system taken by the All Sky Automated survey \citep[ASAS, ][]{asas} but could not see any variation with a upper limit of 44 mmag. So they concluded that the unseen  companion is most likely a WD. 

The \textit{TESS} light curve of this system (see Fig.\ref{lc_GALEXJ0751}) varies with half of the orbital period most likely due to ellipsoidal deformation confirming that the companion is indeed a WD. Our analysis results in an inclination of $i=74\pm10^\circ$ and a mass ratio of $0.85^{+0.09}_{-0.04}$, making the companion a He WD with a mass of $0.31{+0.07}_{-0.03}\,\rm M_\odot$. 

\subsection{HS1741+2133}
HS 1741+2133 was identified as a sdB by \citet[][]{Edelmann2003}. \citet[][]{kupfer:2014} observed this system further and found it to be a close sdB binary with a period of 0.2 d and a semi-amplitude of $K_1=157\,\rm km/s$, giving a minimum companion mass of $0.39\,\rm M_\odot$. As no photometric variability with an upper limit of 6 mmag was found by \citet[][]{dreizler02}, they concluded that the companion must be a WD.

HS1741+2133 was only observed on the full-frame images in sector 26 with an exposure time of 1426 sec resulting in 12 data points per orbital period. Phase-folding it to the orbital period showed an ellipsoidal deformation (Fig. \ref{lc_HS1741+2133}). From the light curve we obtain an inclination of $47^\circ\pm11^\circ$ resulting in a companion mass of $0.58^{+0.3}_{-0.15}\,\rm M_\odot$. So the companion is most likely a CO WD. To exclude smearing effects the analysis should be repeated with a higher cadence light curve.

\subsection{PG 1136-003 }
PG 1136-003 was also found to be a sdB by \citet[][]{green86}. Spectroscopic follow-up by \citet[][]{geier:2011_2} showed a large RV shift over a period of one day and made it a high priority target. \citet[][]{geier:2011} showed that it is a close binary with a period of 0.207536 d. The companion has a minimum mass of $0.42\,\rm M_\odot$ and is most likely a WD.

The 30 min \textit{K2}  light curve of PG 1136-003 (see Fig. \ref{lc_PG1136-003}) clearly shows ellipsoidal modulation as well as beaming due to a massive compact companion confirming that it is most likely a WD companion. As the cadence of the light curve is  10\% of the orbital period, smearing will be quite important and the amplitude of the ellipsoidal modulation will be underestimated. So photometric follow-up has to be obtained for an analysis of the system. Luckily the system was also observed in \textit{TESS} with a 2 min cadence. The analysis of this light curve resulted in an inclination of $75^\circ \pm 11^\circ$, as well as a mass ratio of $0.90^{+0.10}_{-0.04}$ corresponding to a companion mass of $0.45^{+0.08}_{-0.05}\,\rm M_\odot$. With this mass the companion could be a He WD or a low-mass CO WD orbiting a sdB star with $0.501^{+0.096}_{-0.078}\,\rm M_\odot$.

\subsection{GD 687}

GD 687 was classified first as a WD by \citet[][]{guseinov83}. \citet[][]{lisker} revised that classification and determined the star to be a sdB instead. \citet[][]{gd_687}  discovered that it is RV variable with a period of 0.37765 d. Using the assumption of a tidally locked rotation they derive a mass of $0.7\pm0.2\,\rm M_\odot$ for the companion, which would mean it is very likely a CO WD. 

The system was also observed by \textit{TESS}, and the phased light curve (Fig. \ref{lc_GD687}) shows a sinusoidal variation with a period of half of the orbital period, which is most likely due to ellipsoidal deformation. From the fit of the light curve we get an inclination of $58^\circ\pm8^\circ$ with a mass ratio of $1.23^{+0.24}_{-0.14}$, which constrains the white dwarf companion to a He WD with a mass of $0.35^{+0.09}_{-0.06}\,\rm M_\odot$, as the mass of the sdB is only $0.283^{+0.042}_{-0.037}\,\rm M_\odot$. This means that the sdB is more likely a pre-He WD instead of a helium-core burning object. Using the rotational velocity by  \citet[][]{gd_687}, as well as the inclination and sdB radius we derive a rotational period of $P_{\rm rot}=0.39\pm0.05\rm\, d$, agreeing with a tidally locked rotation of the sdB.

\subsection{GALEX J234947.7+384440}
GALEX J234947.7+384440 was identified as an RV variable sdB by  \citet[][]{kawka:2010}. From the lack of photometric variability they suggested that the companion is most likely a WD with a minimum mass of $0.24\,\rm M_\odot$. 

The \textit{TESS} light curve (Fig. \ref{lc_GALEXJ2349}) clearly shows two peaks over one orbital period, indicating Doppler beaming and ellipsoidal modulation. We get a good solution for an inclination of $70^\circ\pm10^\circ$. This corresponds to a He WD companion with a mass of only $0.26\pm0.04\,\rm M_\odot$.

\subsection{PG0101+039}\label{pg0101}


PG0101+039 (Feige 11) was classified as A0p star by \citet[][]{Feige}. Later this classification was revised in the PG survey \citep[][]{green86} and the star identified as sdB star. \citet[][]{maxted:2001} showed that the sdB is in a close binary with period of 0.567 d orbited by a white dwarf companion. \citet[][]{green:2003} discovered that the sdB also shows low-amplitude long-period pulsations. \citet[][]{randall} analyzed the $\sim 400$ h long light curve of PG0101+039 observed with the MOST satellite and found that, in addition to several pulsation modes, a long-period variation is also visible with half the orbital period and likely originates from ellipsoidal deformation of the sdB. \citet[][]{geier08} investigated this system further by trying to model the ellipsoidal deformation by determining the inclination by measuring the rotational velocity of the sdB and assuming synchronization. They found that the amplitude of the variation is of the order of what is expected from ellipsoidal deformation, but the period was not sufficiently well determined to match the light variations with the RV variations.

PG0101+039 was also observed by the \textit{K2}  mission. The light curve phased to the orbital period is shown in Fig. \ref{lc_PG0101+039}. The main variation seems not to be half of the orbital phase, but instead the light curve seems to vary with the full orbital period and is dominated by beaming. However, there is a second, much smaller peak at half of the orbit showing a tiny ellipsoidal deformation. From the light curve we get a good solution for a very high inclination close to $89.4^\circ\pm0.06^\circ$, resulting in a mass ratio of $0.8174^{+0.0001}_{-0.0001}$ corresponding to a He WD companion with $0.34\pm0.04\,\rm M_\odot$. Using the rotational velocity measured by \citet[][]{geier08} we derive a rotational period of $P_{\rm rot}=0.85 \pm 0.09\,\rm d$, which means that the sdB rotation is not tidally locked but the sdB is rotating slower. The light curve model does not agree perfectly, but some residuals remain. They come probably from the effect that the tidal bulge is lacking behind, as the sdB rotates slower than the orbital period. A more detailed analysis considering this fact is beyond the scope of this paper, but should be performed in the future.

\subsection{EC 13332-1424}


EC 13332-1424 was classified as a sdB in the EC survey \citep[][]{kilkenny:1997}. Spectroscopic follow-up by \citet[][]{Copperwheat11} showed that the sdB is in a close binary system with a period of 0.82794 d with a semi-amplitude of the RV curve of $104.1 \pm 3.0\,\rm km\,s^{-1}$. Using this result we can calculate a minimum mass of the companion of  $0.43\,\rm M_\odot$, so it is most likely a white dwarf.

EC 13332-1424 was observed in the \textit{K2}  mission and in the light curve a clear periodic signal with the orbital period is visible (see Fig. \ref{lc_EC13332-1424}). There is one dominating peak visible, which is most likely resulting from Doppler beaming as in PG 1232-136. Moreover, there is a smaller variation apparent most likely due to a tiny ellipsoidal deformation. From the light curve we can constrain the inclination to $i=82^\circ\pm2^\circ$ resulting in a He-WD companion with a mass of $0.4\pm0.06\,\rm M_\odot$ assuming the most probable mass of an sdB with a WD companion ($0.4\,\rm M_\odot$, see paper I).

\section{Radial velocity measurements}

\begin{table}[H]

\caption{RV measurements of TYC5977-517-1}
    \centering
    \begin{tabular}{lll}
    \hline\hline
         BMJD&  RV & RV error\\
 & [km/s]        & [km/s]\\\hline

 58641.48626 & 60.1 & 11.8 \\ 
58641.48675 & 56.6 & 4.1 \\ 
58641.48724 & 51.3 & 12.4 \\ 
58641.48772 & 23.8 & 12.4 \\ 
58641.48821 & 39.3 & 6.3 \\ 
58641.48870 & 49.4 & 6.2 \\ 
58641.48919 & 36.3 & 17.6 \\ 
58641.48968 & 39.4 & 7.5 \\ 
58641.49017 & 27.0 & 8.1 \\ 
58641.49065 & 50.5 & 17.2 \\ 
58641.49163 & 57.3 & 15.5 \\ 
58641.49212 & 41.8 & 8.6 \\ 
58641.49261 & 47.8 & 10.2 \\ 
58641.49310 & 89.3 & 10.3 \\ 
58641.49359 & 71.9 & 10.1 \\ 
58641.49408 & 68.9 & 9.3 \\ 
58641.49456 & 45.7 & 11.2 \\ 
58641.49505 & 54.9 & 8.8 \\ 
58641.49554 & 44.0 & 11.1 \\ 
58641.49603 & 54.3 & 8.4 \\ 
58641.49652 & 37.4 & 7.0 \\ 
58641.49701 & 56.1 & 10.1 \\ 
58641.49750 & 81.7 & 8.3 \\ 
58641.49799 & 60.3 & 11.6 \\ 
58641.49847 & 66.9 & 11.6 \\ 
58641.49896 & 34.3 & 15.4 \\ 
58641.49945 & 82.8 & 7.1 \\ 
58641.49994 & 80.7 & 3.6 \\ 
58641.50043 & 95.2 & 13.7 \\ 
58641.50092 & 68.1 & 15.0 \\ 
58641.50140 & 105.4 & 9.2 \\ 
58641.50189 & 73.0 & 12.6 \\ 
58641.50238 & 81.9 & 12.8 \\ 
58641.50287 & 107.6 & 16.4 \\ 
58641.50336 & 79.6 & 9.2 \\ 
58641.50385 & 88.0 & 11.4 \\ 
58641.50434 & 106.8 & 15.5 \\ 
58641.50483 & 86.7 & 14.6 \\ 
58641.50531 & 73.9 & 22.4 \\ 
58641.50580 & 56.6 & 10.4 \\ 
58641.50629 & 73.4 & 9.3 \\ 
58641.50678 & 77.8 & 9.2 \\ 
58641.50727 & 65.9 & 8.3 \\ 
58641.50776 & 59.1 & 14.5 \\ 
58641.50824 & 66.1 & 11.8 \\ 
58641.50873 & 61.4 & 5.3 \\ 
58641.50922 & 75.3 & 9.9 \\ 
58641.50971 & 62.1 & 9.0 \\ 
58641.51020 & 62.8 & 11.7 \\ 
58641.51069 & 57.2 & 7.3 \\ 
58641.51118 & 48.0 & 13.3 \\ 
58641.51167 & 54.1 & 13.6 \\ 
58641.51215 & 63.2 & 40.9 \\ 
58641.51264 & 74.8 & 14.4 \\ 
58641.51313 & 61.8 & 7.7 \\ 
58641.51411 & 84.6 & 14.0 \\ 
58641.51460 & 49.9 & 17.3 \\ 
58641.51509 & 103.9 & 10.8 \\ 
58641.51557 & 78.9 & 10.9 \\ 
58641.51606 & 88.8 & 9.4 \\ 
58641.51655 & 65.1 & 17.0 \\ 
58641.51704 & 91.4 & 15.1 \\ 
58642.45281 & -94.3 & 7.2 \\ 
58642.45330 & -83.9 & 7.6 \\ 
58642.45378 & -71.1 & 4.8 \\ 
58642.45427 & -79.5 & 5.5 \\ 
58642.45476 & -72.0 & 3.8 \\ 
\hline
    \label{RV_tyc}
\end{tabular}
\end{table} 
\begin{table}[H]\ContinuedFloat
\caption{RV measurements of TYC5977-517-1 (continued)}
    \centering
    \begin{tabular}{lll}
    \hline\hline
         BMJD&  RV & RV error\\
 & [km/s]        & [km/s]\\\hline

58642.46040 & -53.7 & 7.2 \\ 
58642.46089 & -65.7 & 5.2 \\ 
58642.46138 & -53.9 & 7.0 \\ 
58642.46187 & -54.8 & 5.7 \\ 
58642.46236 & -49.7 & 7.9 \\ 
58642.46284 & -37.8 & 7.8 \\ 
58642.46333 & -41.7 & 10.3 \\ 
58642.46382 & -51.8 & 3.4 \\ 
58642.46431 & -31.6 & 9.0 \\ 
58642.46480 & -51.6 & 5.9 \\ 
58642.46529 & -37.2 & 9.8 \\ 
58642.46578 & -48.2 & 6.1 \\ 
58642.46627 & -37.7 & 5.2 \\ 
58642.46675 & -44.2 & 4.1 \\ 
58642.46724 & -24.4 & 6.2 \\ 
58642.46773 & -30.6 & 6.0 \\ 
58642.46822 & -25.9 & 4.6 \\ 
58642.46871 & -26.3 & 4.6 \\ 
58642.46920 & -18.1 & 4.0 \\ 
58642.46969 & -14.1 & 8.1 \\ 
58642.47018 & -27.4 & 4.7 \\ 
58642.47066 & -21.5 & 2.1 \\ 
58642.47115 & -14.3 & 7.4 \\ 
58642.47164 & -12.6 & 4.3 \\ 
58642.47213 & -28.6 & 1.5 \\ 
58642.47262 & -19.0 & 3.4 \\ 
58642.47311 & -0.5 & 11.6 \\ 
58642.47360 & -20.5 & 6.4 \\ 
58642.47409 & -9.8 & 6.9 \\ 
58642.47457 & -8.9 & 5.7 \\ 
58642.47506 & -10.3 & 3.4 \\ 
58642.47555 & 0.7 & 11.5 \\ 
58642.47604 & 7.5 & 16.4 \\ 
58642.47653 & 11.8 & 9.0 \\ 
58642.47702 & -2.4 & 5.7 \\ 
58642.47751 & -6.8 & 7.1 \\ 
58642.47800 & -0.0 & 9.8 \\ 
58642.47848 & -0.8 & 5.7 \\ 
58642.47897 & -9.4 & 3.3 \\ 
58642.47946 & -2.8 & 10.7 \\ 
58642.47995 & 3.5 & 7.5 \\ 
58642.48044 & 1.1 & 9.1 \\ 
58642.48093 & 14.3 & 9.1 \\ 
58642.48142 & 21.9 & 6.9 \\ 
58642.48190 & 28.4 & 8.3 \\ 
 58642.48239 & 23.7 & 13.5 \\ 
58642.48288 & 14.2 & 5.0 \\ 
58642.48337 & 43.0 & 15.5 \\ 
58642.48386 & 26.9 & 6.2 \\ 
58642.48435 & 36.2 & 4.8 \\ 
58642.48484 & 33.9 & 7.1 \\ 
58642.48533 & 22.4 & 5.5 \\ 
58642.48581 & 11.6 & 11.3 \\ 
58642.48630 & 17.0 & 6.4 \\ 
58642.48679 & 27.1 & 6.4 \\ 
58642.48728 & 39.2 & 7.3 \\ 
58642.48777 & 44.9 & 4.8 \\ 
58642.48826 & 46.1 & 5.8 \\ 
58642.48875 & 43.1 & 8.9 \\ 
58642.48924 & 28.3 & 3.9 \\ 
58642.48972 & 31.5 & 8.6 \\ 
58642.49021 & 54.6 & 6.2 \\ 
58642.49070 & 56.2 & 6.1 \\ 
58642.49119 & 75.2 & 12.2 \\ 
58642.49168 & 49.2 & 10.9 \\ 
58642.49217 & 55.6 & 16.0 \\ 
58642.49266 & 70.0 & 6.2 \\ 

\hline
\end{tabular}
\end{table} 
\begin{table}[H]\ContinuedFloat
\caption{RV measurements of TYC5977-517-1 (continued)}
    \centering
    \begin{tabular}{lll}
    \hline\hline
         BMJD&  RV & RV error\\
 & [km/s]        & [km/s]\\\hline

58642.49315 & 49.8 & 14.1 \\ 
58642.49363 & 33.3 & 10.4 \\ 
58642.49412 & 61.7 & 4.7 \\ 
58642.49461 & 81.1 & 11.4 \\ 
58642.49510 & 43.7 & 6.4 \\ 
58642.49559 & 54.8 & 5.2 \\ 
58642.49608 & 49.0 & 5.7 \\ 
58642.49657 & 53.7 & 7.0 \\ 
58642.49705 & 62.3 & 7.8 \\ 
58642.49754 & 58.5 & 4.2 \\ 
58642.49803 & 82.1 & 9.6 \\ 
58642.49852 & 72.3 & 6.6 \\ 
58642.49901 & 73.6 & 7.0 \\ 
58642.49950 & 65.5 & 8.3 \\ 
58642.49999 & 83.4 & 13.7 \\ 
58642.50048 & 71.0 & 5.7 \\ 
58642.50096 & 76.7 & 7.1 \\ 
58642.50145 & 66.0 & 3.1 \\ 
58642.50194 & 80.2 & 8.7 \\ 
58642.50243 & 77.3 & 6.5 \\ 
58642.50292 & 79.4 & 3.1 \\ 
58642.50341 & 78.8 & 6.0 \\ 
58642.50390 & 83.3 & 6.3 \\ 
58642.50439 & 86.1 & 6.2 \\ 
58642.50487 & 79.1 & 3.3 \\ 
58642.50536 & 93.7 & 6.8 \\ 
58642.50585 & 81.1 & 3.1 \\ 
58642.50634 & 84.0 & 8.7 \\ 
58642.50683 & 107.0 & 10.3 \\ 
58642.50732 & 103.8 & 8.0 \\ 
58642.50781 & 91.0 & 8.8 \\ 
58642.50830 & 93.8 & 3.5 \\ 
58642.50879 & 84.8 & 8.3 \\ 
58642.50927 & 113.5 & 11.0 \\ 
58642.50976 & 98.8 & 9.6 \\ 
58642.51025 & 98.6 & 6.5 \\ 
58642.51074 & 89.0 & 5.6 \\ 
58642.51123 & 100.5 & 7.9 \\ 
58642.51172 & 107.0 & 11.2 \\ 
58642.51221 & 100.8 & 9.6 \\ 
58642.51269 & 97.7 & 4.6 \\ 
58642.51318 & 104.3 & 12.3 \\ 
58642.51367 & 98.6 & 4.6 \\ 
58642.51416 & 94.6 & 5.2 \\ 
58642.51465 & 78.4 & 5.5 \\ 
58642.51514 & 99.0 & 4.6 \\ 
58642.51563 & 121.1 & 11.1 \\ 
58642.51612 & 108.5 & 7.1 \\ 
58643.45344 & -82.9 & 3.1 \\ 
58643.45393 & -79.4 & 6.5 \\ 
58643.45442 & -86.6 & 4.0 \\ 
58643.45491 & -86.1 & 4.7 \\ 
58643.45539 & -84.8 & 4.3 \\ 
58643.45588 & -86.8 & 4.9 \\ 
58643.45637 & -80.1 & 6.0 \\ 
58643.45686 & -79.5 & 4.3 \\ 
58643.45735 & -80.9 & 4.3 \\ 
58643.45784 & -77.7 & 5.9 \\ 
58643.45833 & -80.6 & 4.3 \\ 
58643.45882 & -83.9 & 6.9 \\ 
58643.45931 & -81.9 & 3.6 \\ 
58643.45979 & -68.2 & 6.8 \\ 
58643.46028 & -78.2 & 4.8 \\ 
58643.46077 & -71.3 & 5.3 \\ 
58643.46126 & -85.2 & 5.8 \\ 
58643.46175 & -68.6 & 6.1 \\ 

\hline
\end{tabular}
\end{table} 
\begin{table}[H]\ContinuedFloat
\caption{RV measurements of TYC5977-517-1 (continued)}
    \centering
    \begin{tabular}{lll}
    \hline\hline
         BMJD&  RV & RV error\\
 & [km/s]        & [km/s]\\\hline

58643.46224 & -69.0 & 5.5 \\ 
58643.46273 & -72.0 & 7.0 \\ 
58643.46322 & -69.7 & 5.3 \\ 
58643.46370 & -62.5 & 5.8 \\ 
58643.46419 & -69.5 & 4.1 \\ 
58643.46468 & -65.6 & 5.3 \\ 
58643.46517 & -69.8 & 7.4 \\ 
58643.46566 & -60.0 & 4.9 \\ 
58643.46615 & -59.1 & 3.4 \\ 
58643.46664 & -56.7 & 5.9 \\ 
58643.46713 & -53.7 & 5.1 \\ 
58643.46761 & -49.7 & 4.4 \\ 
58643.46810 & -46.7 & 3.5 \\ 
58643.46859 & -52.2 & 6.0 \\ 
58643.46908 & -44.1 & 5.3 \\ 
58643.46957 & -57.8 & 3.6 \\ 
58643.47006 & -47.2 & 3.7 \\ 
58643.47054 & -44.8 & 5.9 \\ 
58643.47103 & -46.1 & 5.2 \\ 
58643.47152 & -43.2 & 6.8 \\ 
58643.47201 & -36.2 & 3.9 \\ 
58643.47250 & -37.0 & 6.2 \\ 
58643.47299 & -38.1 & 4.7 \\ 
58643.47348 & -34.1 & 2.8 \\ 
58643.47397 & -38.8 & 2.8 \\ 
58643.47446 & -36.2 & 6.3 \\ 
58643.47494 & -39.6 & 3.7 \\ 
58643.47543 & -30.8 & 3.6 \\ 
58643.47592 & -28.5 & 4.0 \\ 
58643.47641 & -23.5 & 4.0 \\ 
58643.47690 & -30.3 & 4.0 \\ 
58643.47739 & -29.4 & 2.2 \\ 
58643.47788 & -28.2 & 5.6 \\ 
58643.47836 & -28.2 & 3.7 \\ 
58643.47885 & -15.3 & 5.4 \\ 
58643.47934 & -22.5 & 5.1 \\ 
58643.47983 & -17.1 & 4.0 \\ 
58643.48032 & -3.2 & 7.3 \\ 
58643.48081 & -8.8 & 7.1 \\ 
58643.48130 & -13.9 & 3.0 \\ 
58643.48179 & -1.5 & 3.2 \\ 
58643.48227 & -5.0 & 4.7 \\ 
58643.48276 & -1.2 & 5.5 \\ 
58643.48325 & -4.5 & 6.6 \\ 
58643.48374 & -2.0 & 6.1 \\ 
58643.48423 & 4.8 & 3.3 \\ 
58643.48472 & 2.6 & 7.5 \\ 
58643.48521 & 16.1 & 7.0 \\ 
58643.48570 & 8.9 & 5.4 \\ 
58643.48619 & 11.1 & 6.5 \\ 
58643.48667 & 20.0 & 3.9 \\ 
58643.48716 & 13.1 & 4.6 \\ 
58643.48765 & 15.0 & 5.9 \\ 
58643.48814 & 24.8 & 8.4 \\ 
58643.48863 & 15.2 & 5.6 \\ 
58643.48912 & 22.6 & 3.8 \\ 
58643.48961 & 30.5 & 8.7 \\ 
58643.49010 & 21.3 & 5.5 \\ 
58643.49058 & 26.3 & 2.8 \\ 
58643.49107 & 21.9 & 3.2 \\ 
58643.49156 & 24.2 & 4.4 \\ 
58643.49205 & 28.3 & 6.9 \\ 
58643.49254 & 27.7 & 3.8 \\ 
58643.49303 & 32.2 & 5.7 \\ 
58643.49352 & 39.1 & 7.6 \\ 
58643.49401 & 25.2 & 2.6 \\ 
58643.49449 & 35.8 & 3.6 \\

\hline
\end{tabular}
\end{table} 
\begin{table}[H]\ContinuedFloat
\caption{RV measurements of TYC5977-517-1 (continued)}
    \centering
    \begin{tabular}{lll}
    \hline\hline
         BMJD&  RV & RV error\\
 & [km/s]        & [km/s]\\\hline

58643.49498 & 40.9 & 3.6 \\ 
58643.49547 & 38.7 & 4.5 \\ 
58643.49596 & 49.1 & 6.8 \\ 
58643.49645 & 45.8 & 4.5 \\ 
58643.49694 & 45.3 & 2.3 \\ 
58643.49743 & 40.1 & 3.2 \\ 
58643.49791 & 42.9 & 1.6 \\ 
58643.49840 & 41.6 & 2.3 \\ 
58643.49889 & 48.3 & 7.5 \\ 
58643.49938 & 58.9 & 3.3 \\ 
58643.49987 & 56.4 & 5.4 \\ 
58643.50036 & 41.6 & 6.8 \\ 
58643.50085 & 54.2 & 2.2 \\ 
58643.50134 & 49.3 & 4.1 \\ 
58643.50182 & 57.8 & 4.7 \\ 
58643.50231 & 70.1 & 3.2 \\ 
58643.50280 & 59.2 & 5.4 \\ 
58643.50329 & 71.2 & 2.5 \\ 
58643.50378 & 75.6 & 5.4 \\ 
58643.50427 & 70.6 & 3.1 \\ 
58643.50476 & 69.1 & 4.0 \\ 
58643.50525 & 78.9 & 8.8 \\ 
58643.50574 & 68.4 & 4.0 \\ 
58643.50622 & 78.3 & 2.4 \\ 
58643.50671 & 89.7 & 6.0 \\ 
58643.50720 & 75.8 & 5.7 \\ 
58643.50769 & 89.3 & 3.0 \\ 
58643.50818 & 89.3 & 1.9 \\ 

\hline
\end{tabular}
\end{table}

\begin{table}[H]
\caption{RV measurements of EC 01578-1743}
    \centering
    \begin{tabular}{lll}\hline\hline
         BMJD&  RV & RV error\\
 & [km/s]        & [km/s]\\\hline
58117.61897 & -97.5 & 1.3 \\ 
58136.59910 & 63.0 & 2.0 \\ 
58144.57225 & 31.1 & 1.7 \\ 
58149.54667 & 41.0 & 1.5 \\ 
58149.56061 & 19.9 & 1.3 \\ 
58174.50767 & 25.6 & 1.4 \\ 
58176.51123 & -90.2 & 1.8 \\ 
58329.90219 & 50.4 & 1.2 \\ 
58329.91613 & 61.4 & 1.1 \\ 
58329.93007 & 67.8 & 1.0 \\ 
58332.86085 & -75.1 & 1.2 \\ 
58332.87480 & -92.2 & 2.8 \\ 
58332.88874 & -105.0 & 2.4 \\ 
58344.82524 & -39.3 & 1.3 \\ 
58344.83918 & -14.7 & 1.4 \\ 
58344.85313 & 13.2 & 1.7 \\ 
58344.86707 & 43.0 & 1.2 \\ 
58344.88101 & 58.0 & 1.2 \\ 
58345.89400 & 32.3 & 1.5 \\ 
58345.90800 & 52.1 & 1.0 \\ 
58345.92199 & 64.2 & 1.0 \\ 
58345.93598 & 65.3 & 0.9 \\ 
58357.80549 & 67.1 & 1.1 \\ 
58357.81943 & 60.6 & 1.3 \\ 
58357.83338 & 46.1 & 1.0 \\ 
58357.84732 & 24.7 & 0.9 \\ 
58357.86126 & -2.4 & 1.0 \\ 
58358.80734 & 44.9 & 1.1 \\ 
58358.82128 & 60.3 & 1.1 \\ 
58358.83523 & 66.9 & 1.2 \\ 
58358.84917 & 62.6 & 1.0 \\ 
58358.86311 & 50.7 & 1.1 \\ 
58379.88558 & -102.6 & 3.1 \\ 
58382.80772 & 41.2 & 1.7 \\ 
58382.82166 & 57.9 & 1.5 \\ 
58383.75839 & -106.0 & 4.3 \\ 
58383.77257 & -87.5 & 1.7 \\ 
58384.72334 & -44.9 & 1.3 \\ 
58384.73941 & -72.7 & 1.4 \\ \hline
    \end{tabular}
    \label{RV_ec}
\end{table}

\begin{table}[H]
\caption{RV measurements of KPD 0629-0016}
    \centering
    \begin{tabular}{llll}
    \hline\hline
         BMJD&  RV & RV error & instrument\\
 & [km/s]        & [km/s]\\\hline

54476.69426 &  37.1 & 8.6 & EMMI\\
54477.61197 &  -1.3 & 8.0 &EMMI\\
54477.68846 & -13.4 & 3.4 &EMMI\\
54478.62101 & -34.4 & 7.1 &EMMI\\
54478.67763 & -58.2 & 3.7 &EMMI\\
54478.77402 & -73.3 & 4.3 &EMMI\\
54479.62414 & -75.1 & 8.9 &EMMI\\
54479.63799 & -46.8 & 10.1 &EMMI\\
54479.66199 & -53.9 & 8.3 &EMMI\\
54479.67306 & -63.6 & 8.1 &EMMI\\
54479.68573 & -74.7 & 8.2 &EMMI\\
54479.69438 & -47.8 & 7.7 &EMMI\\
54755.78643225&  57.23 &   13.05 & EFOSC2\\
54756.78528812&  64.36 &   12.14 & EFOSC2\\
56690.53874 &    51.89 &   9.95   & EFOSC2\\
56690.61981 &    44.66 &   9.95   & EFOSC2\\
56691.52766  &   -8.88 &    11.04 & EFOSC2\\

\hline
    \label{RV_kpd}
\end{tabular}
\end{table}

\onecolumn
\section{Light curves of the reflection effect, ellipsoidal and Doppler beaming systems}
\begin{figure}[H]
\begin{subfigure}{0.5\linewidth}
    \includegraphics[width=\linewidth]{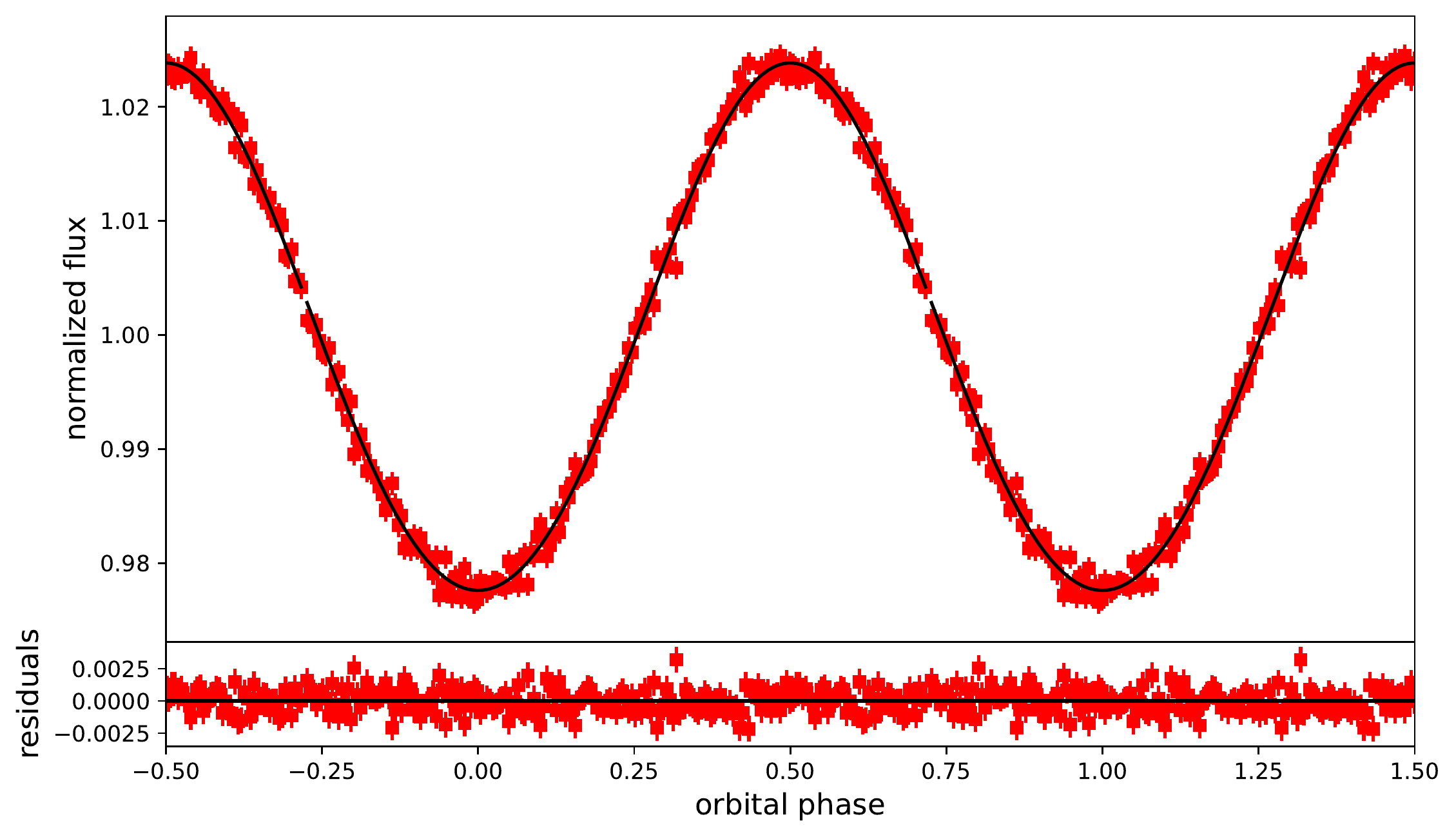}
    \caption{BPSCS 22169-1} 
    \label{lc_bpsc}
\end{subfigure}\hfill
\begin{subfigure}{0.5\linewidth}
\includegraphics[width=\linewidth]{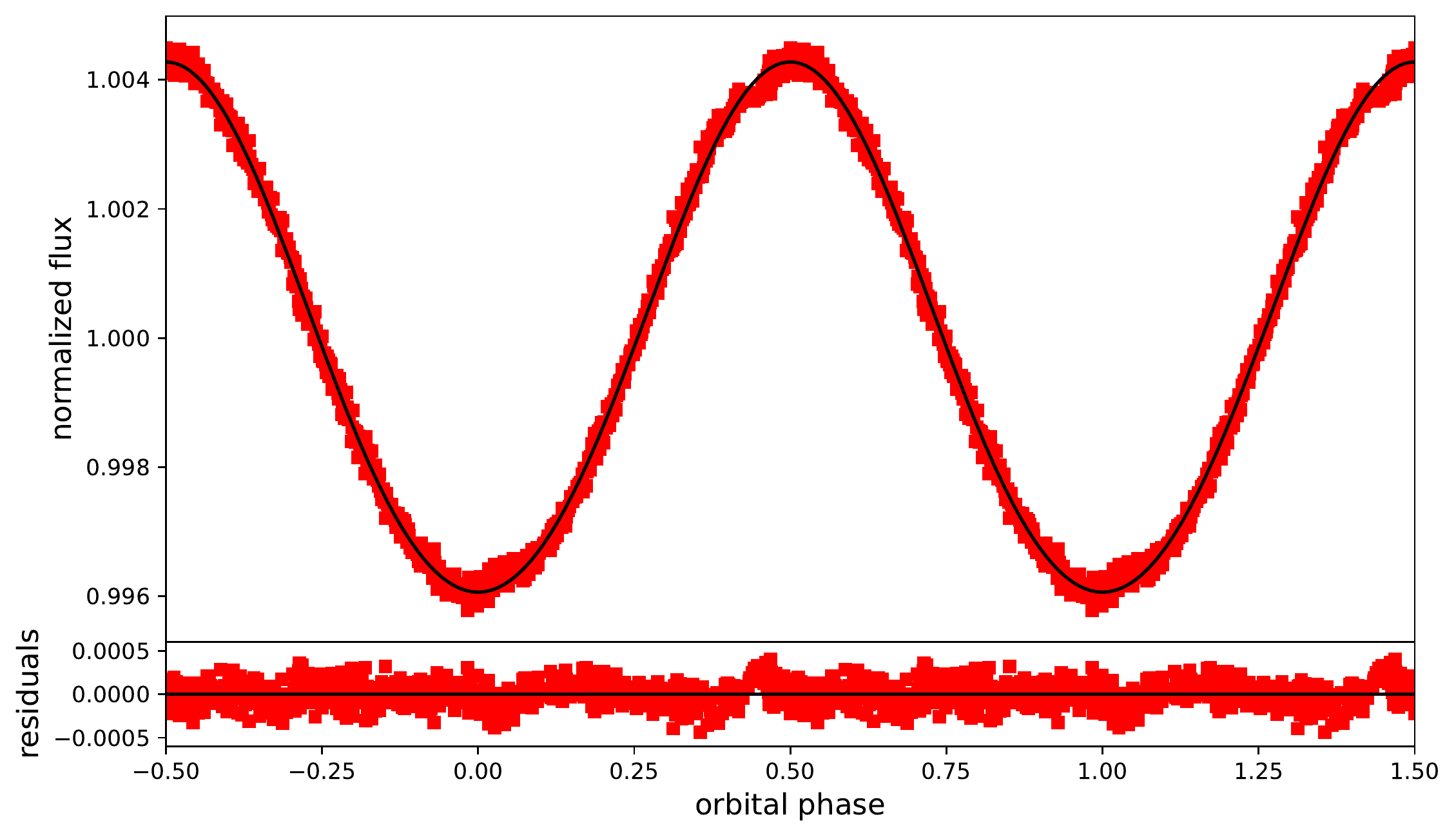}
\caption{PHL457} 
\label{lc_phl}
\end{subfigure}\hfill
\begin{subfigure}{0.5\linewidth}
    \includegraphics[width=\linewidth]{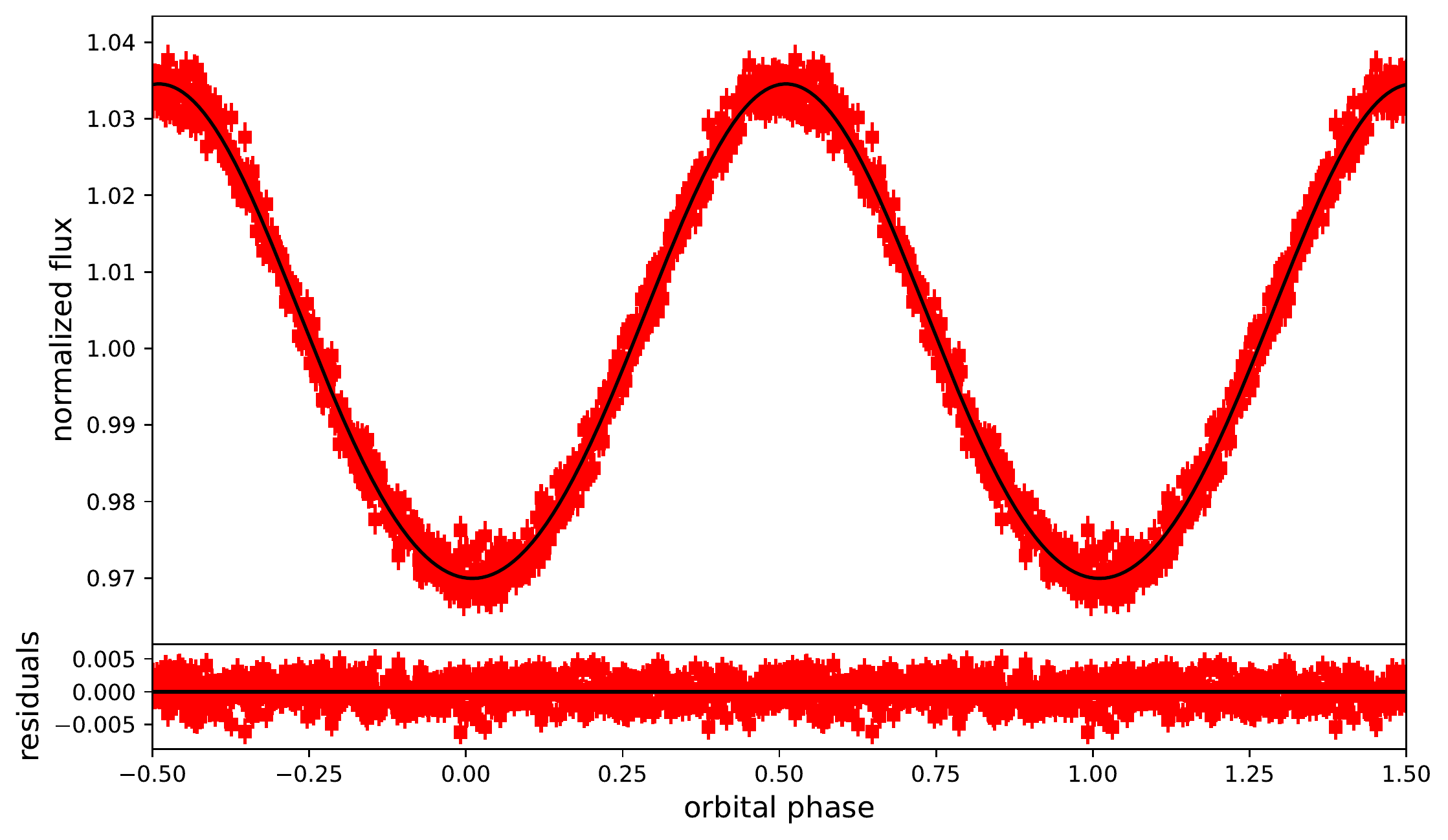}
    \caption{KBS13} 
\label{lc_kbs13}
\end{subfigure}\hfill
\begin{subfigure}{0.5\linewidth}
\includegraphics[width=\linewidth]{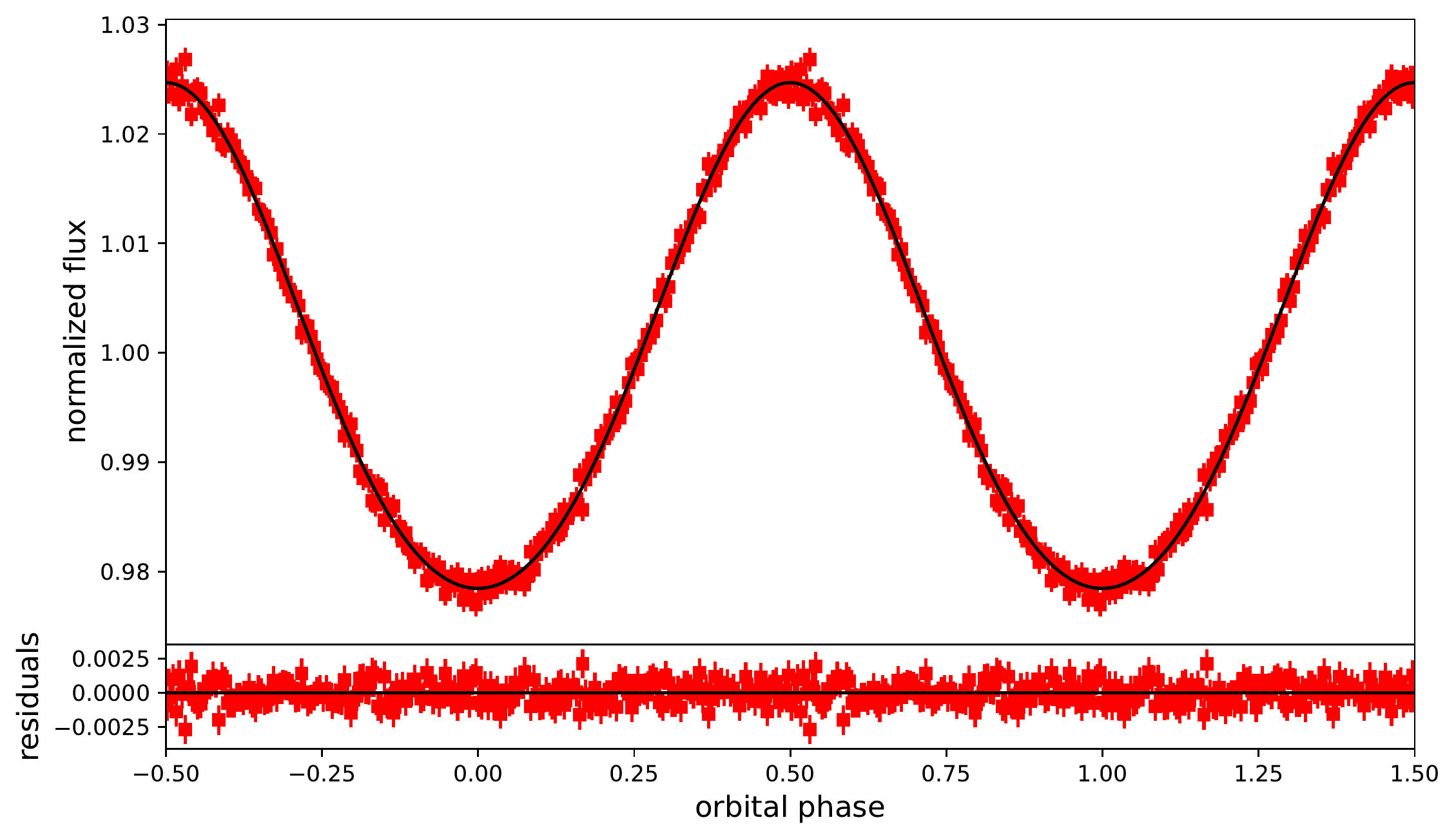}
\caption{Feige 48} 
\label{lc_feige48}
\end{subfigure}\hfill
\begin{subfigure}{0.5\linewidth}
\includegraphics[width=\linewidth]{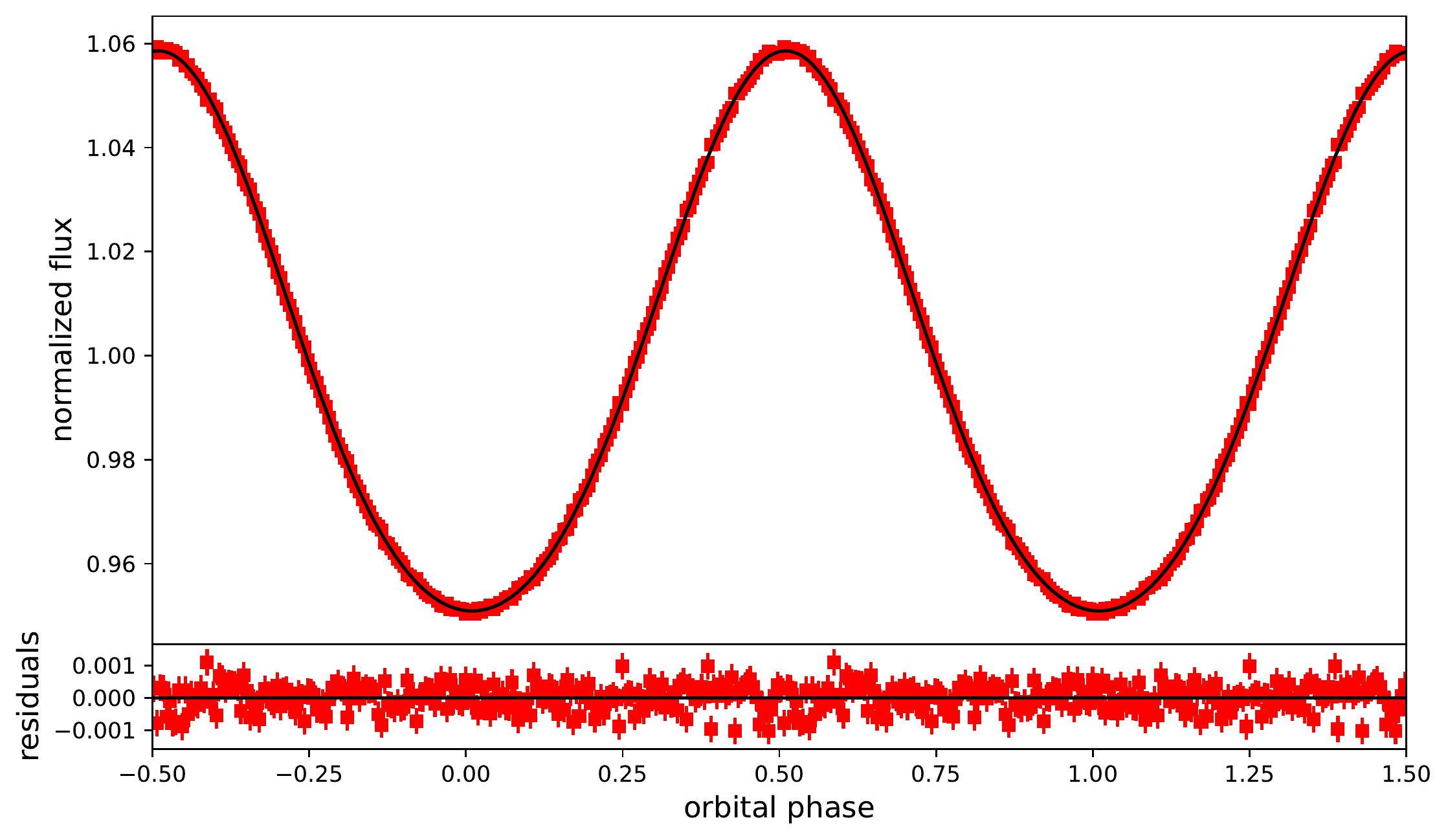}
\caption{GALEX J2205-3141} 
\label{lc_galexj22}
\end{subfigure}\hfill
\begin{subfigure}{0.5\linewidth}
\includegraphics[width=\linewidth]{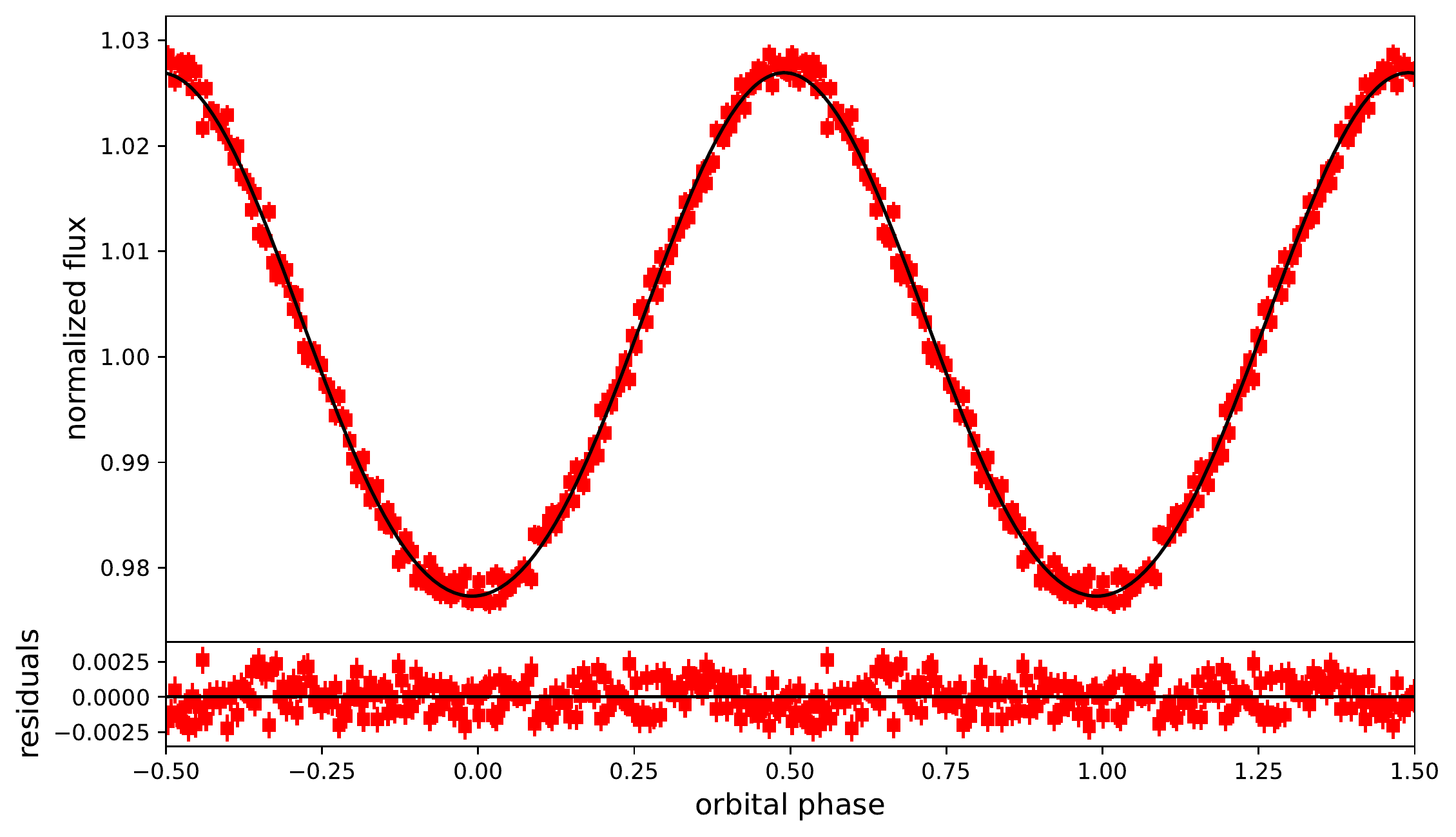}
\caption{GALEX J09348-2512} 
\label{lc_galexj09}
\end{subfigure}\hfill
\begin{subfigure}{0.5\linewidth}
\includegraphics[width=\linewidth]{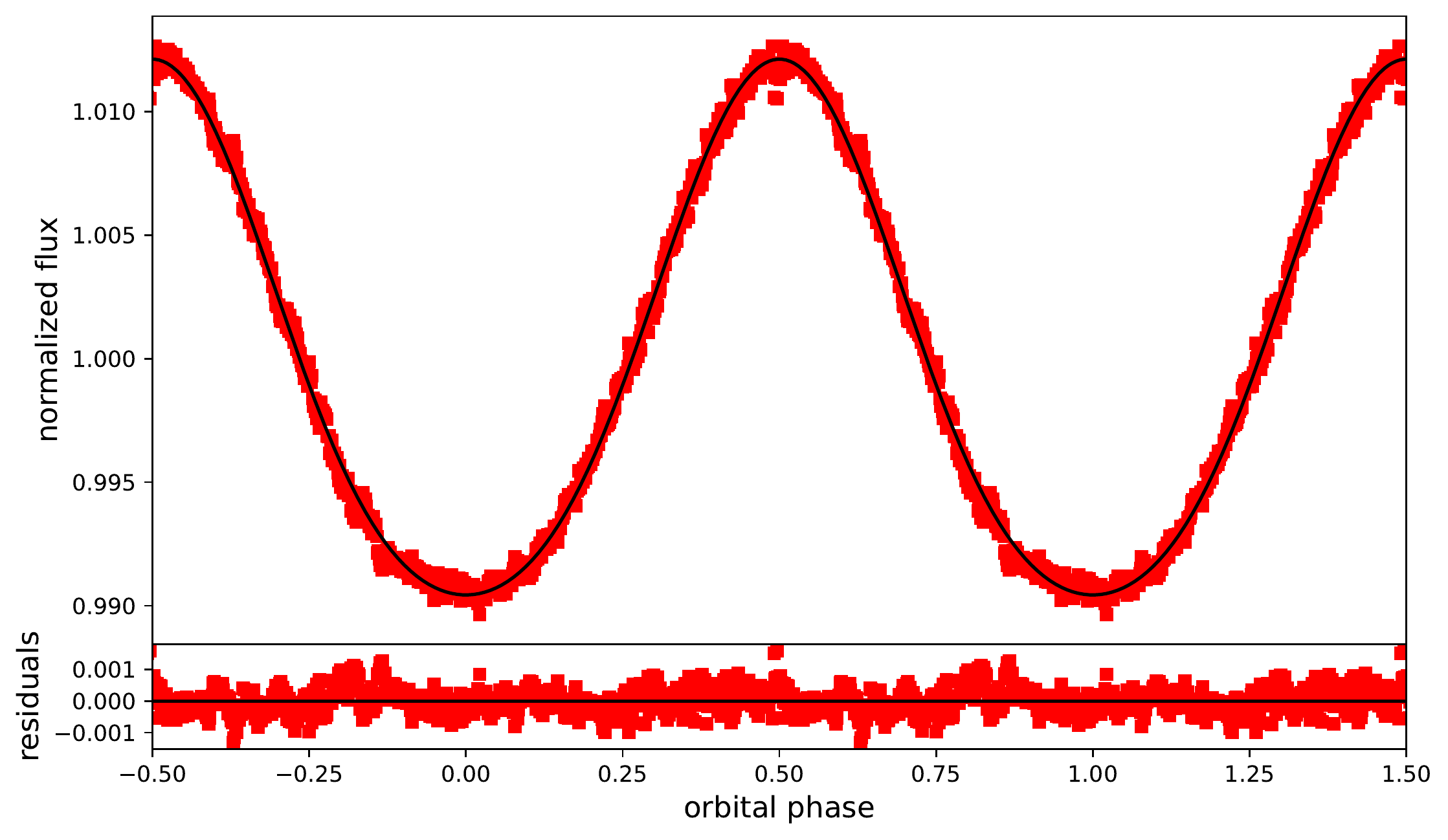}
\caption{EQ Psc} 
\label{lc_eqpsc}
\end{subfigure}\hfill
\begin{subfigure}{0.5\linewidth}
\includegraphics[width=\linewidth]{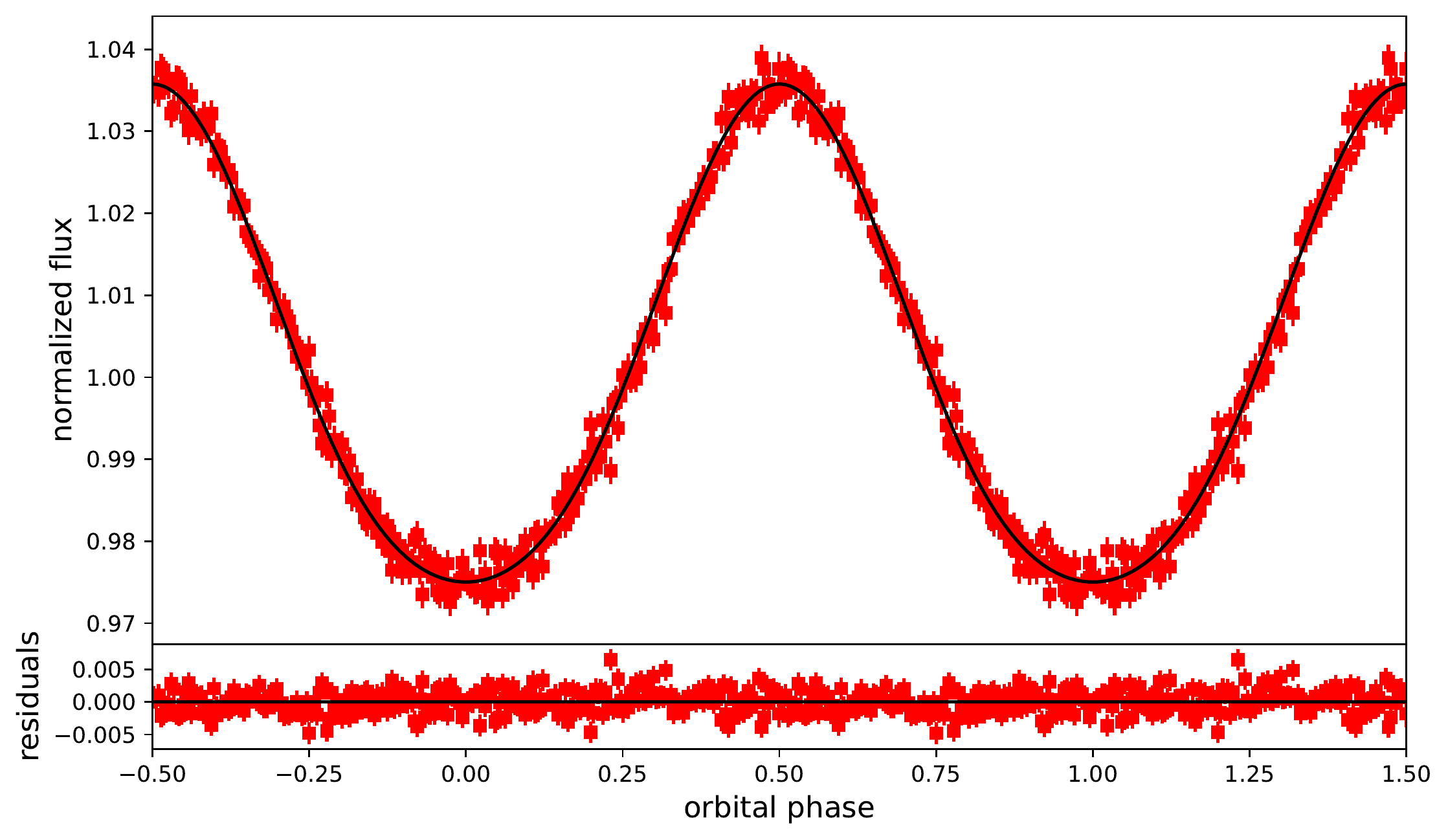}
\caption{PG 1329+159} 
\label{lc_pg1329}
\end{subfigure}\hfill
\vspace{-0.25cm}
\caption{Phased light curve (given by the red squares) together with the best-fit model given by the black line. The lower panel shows the residuals.} 
\label{refl1}
\end{figure}

\begin{figure}\ContinuedFloat

\begin{subfigure}{0.5\linewidth}
\includegraphics[width=\linewidth]{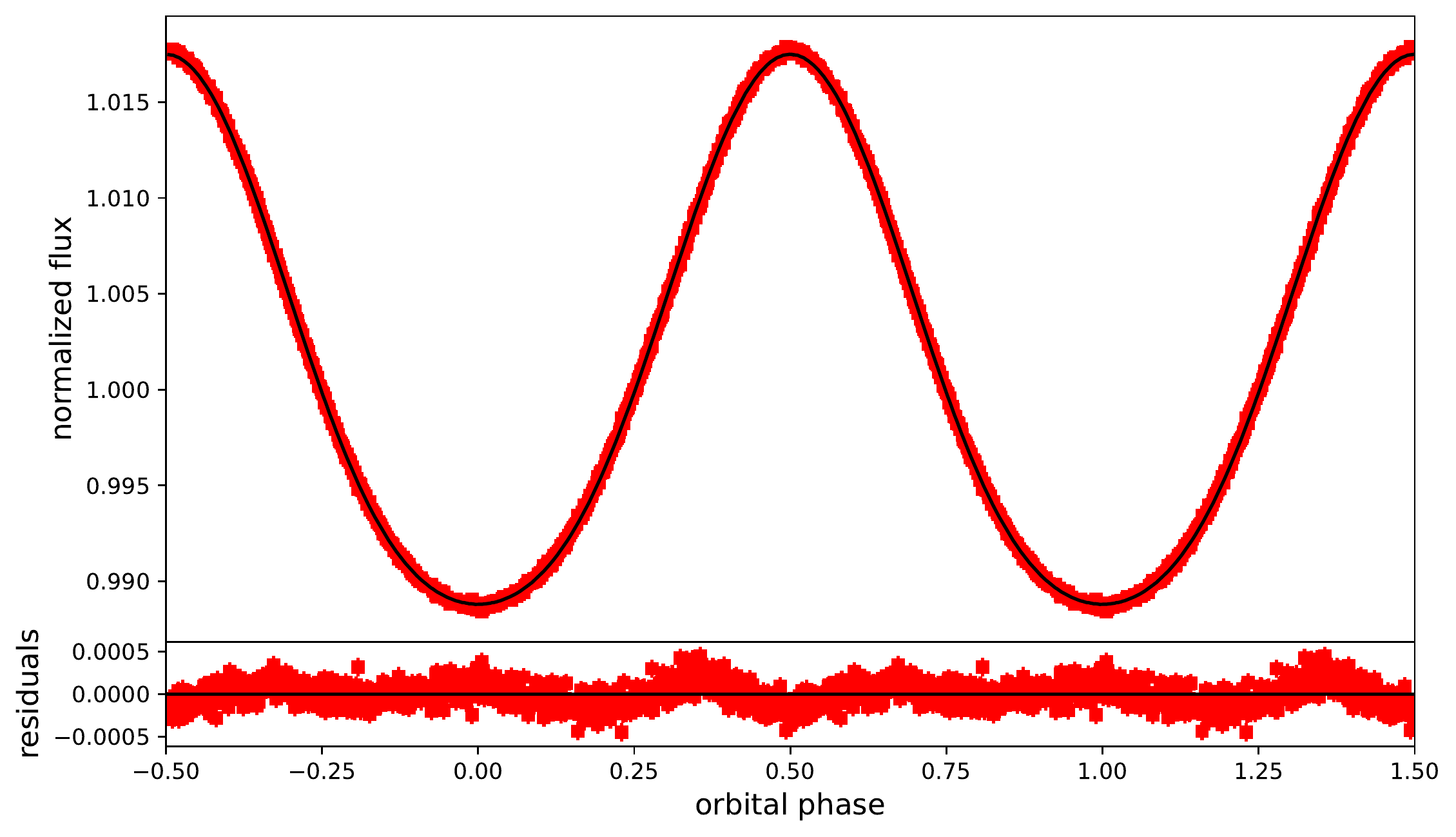}
\caption{CPD-64 481} 
\label{lc_cpd}
\end{subfigure}\hfill
\begin{subfigure}{0.5\linewidth}
\includegraphics[width=\linewidth]{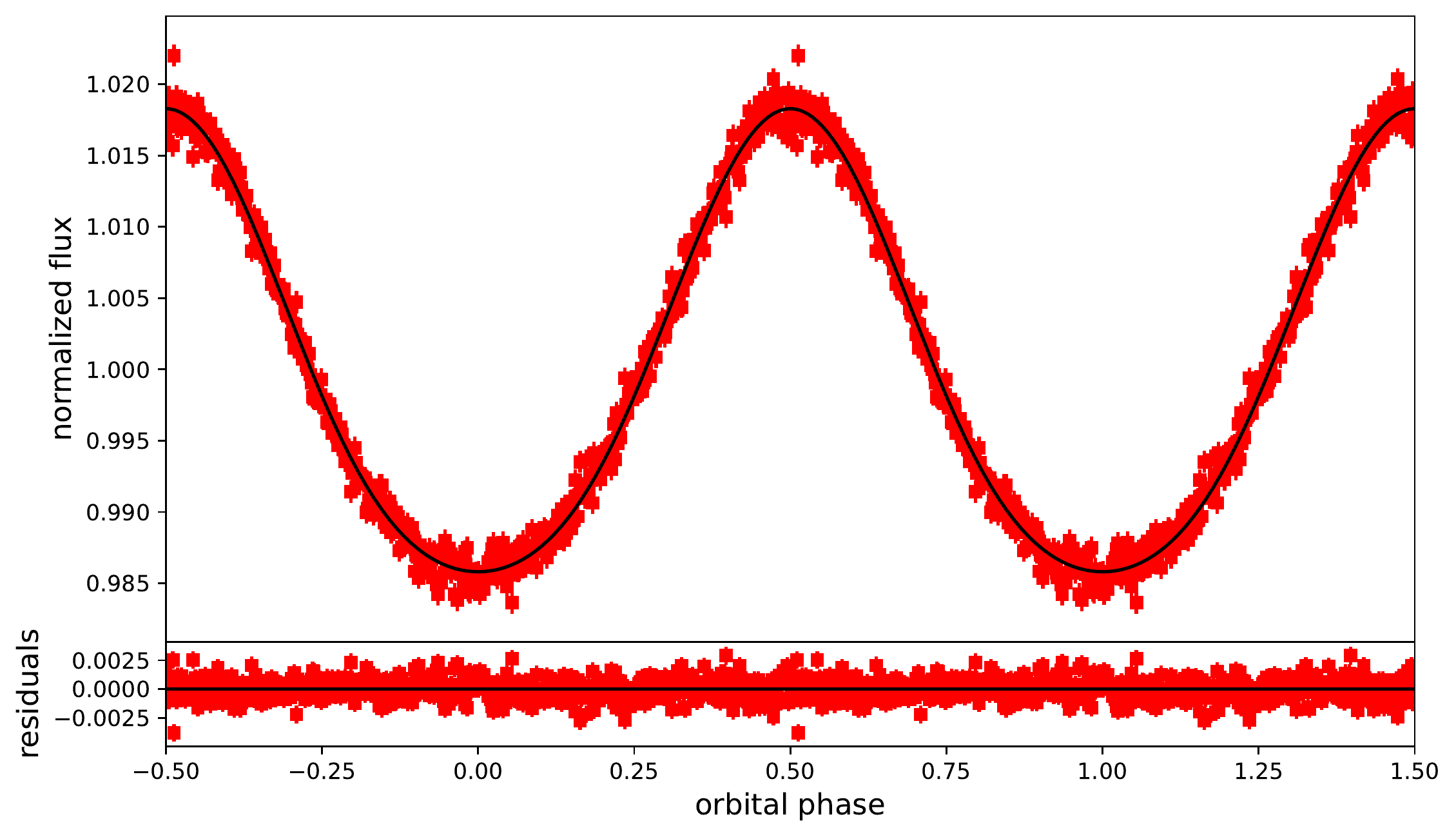}
\caption{JL 82} 
\label{lc_jl82}
\end{subfigure}\hfill
\begin{subfigure}{0.5\linewidth}
\includegraphics[width=\linewidth]{lc_TYC.pdf}
\caption{TYC5977-517-1} 
\label{lc_tyc}
\end{subfigure}\hfill
\begin{subfigure}{0.5\linewidth}
\includegraphics[width=\linewidth]{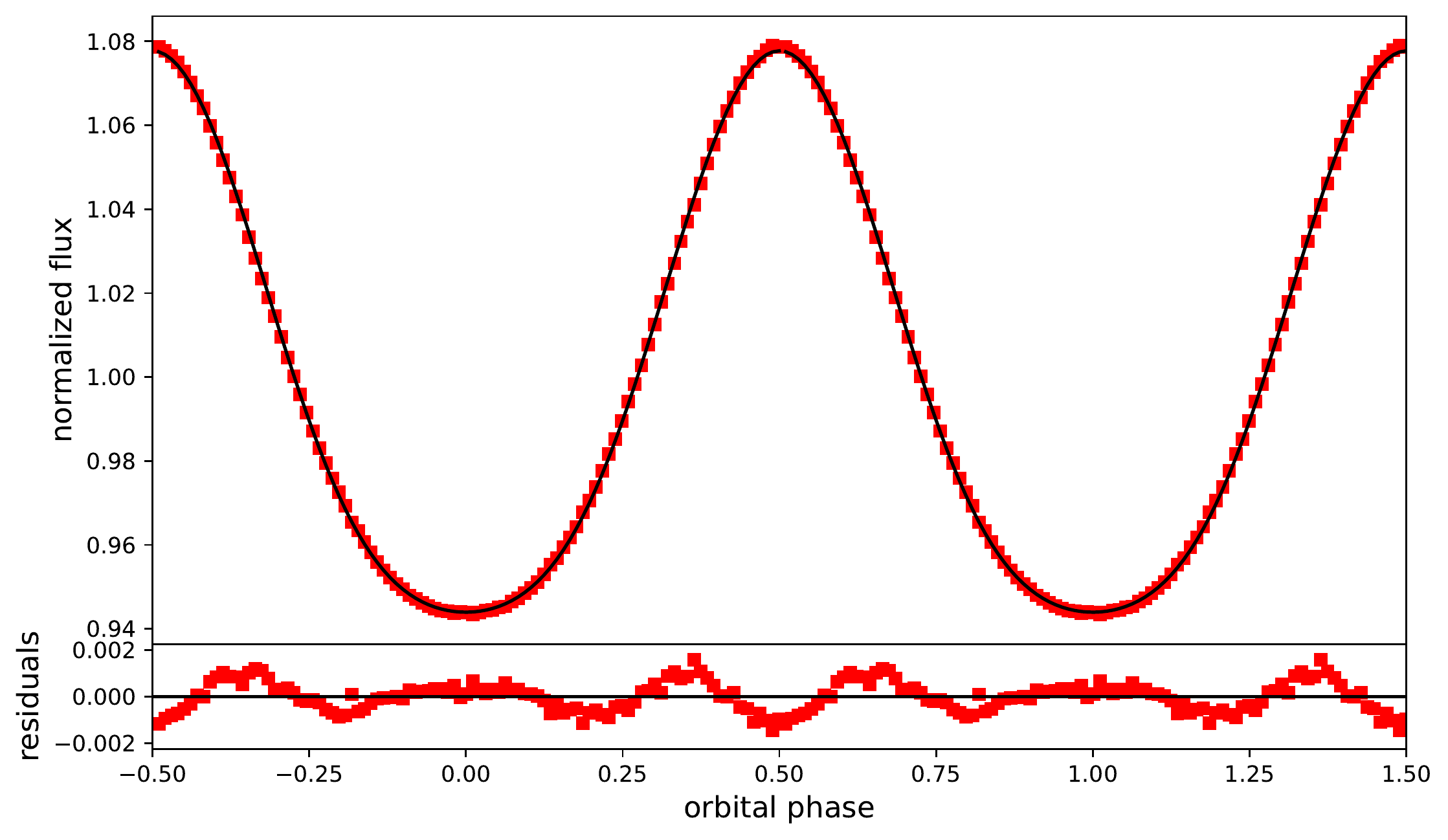}
\caption{GALEX J0321+4727} 
\label{lc_galexj03}
\end{subfigure}\hfill
\begin{subfigure}{0.5\linewidth}
\includegraphics[width=\linewidth]{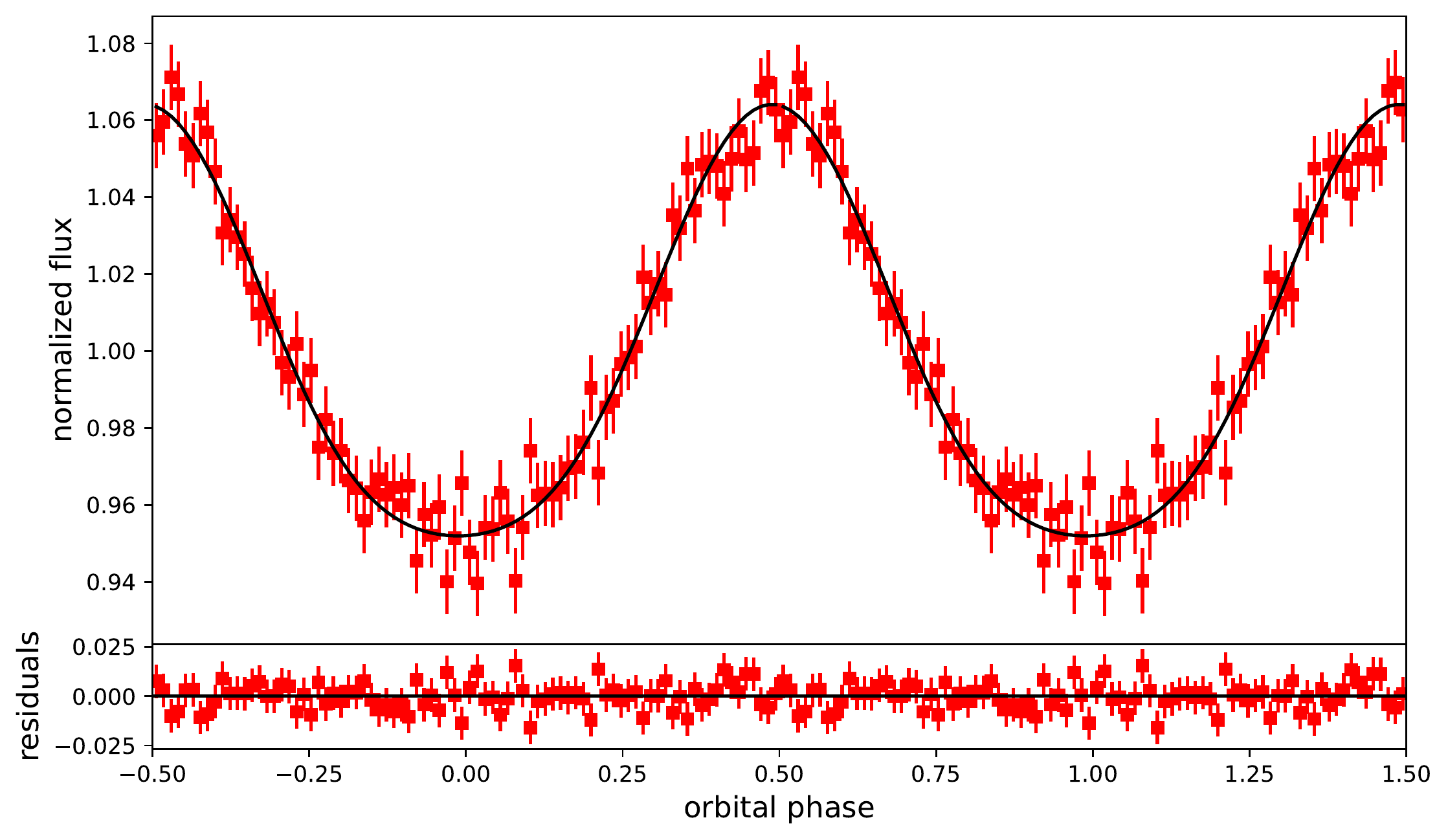}
\caption{SDSS J012022+395059} 
\label{lc_j0102}
\end{subfigure}\hfill
\begin{subfigure}{0.5\linewidth}
\includegraphics[width=\linewidth]{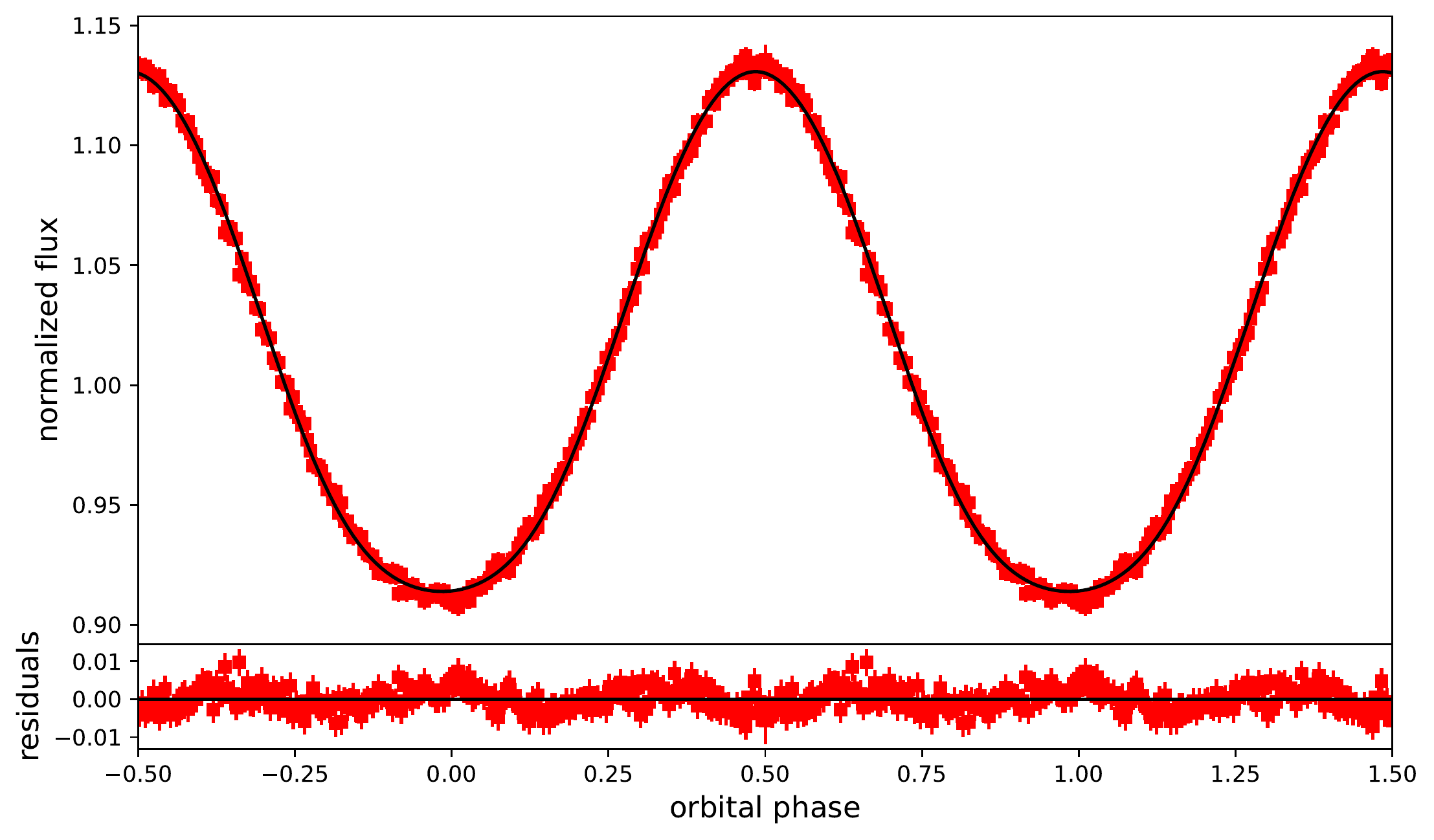}
\caption{UVEX 0328+5035} 
\label{lc_uvex}
\end{subfigure}\hfill
\begin{subfigure}{0.5\linewidth}
\includegraphics[width=\linewidth]{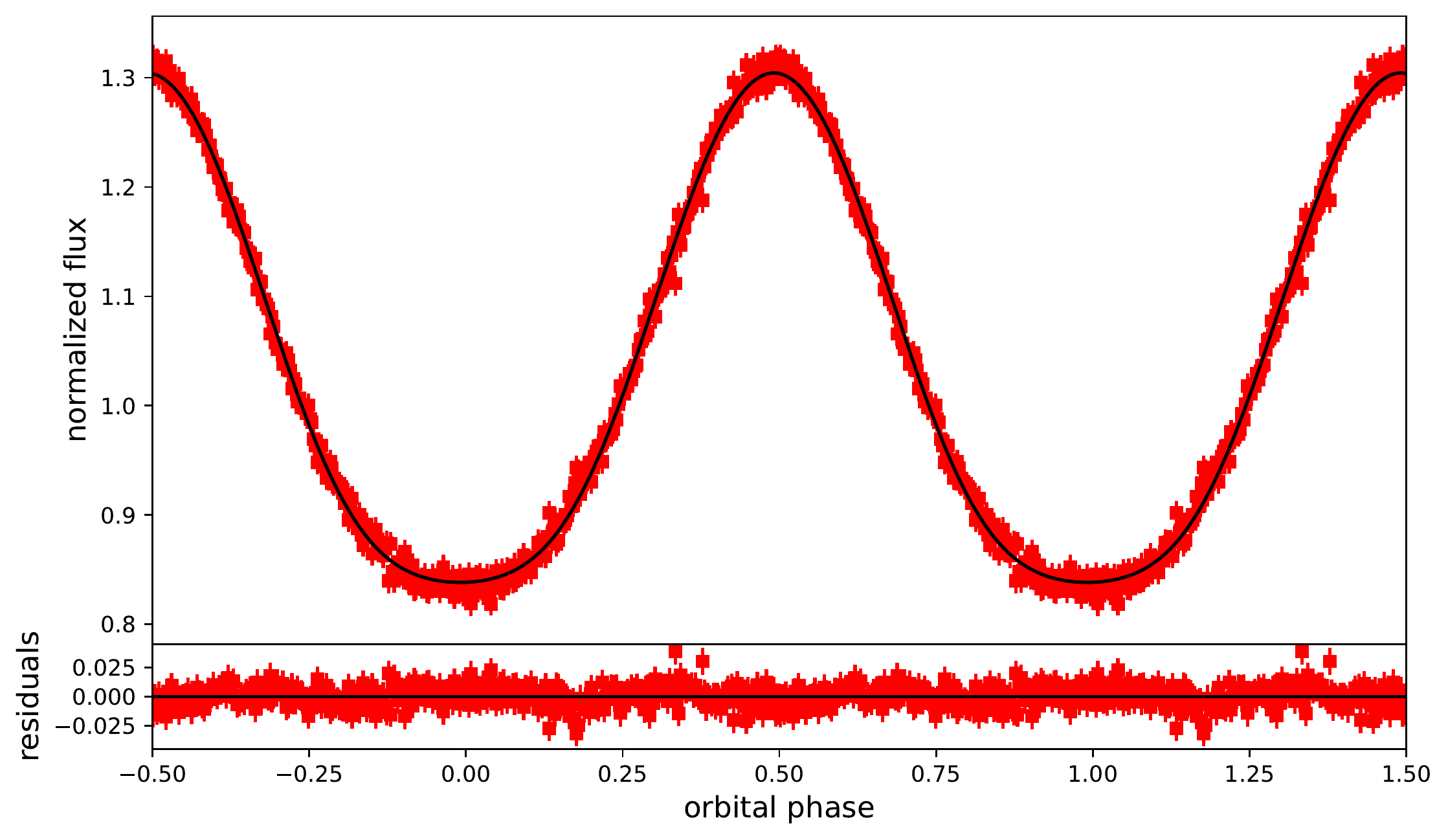}
\caption{HS 2333+3927} 
\label{lc_hs2223}
\end{subfigure}\hfill
\begin{subfigure}{0.5\linewidth}
\includegraphics[width=\linewidth]{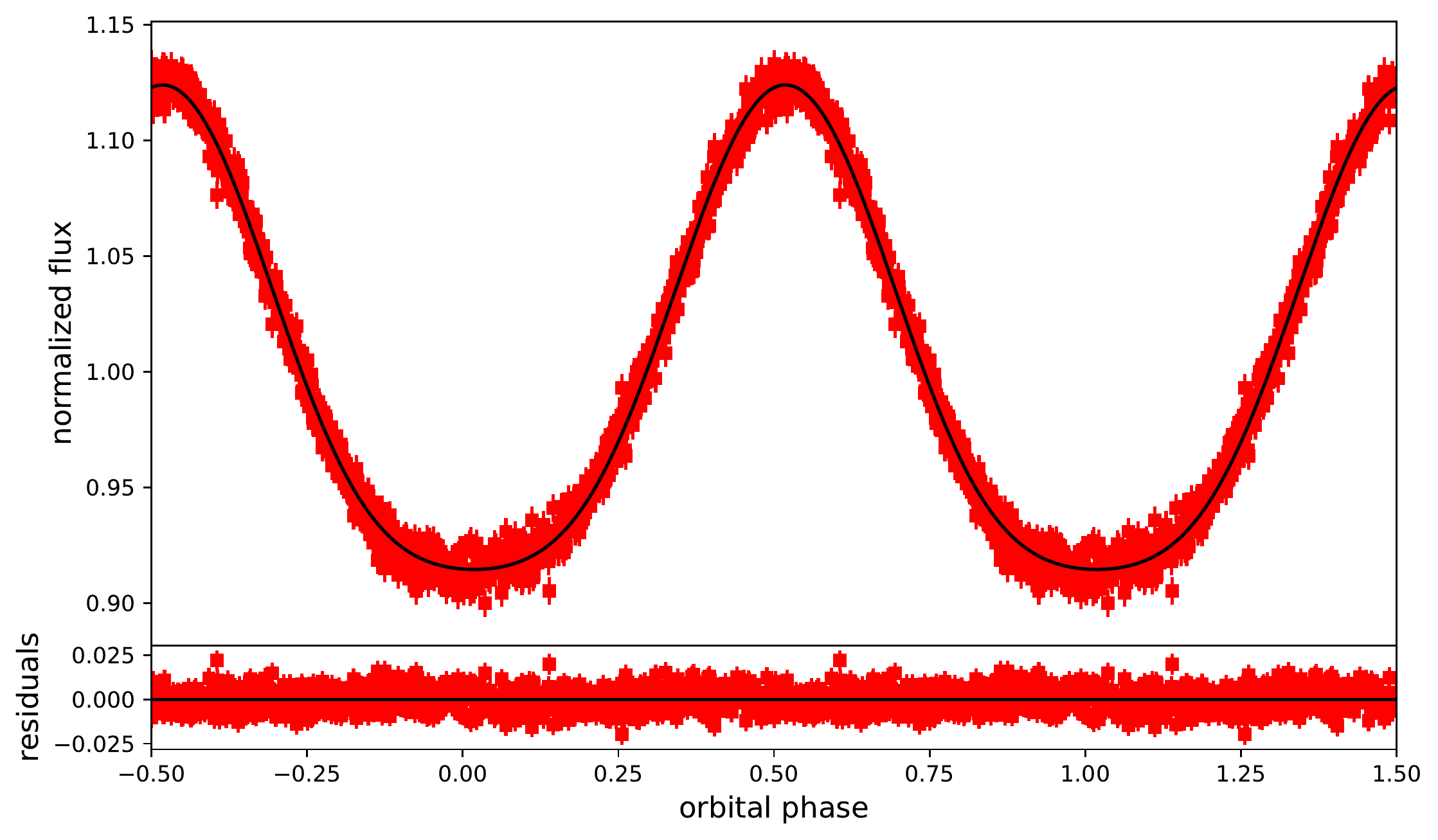}
\caption{V1405 Ori} 
\label{lc_vori}
\end{subfigure}\hfill

\caption{ Phased light curve (given by the red squares) together with the best-fit model given by the black line. The lower panel shows the residuals (continued).} 
\label{refl2}
\end{figure}

\begin{figure}\ContinuedFloat
\begin{subfigure}{0.5\linewidth}
\includegraphics[width=\linewidth]{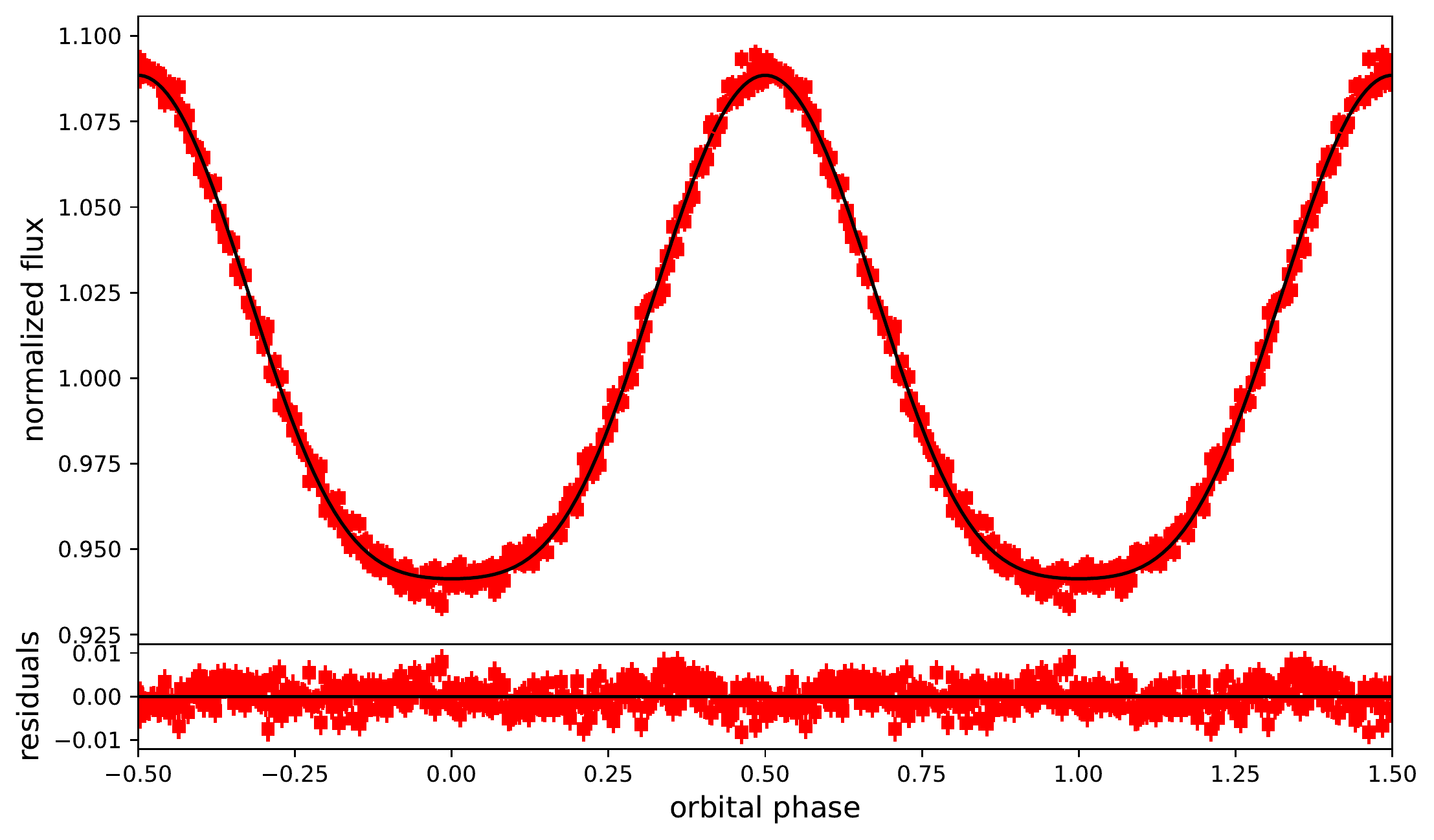}
\caption{HE 1318-2111} 
\label{lc_he1318}
\end{subfigure}\hfill
\begin{subfigure}{0.5\linewidth}
\includegraphics[width=\linewidth]{lc_EC01578.pdf}
\caption{EC 01578-1743} 
\label{lc_ec01678}
\end{subfigure}\hfill
\begin{subfigure}{0.5\linewidth}
\includegraphics[width=\linewidth]{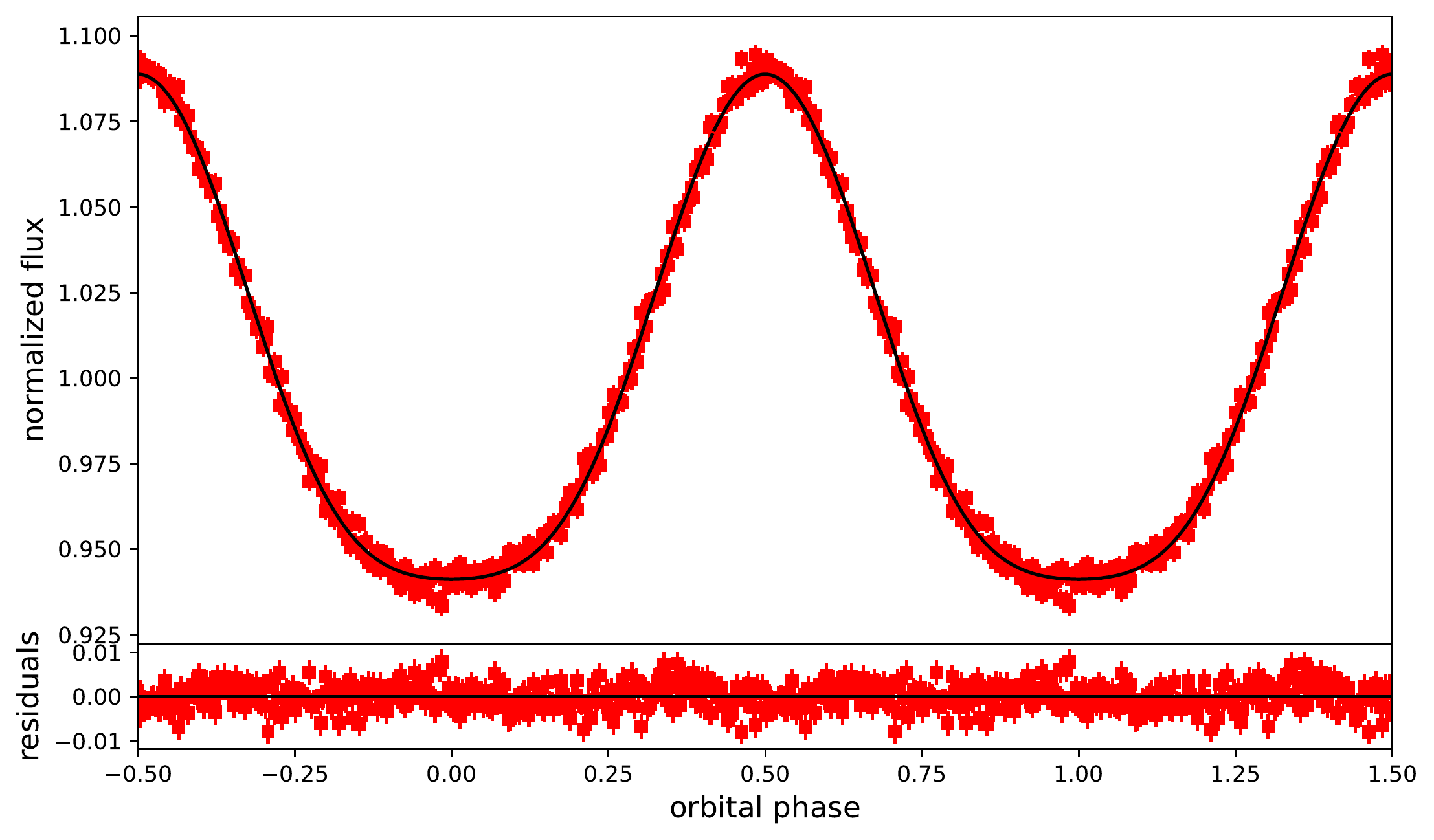}
\caption{HE 0230-4323} 
\label{lc_he0230}
\end{subfigure}\hfill

\caption{Continuation: Phased light curve (given by the red squares) together with the best-fit model given by the black line. The lower panel shows the residuals (continued).} 
\label{refl3}
\end{figure}



\begin{figure*}
    \centering
    \begin{subfigure}{0.5\textwidth}
        \includegraphics[width=\linewidth]{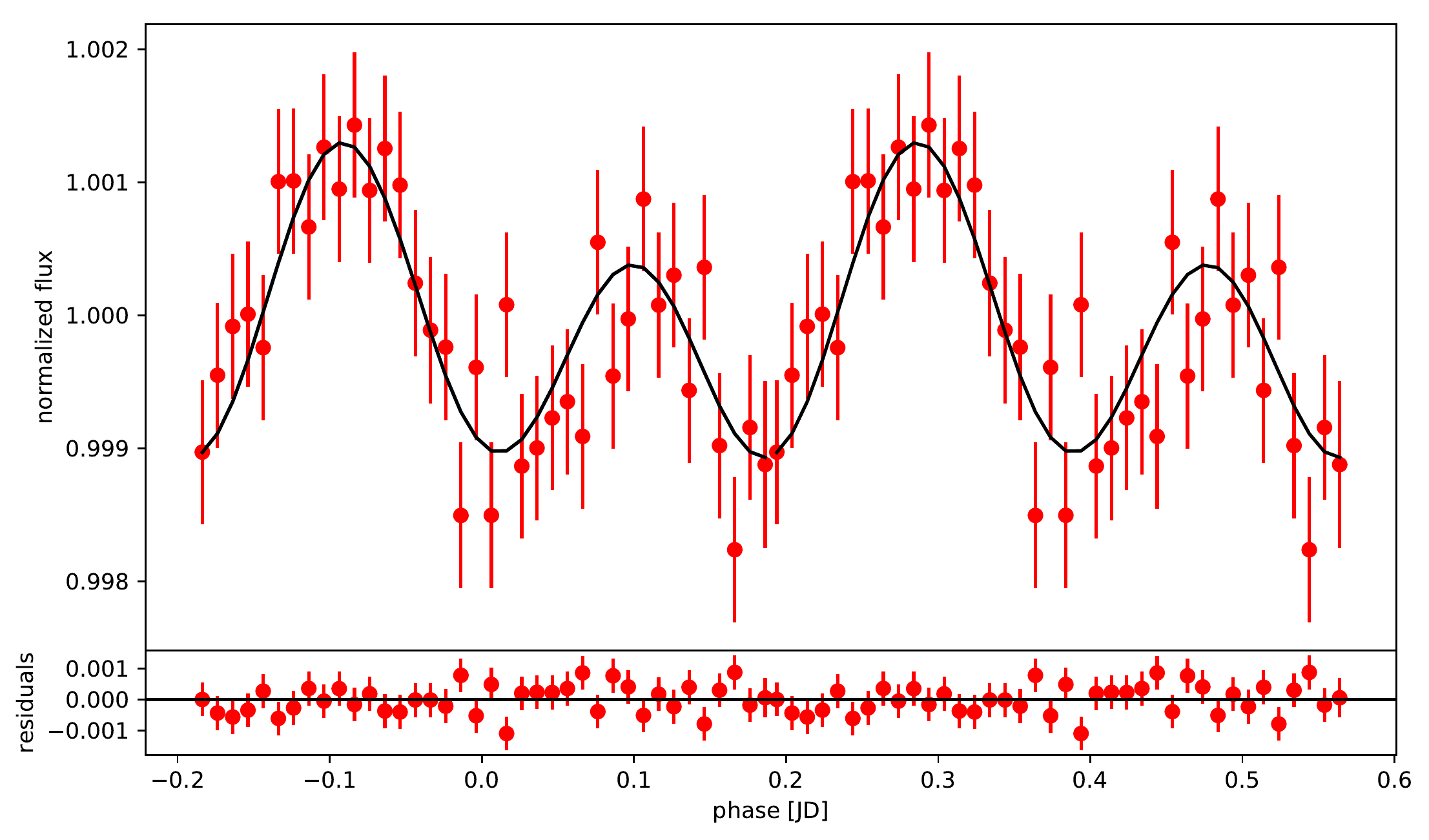}
        \caption{GD687}
        \label{lc_GD687}
    \end{subfigure}\hfill
        \begin{subfigure}{0.5\textwidth}
        \includegraphics[width=\linewidth]{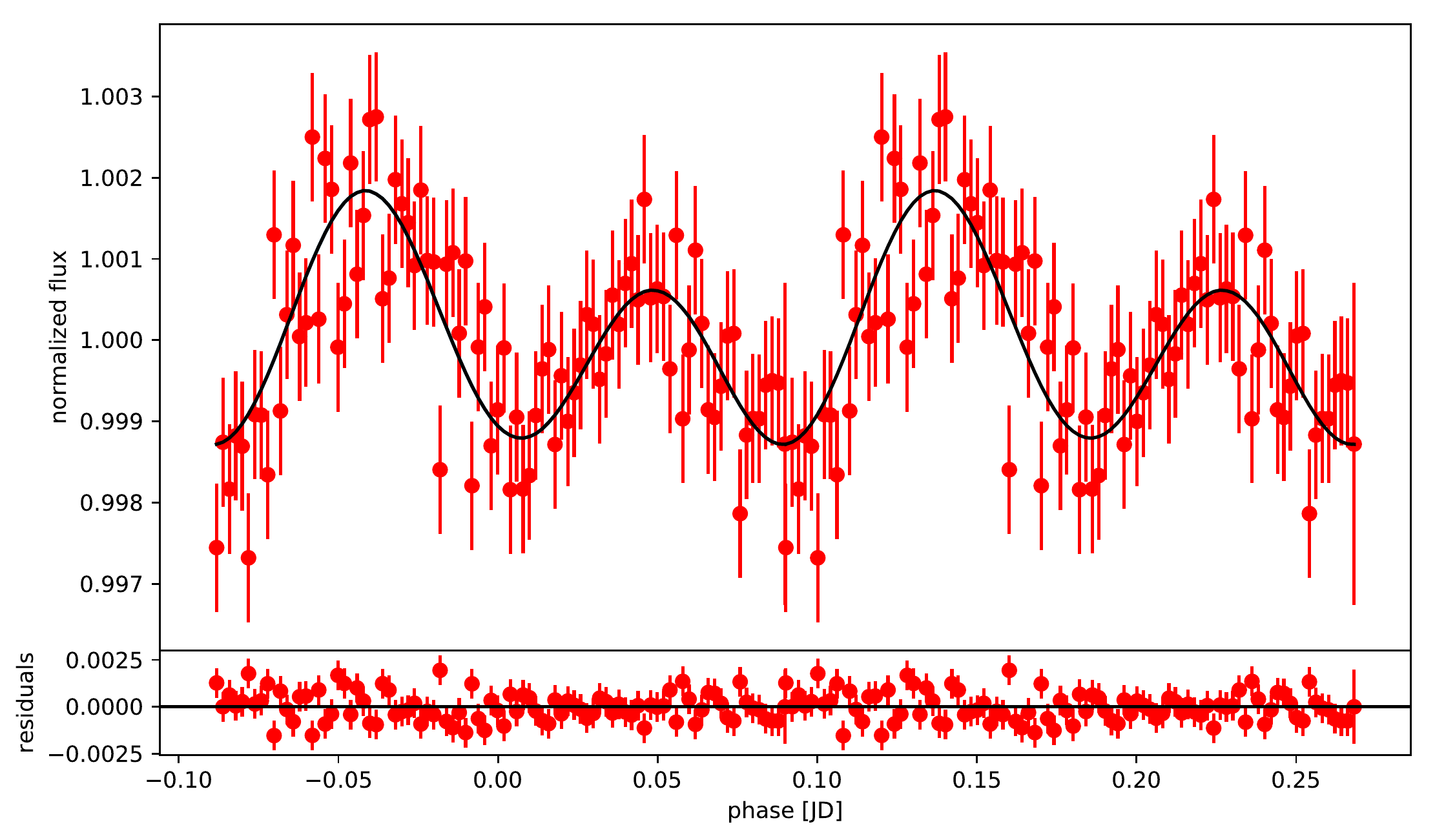}
        \caption{GALEX J075147.0+092526}
        \label{lc_GALEXJ0751}
    \end{subfigure}\hfill
    \begin{subfigure}{0.5\textwidth}
        \includegraphics[width=\linewidth]{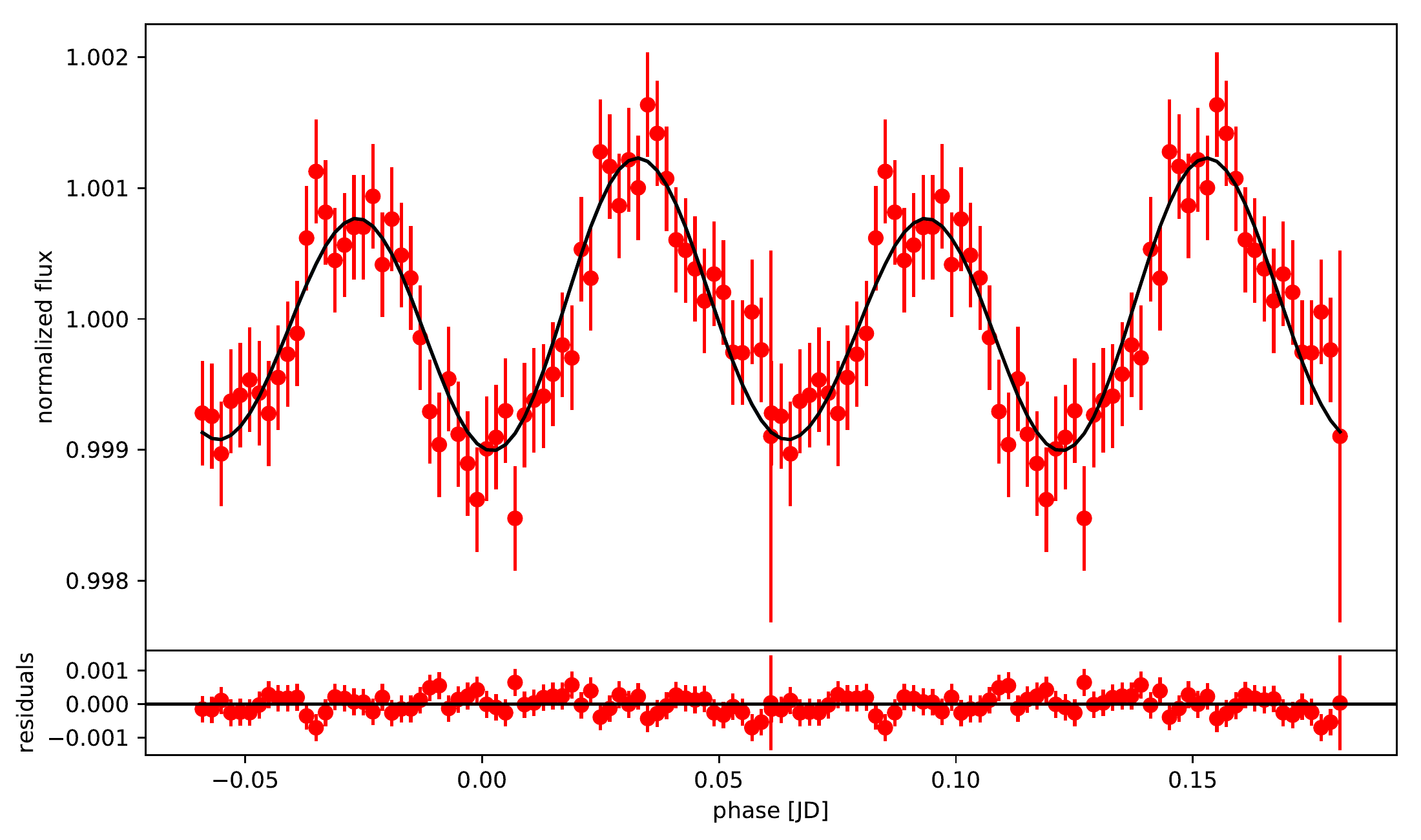}
        \caption{PG1043+760}
        \label{lc_PG1043+760}
    \end{subfigure}\hfill
    \begin{subfigure}{0.5\textwidth}
        \includegraphics[width=\linewidth]{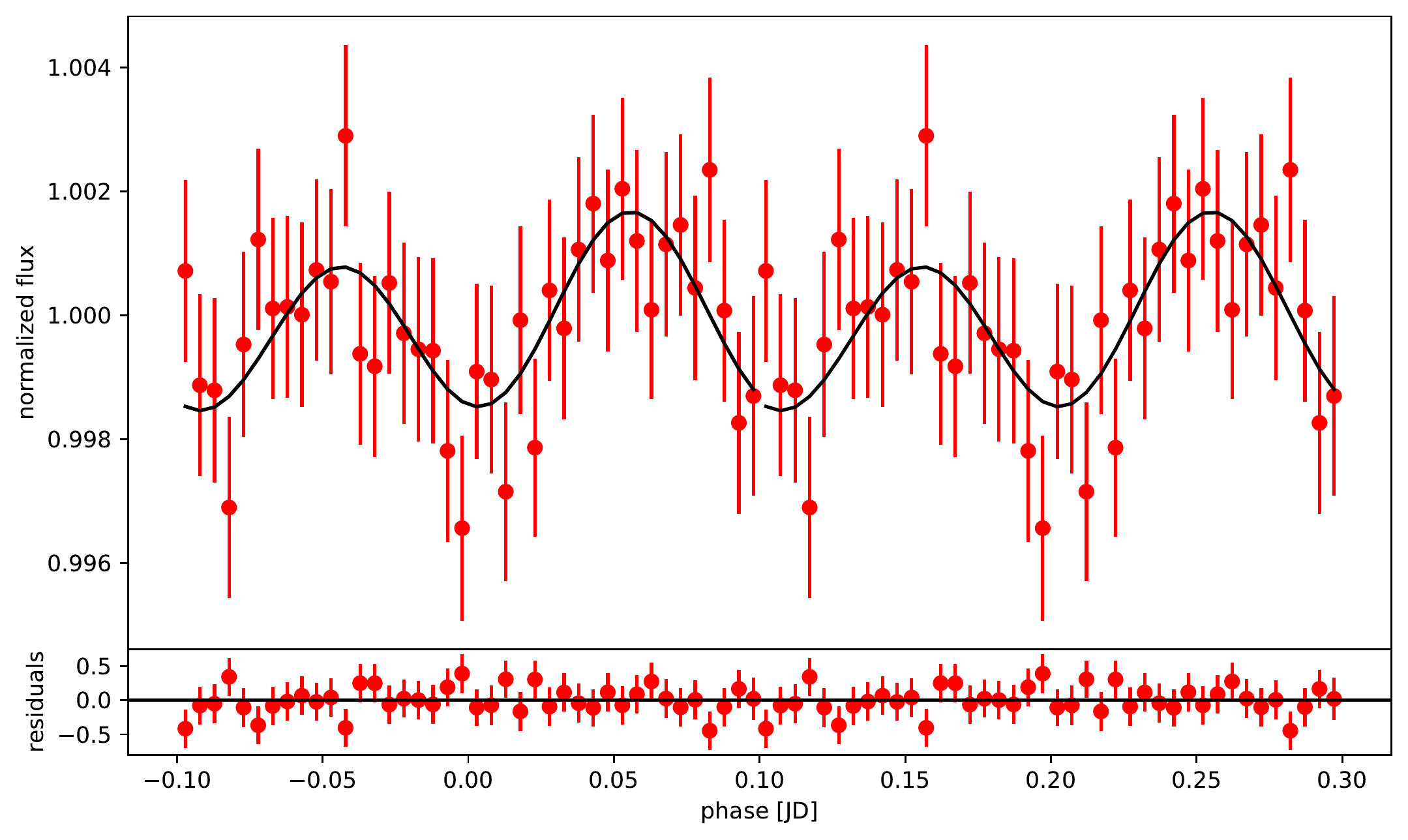}
        \caption{HS1741+2133}
        \label{lc_HS1741+2133}
    \end{subfigure}\hfill
        \begin{subfigure}{0.5\textwidth}
        \includegraphics[width=\linewidth]{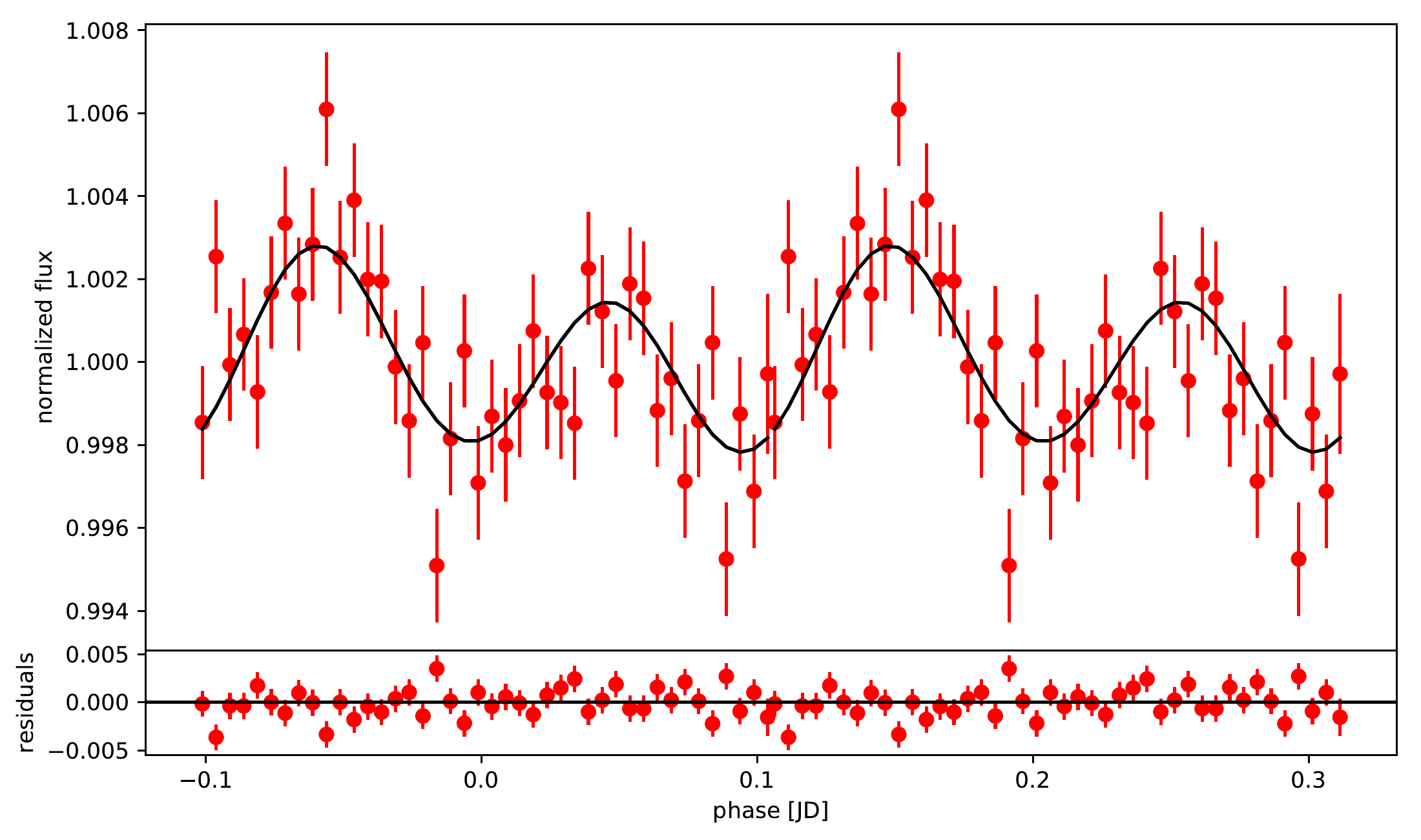}
        \caption{PG1136-003}
        \label{lc_PG1136-003}
    \end{subfigure}\hfill
            \begin{subfigure}{0.5\textwidth}
        \includegraphics[width=\linewidth]{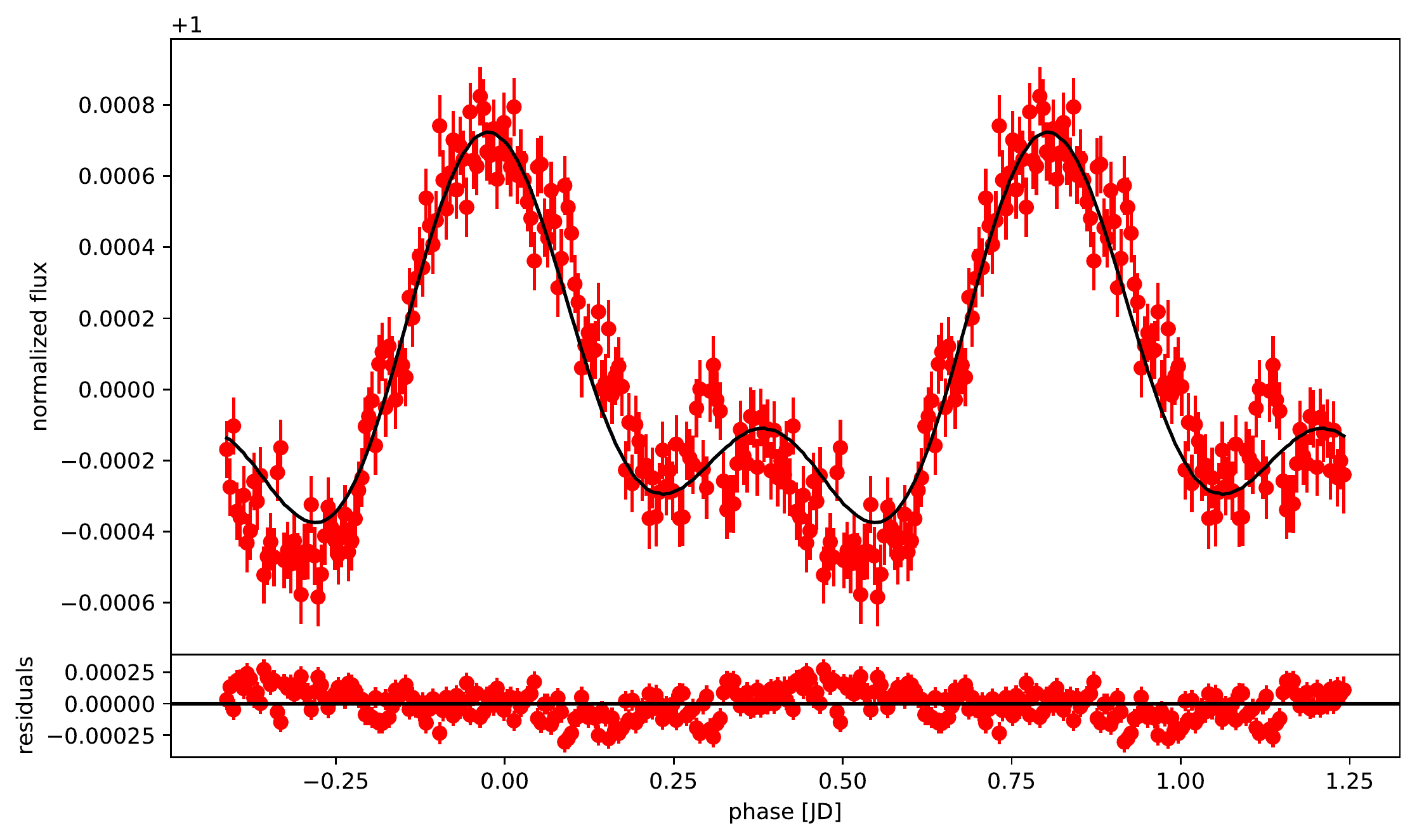}
        \caption{EC13332-1424}
        \label{lc_EC13332-1424}
    \end{subfigure}\hfill
     \begin{subfigure}{0.5\textwidth}
        \includegraphics[width=\linewidth]{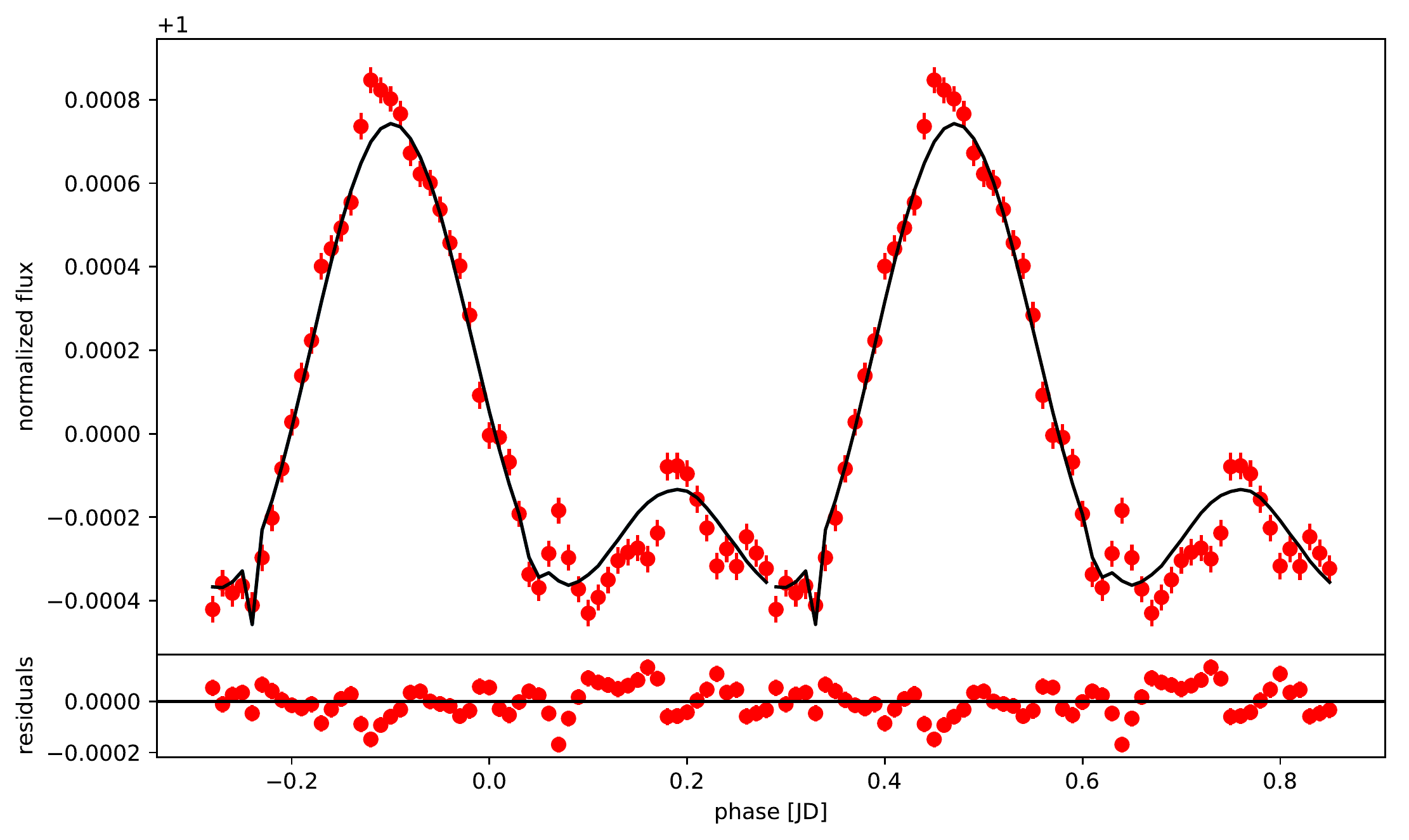}
        \caption{PG0101+039}
        \label{lc_PG0101+039}
    \end{subfigure}\hfill   
        \begin{subfigure}{0.5\textwidth}
        \includegraphics[width=\linewidth]{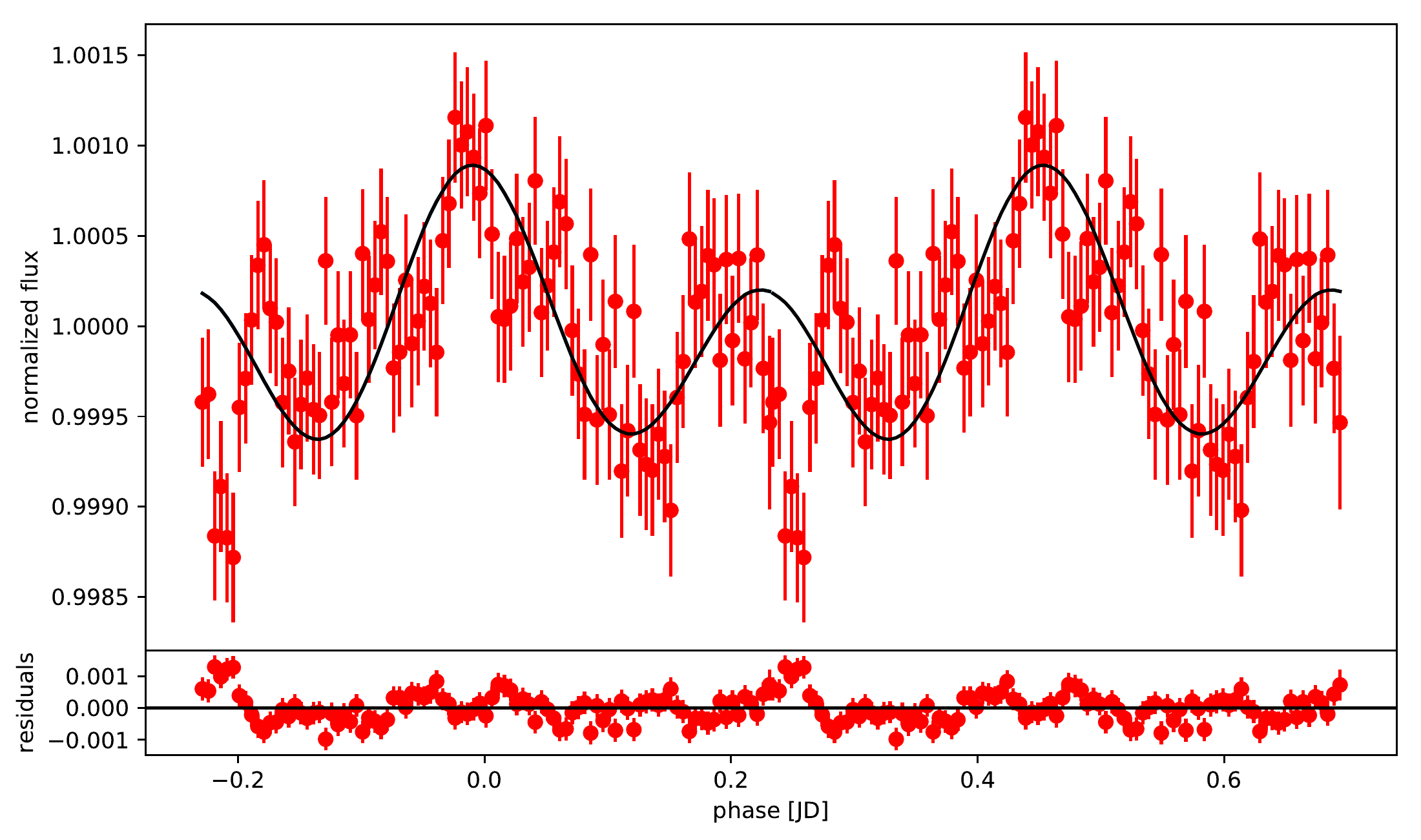}
        \caption{GALEX J234947.7+384440}
        \label{lc_GALEXJ2349}
    \end{subfigure}\hfill 
    \caption{Binned light curve of the newly confirmed sdB+WD systems with best model fit shown with the black line and the residuals in the lower panel.}
    \label{ell1}
\end{figure*}

\begin{figure*}\ContinuedFloat
    \centering
    \begin{subfigure}{0.5\textwidth}
        \includegraphics[width=\linewidth]{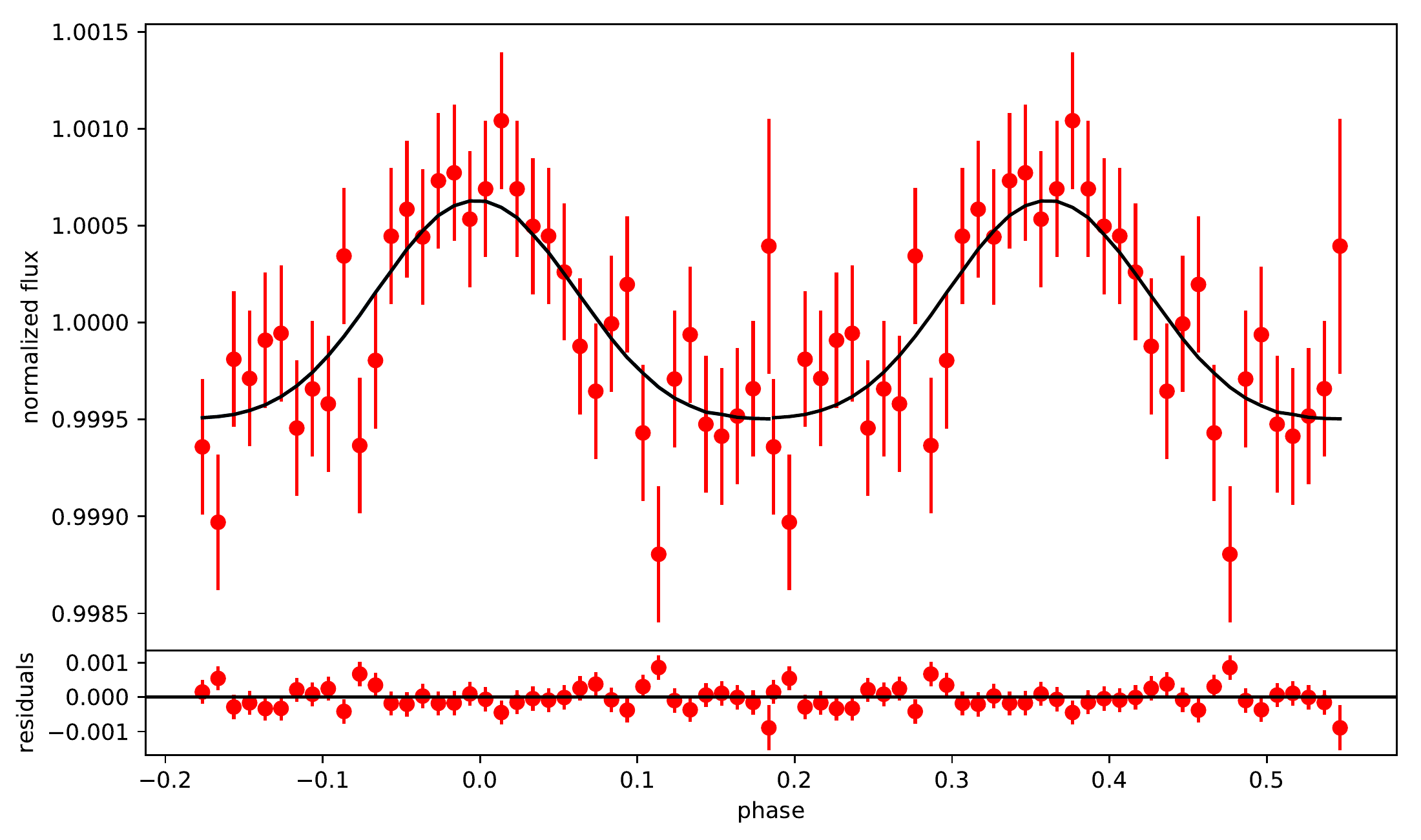}
        \caption{PG1232-136}
    \end{subfigure}\hfill
    \begin{subfigure}{0.5\textwidth}
        \includegraphics[width=\linewidth]{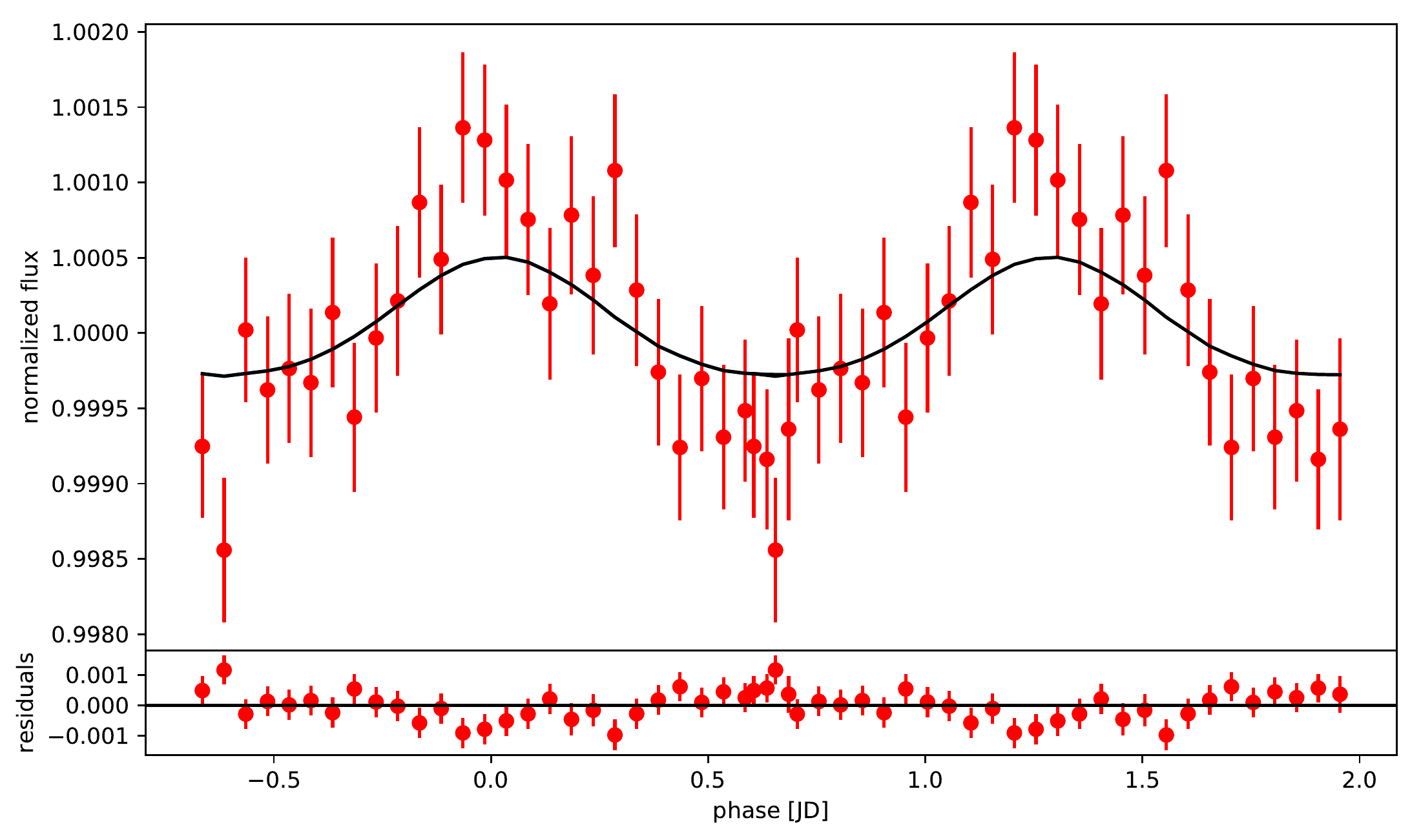}
        \caption{PG1512+244}
    \end{subfigure}\hfill
    \begin{subfigure}{0.5\textwidth}
        \includegraphics[width=\linewidth]{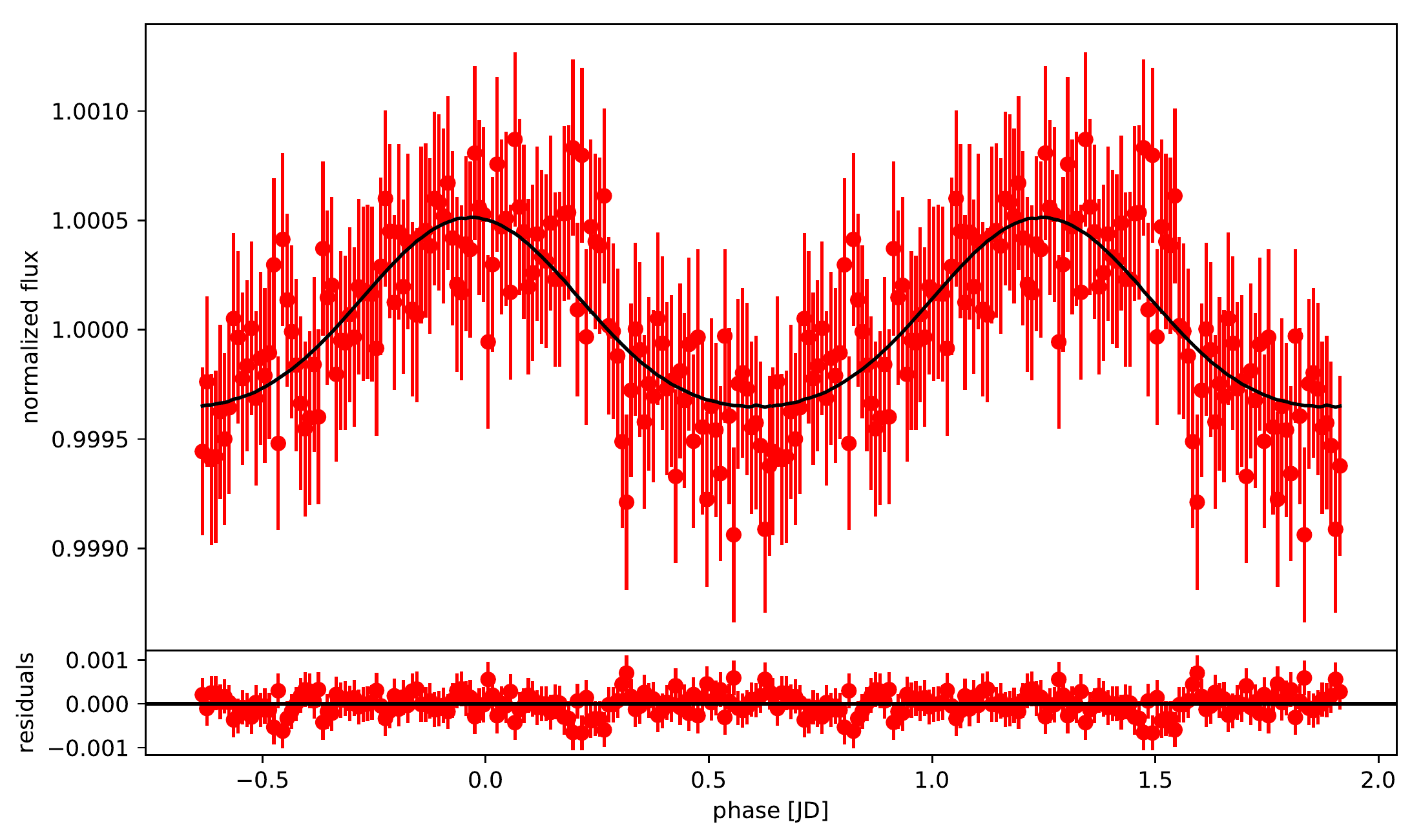}
        \caption{UVO1735+22}
    \end{subfigure}\hfill
    \begin{subfigure}{0.5\textwidth}
        \includegraphics[width=\linewidth]{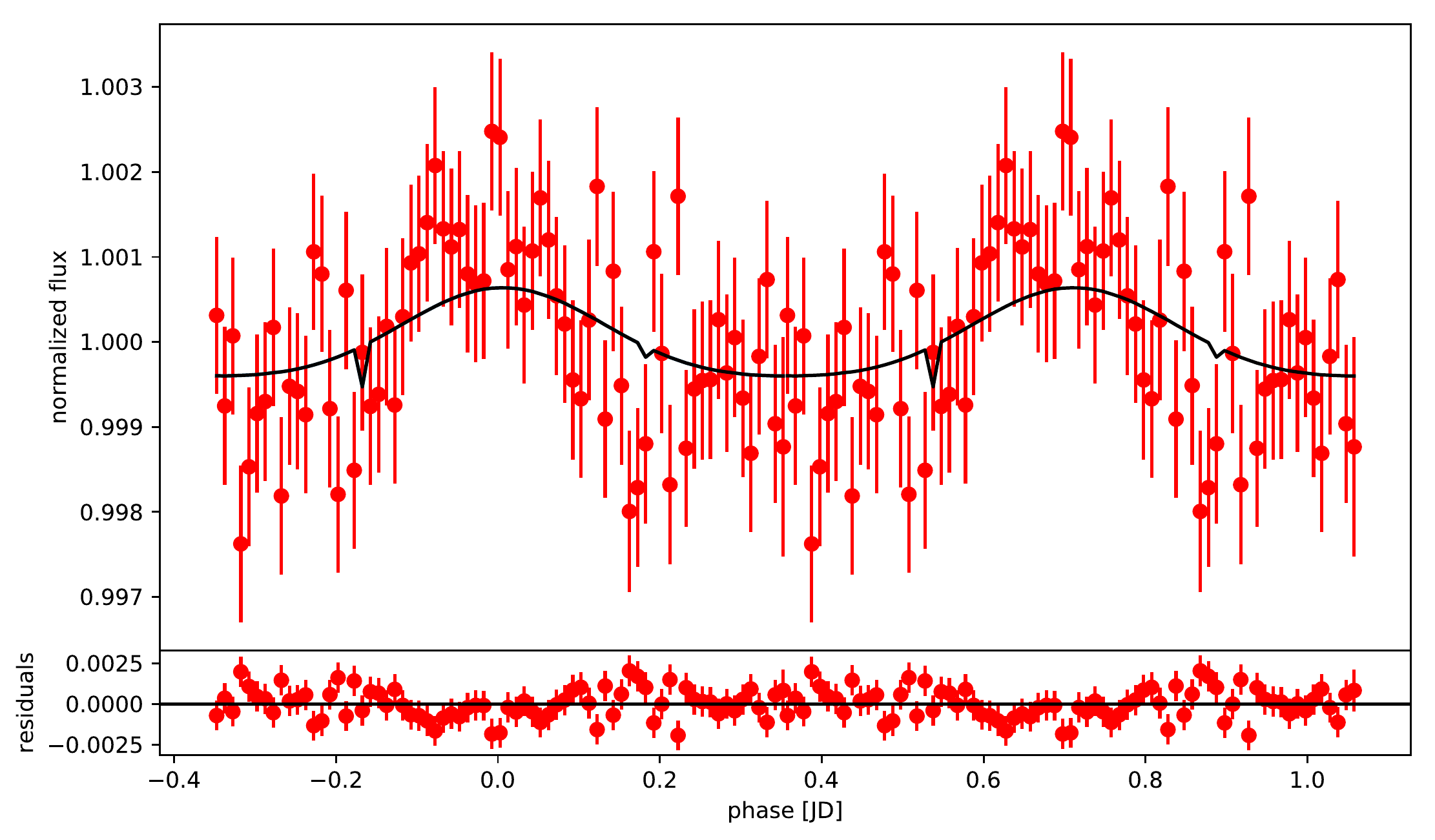}
        \caption{EC22202-1834}
    \end{subfigure}\hfill    
        \begin{subfigure}{0.5\textwidth}
        \includegraphics[width=\linewidth]{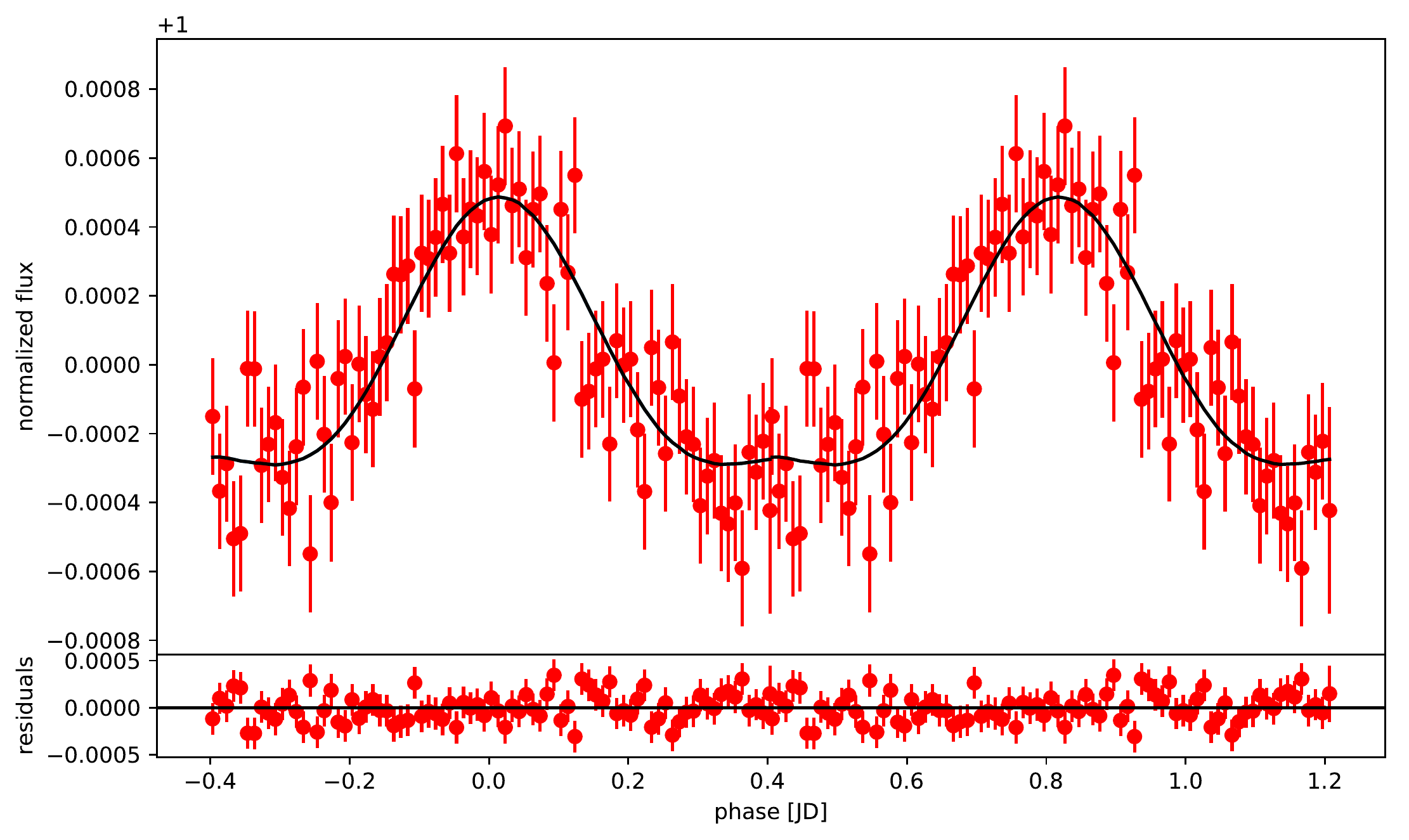}
        \caption{EC02200-2338}
    \end{subfigure}\hfill 
    \begin{subfigure}{0.5\textwidth}
        \includegraphics[width=\linewidth]{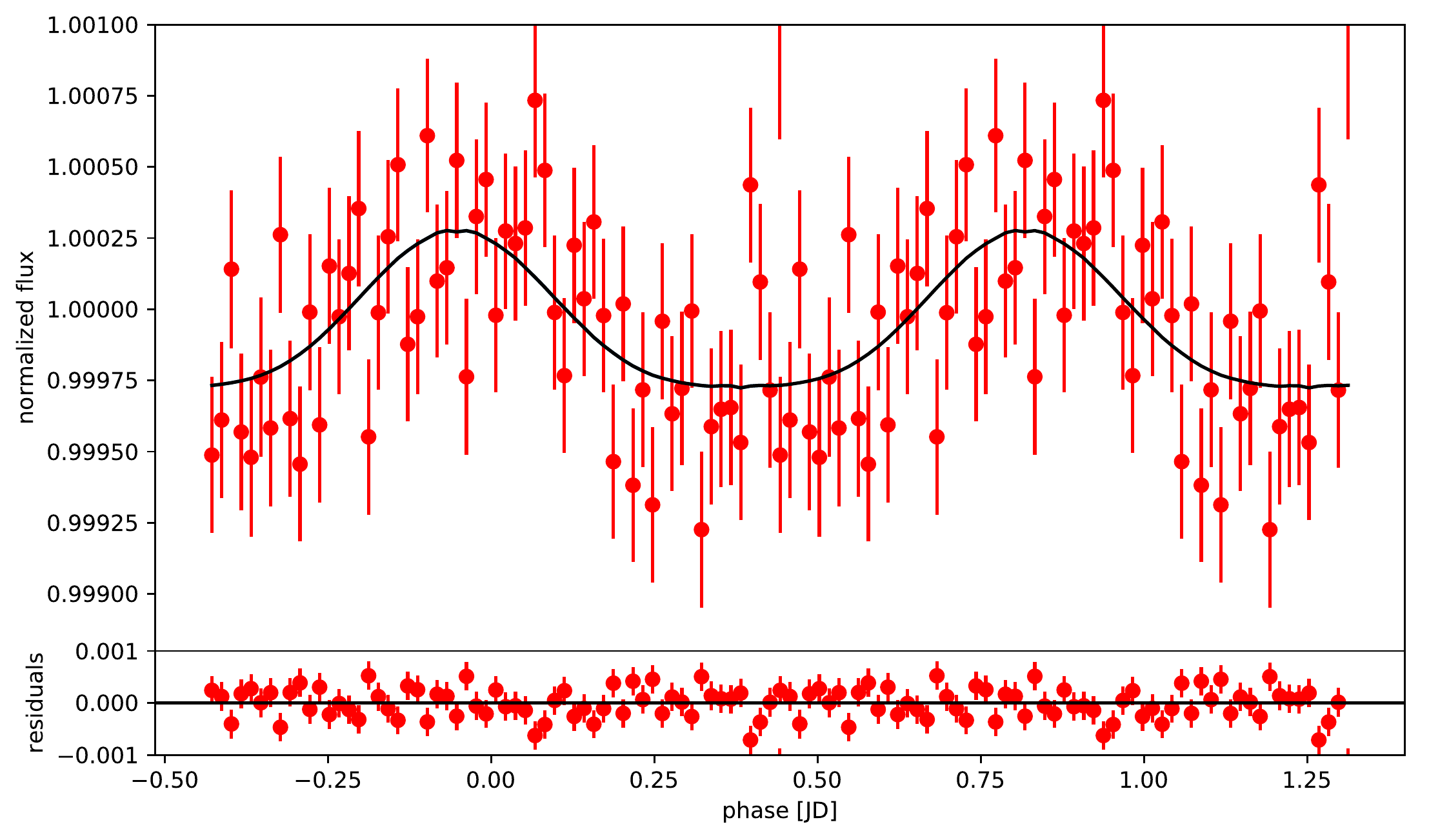}
        \caption{EC21556-5552}
    \end{subfigure}\hfill 
        \begin{subfigure}{0.5\textwidth}
        \includegraphics[width=\linewidth]{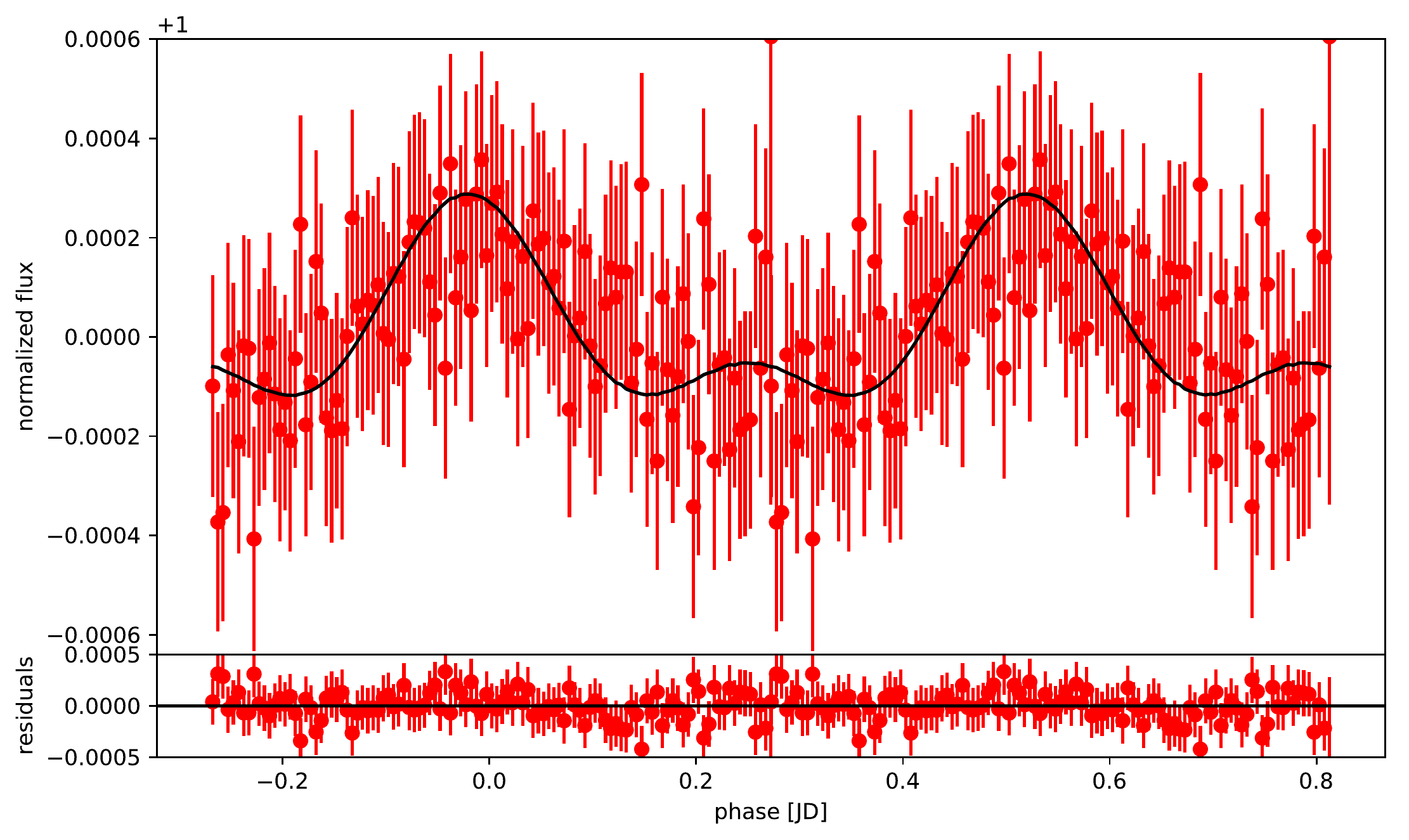}
        \caption{PG1519+640}
    \end{subfigure}\hfill 
    \begin{subfigure}{0.5\textwidth}
        \includegraphics[width=\linewidth]{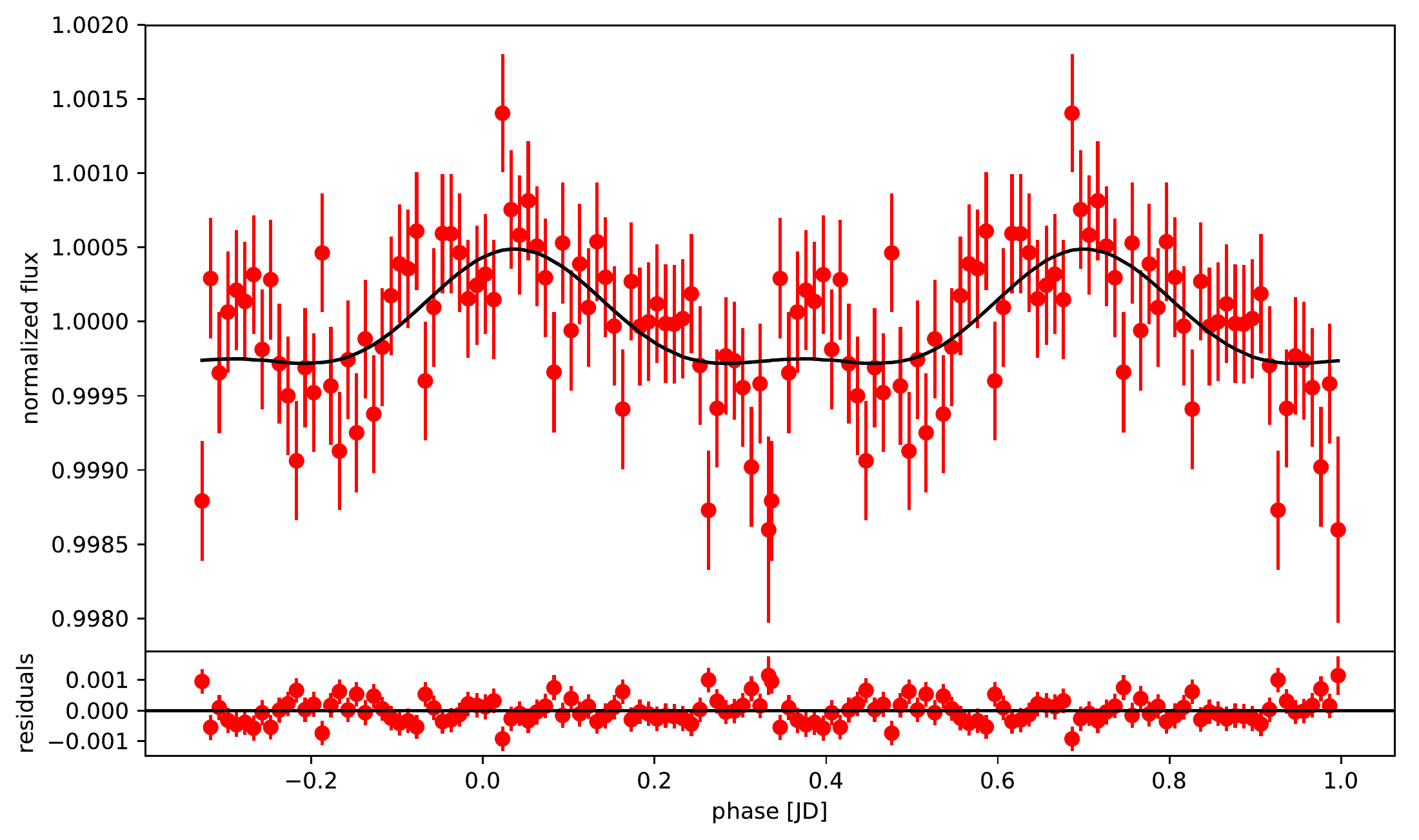}
        \caption{GALEX J025023.8-040611}
    \end{subfigure}\hfill  
    \caption{Binned light curve of the newly confirmed sdB+WD systems with best model fit shown with the black line and the residuals in the lower panel (continued).}
        \label{ell2}
\end{figure*}
\begin{figure*}\ContinuedFloat
    \centering
    \begin{subfigure}{0.5\textwidth}
        \includegraphics[width=\linewidth]{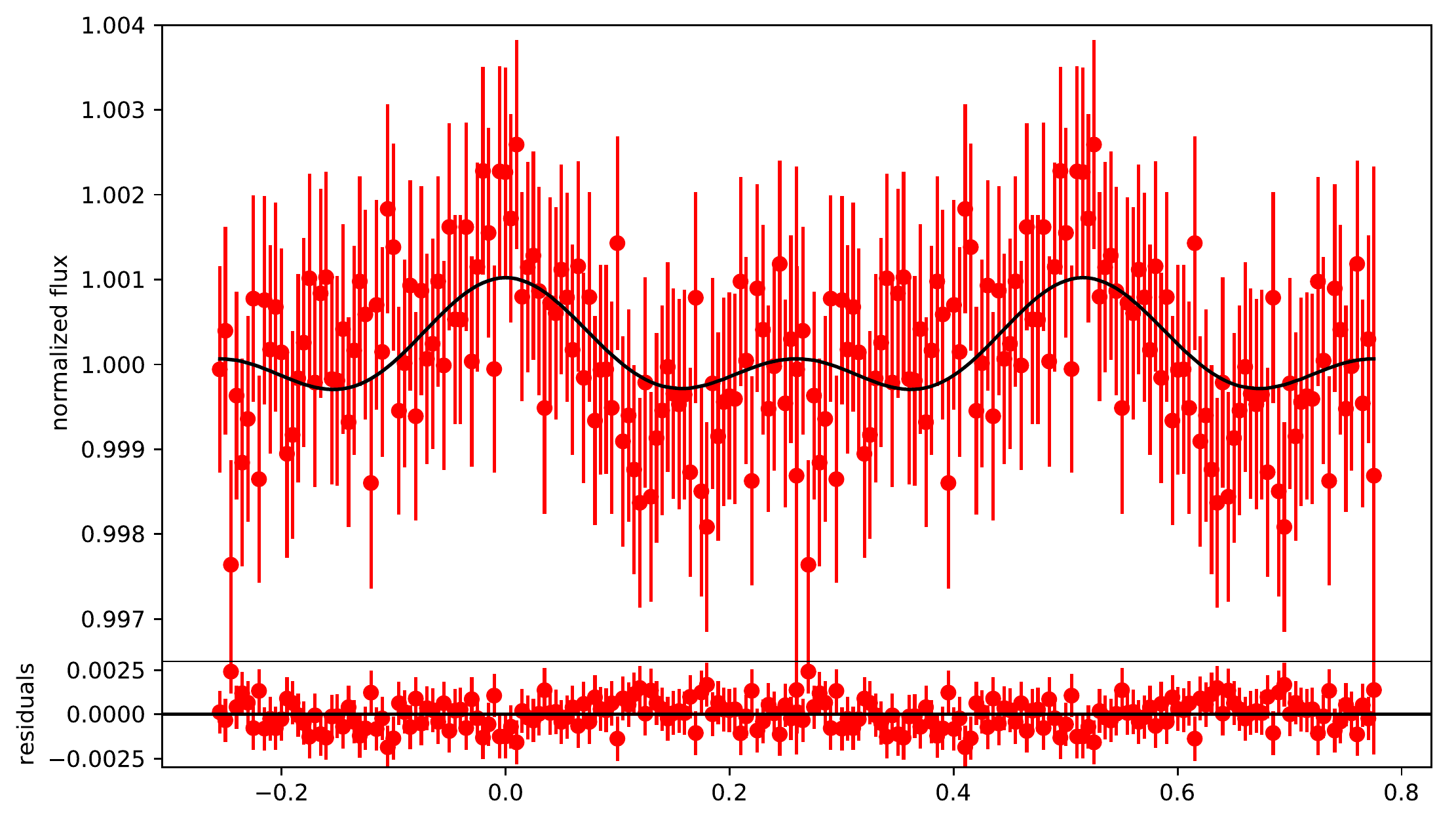}
        \caption{PG1743+477}
    \end{subfigure}\hfill
        \begin{subfigure}{0.5\textwidth}
        \includegraphics[width=\linewidth]{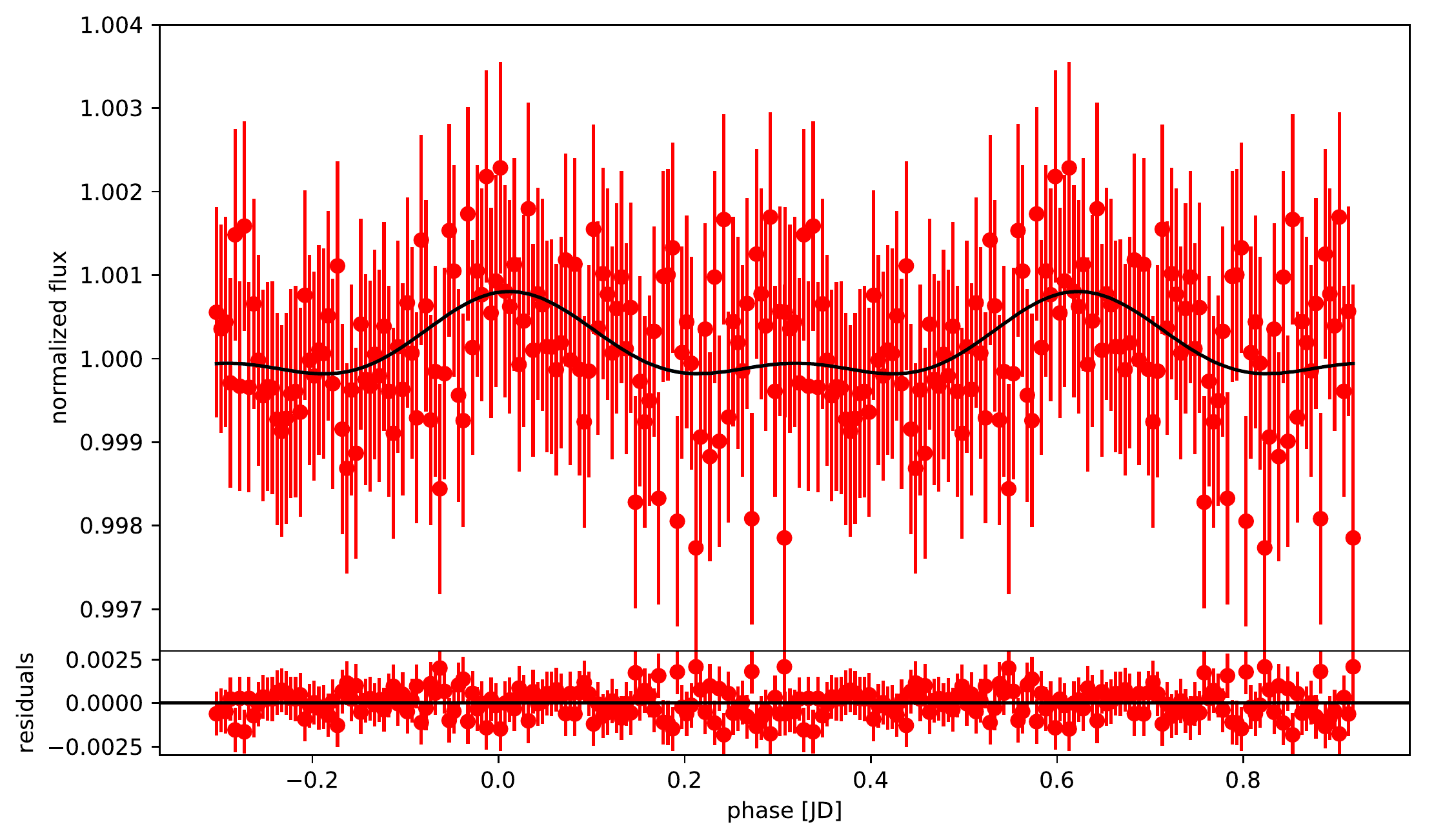}
        \caption{PG1648+536}
    \end{subfigure}\hfill
    \begin{subfigure}{0.5\textwidth}
        \includegraphics[width=\linewidth]{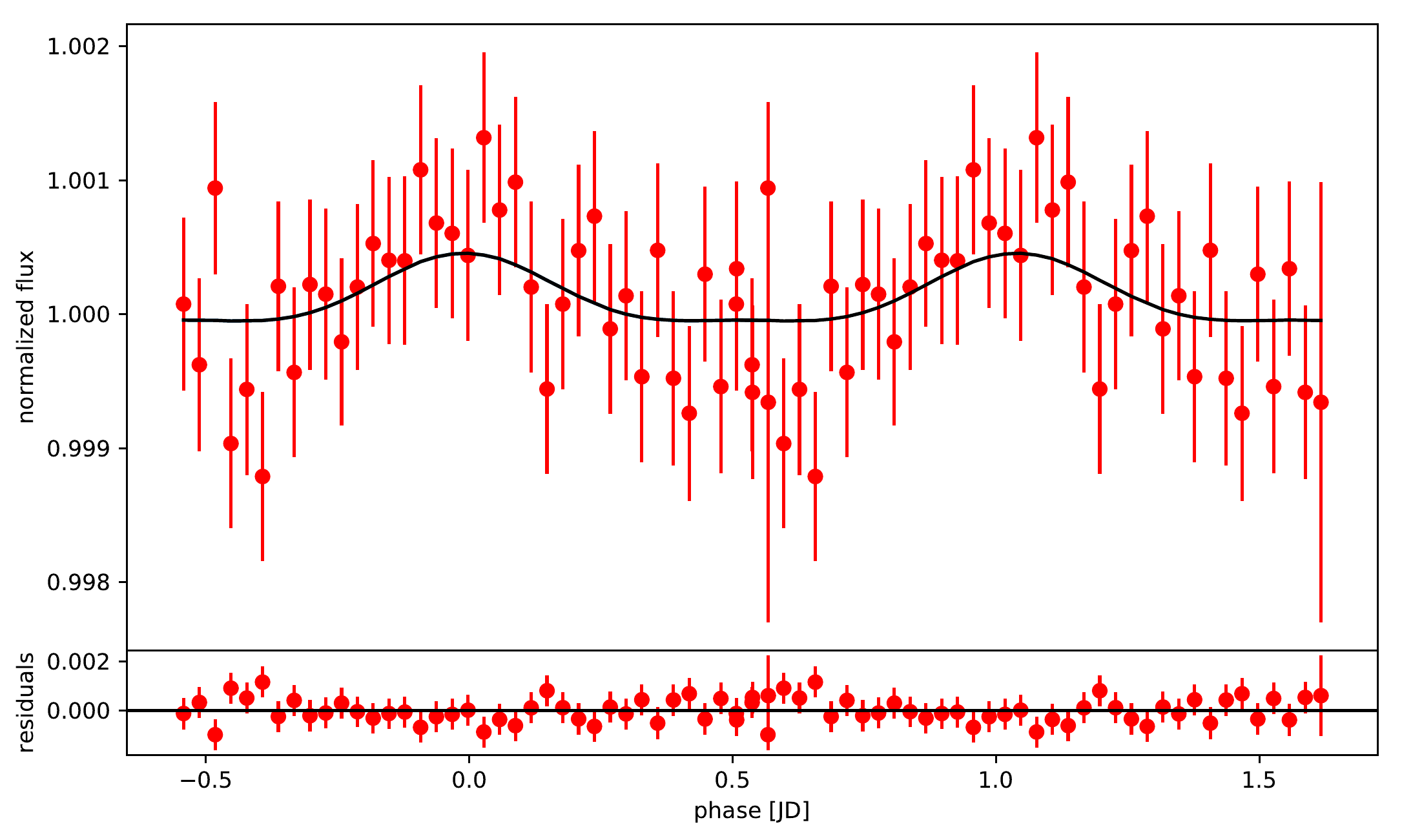}
        \caption{PG1000+408}
    \end{subfigure}\hfill 
    \begin{subfigure}{0.5\textwidth}
        \includegraphics[width=\linewidth]{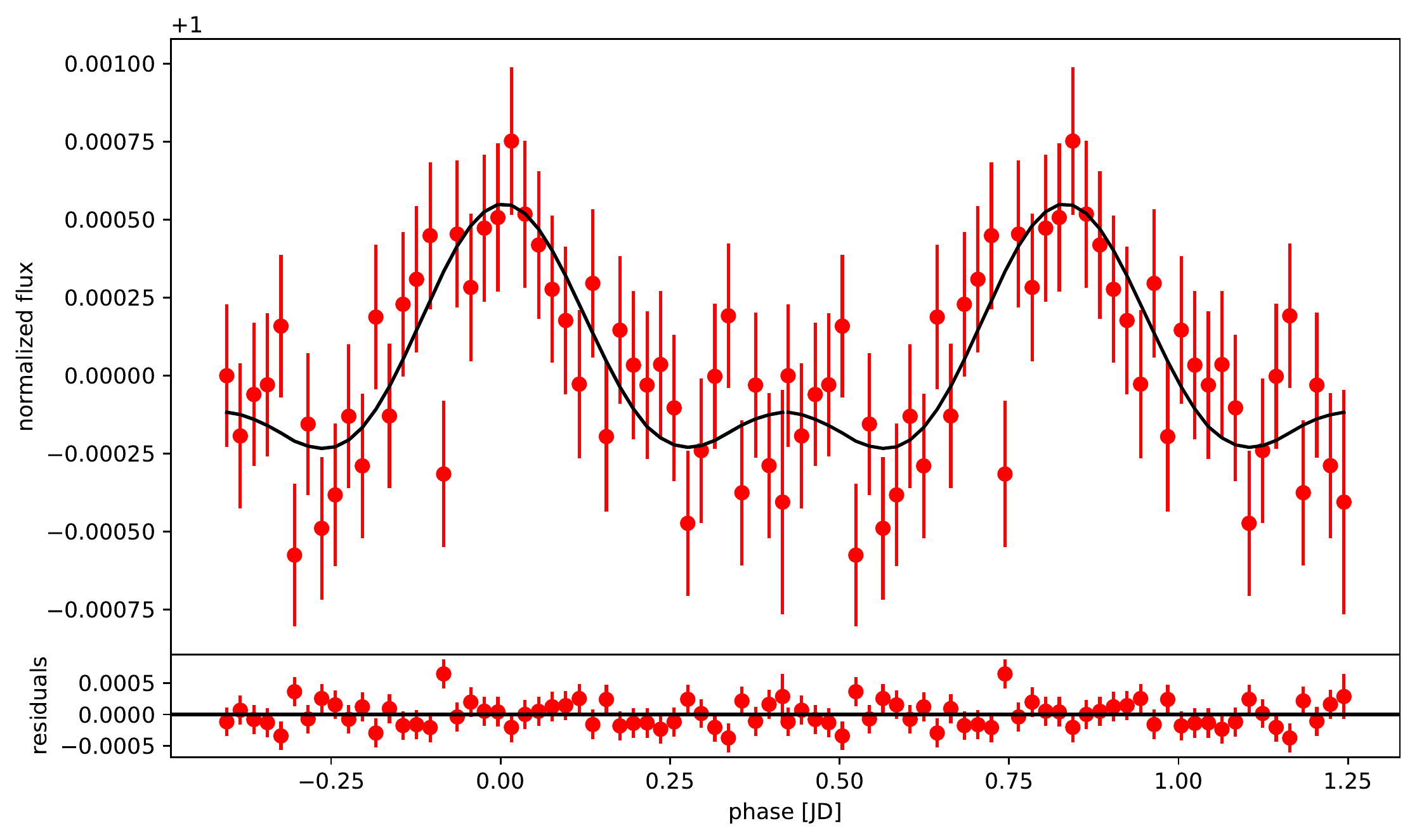}
        \caption{TON S183}
    \end{subfigure}\hfill   
        \begin{subfigure}{0.5\textwidth}
        \includegraphics[width=\linewidth]{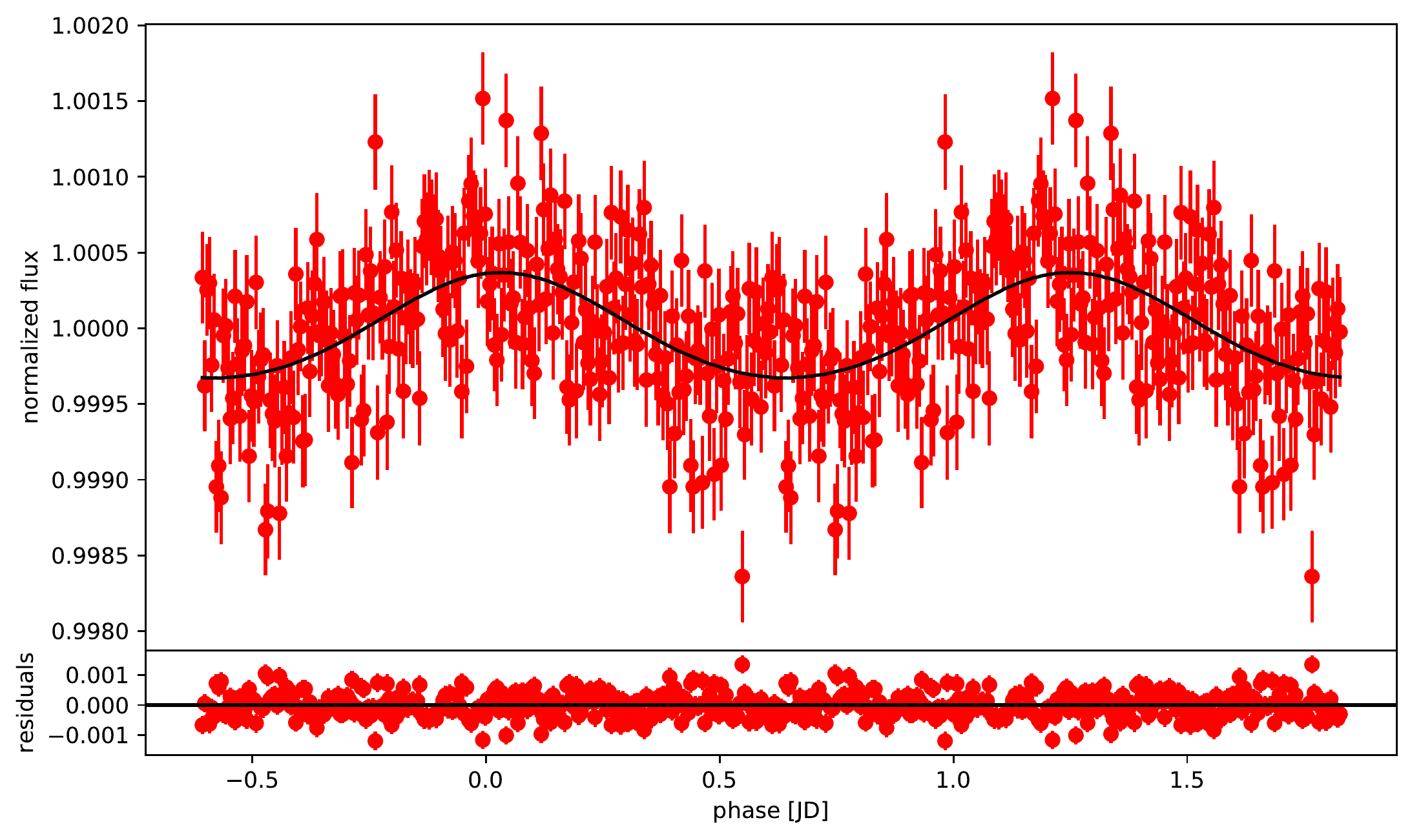}
        \caption{GALEX J225444.1-551505}
    \end{subfigure}\hfill
    \begin{subfigure}{0.5\textwidth}
        \includegraphics[width=\linewidth]{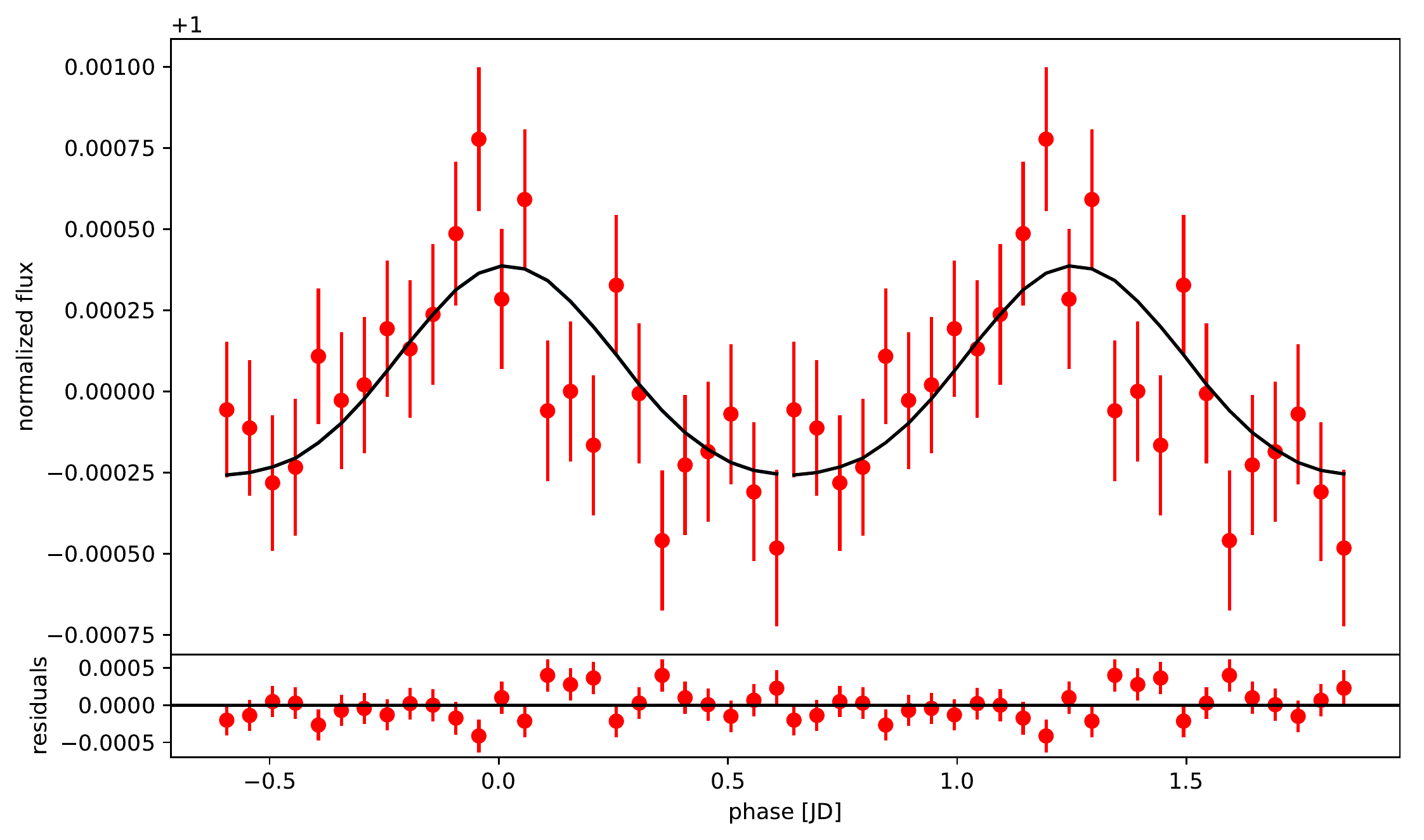}
        \caption{PG0133+114}
    \end{subfigure}\hfill      
    \begin{subfigure}{0.5\textwidth}
        \includegraphics[width=\linewidth]{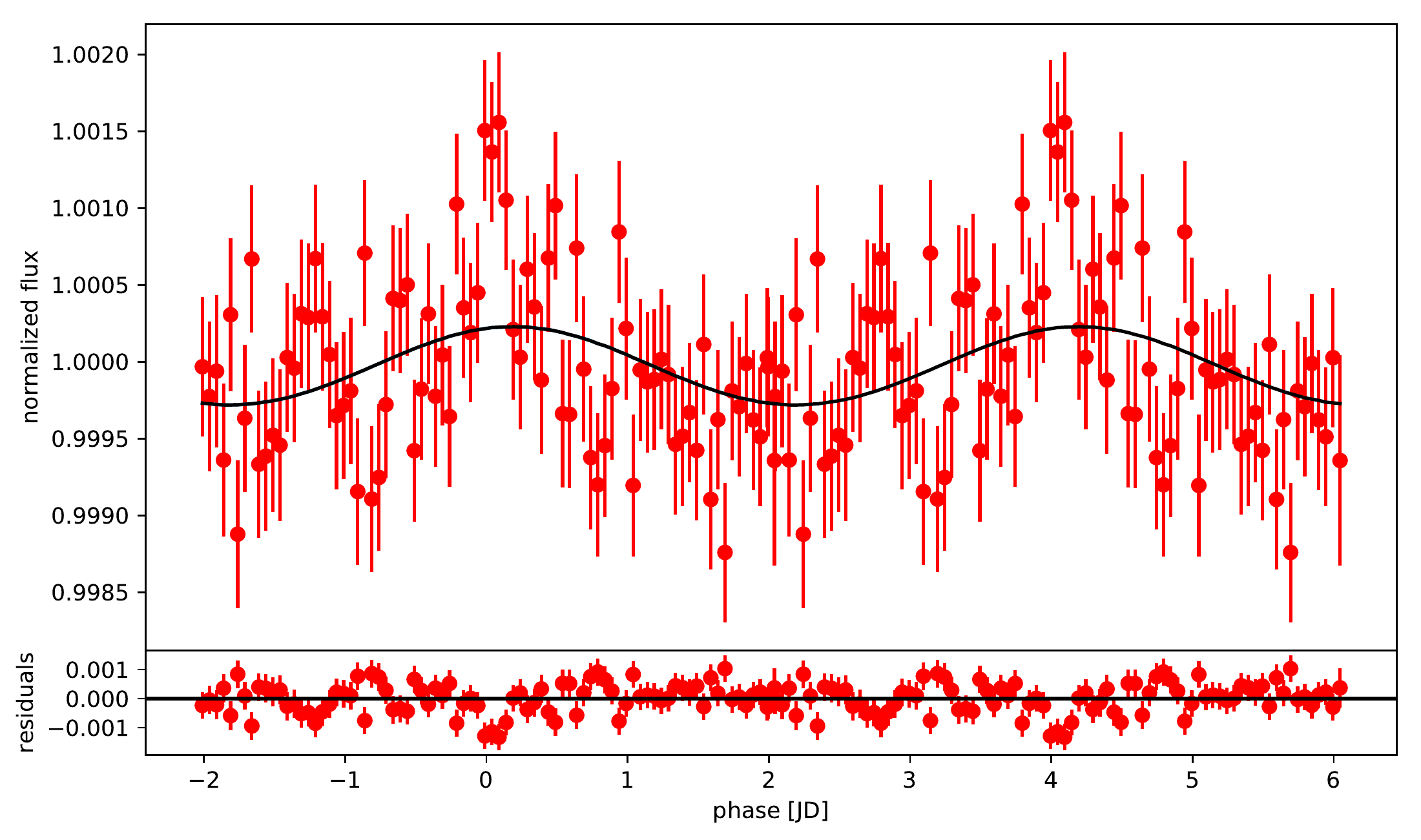}
        \caption{PG0934+186}
    \end{subfigure}\hfill
    \begin{subfigure}{0.5\textwidth}
        \includegraphics[width=\linewidth]{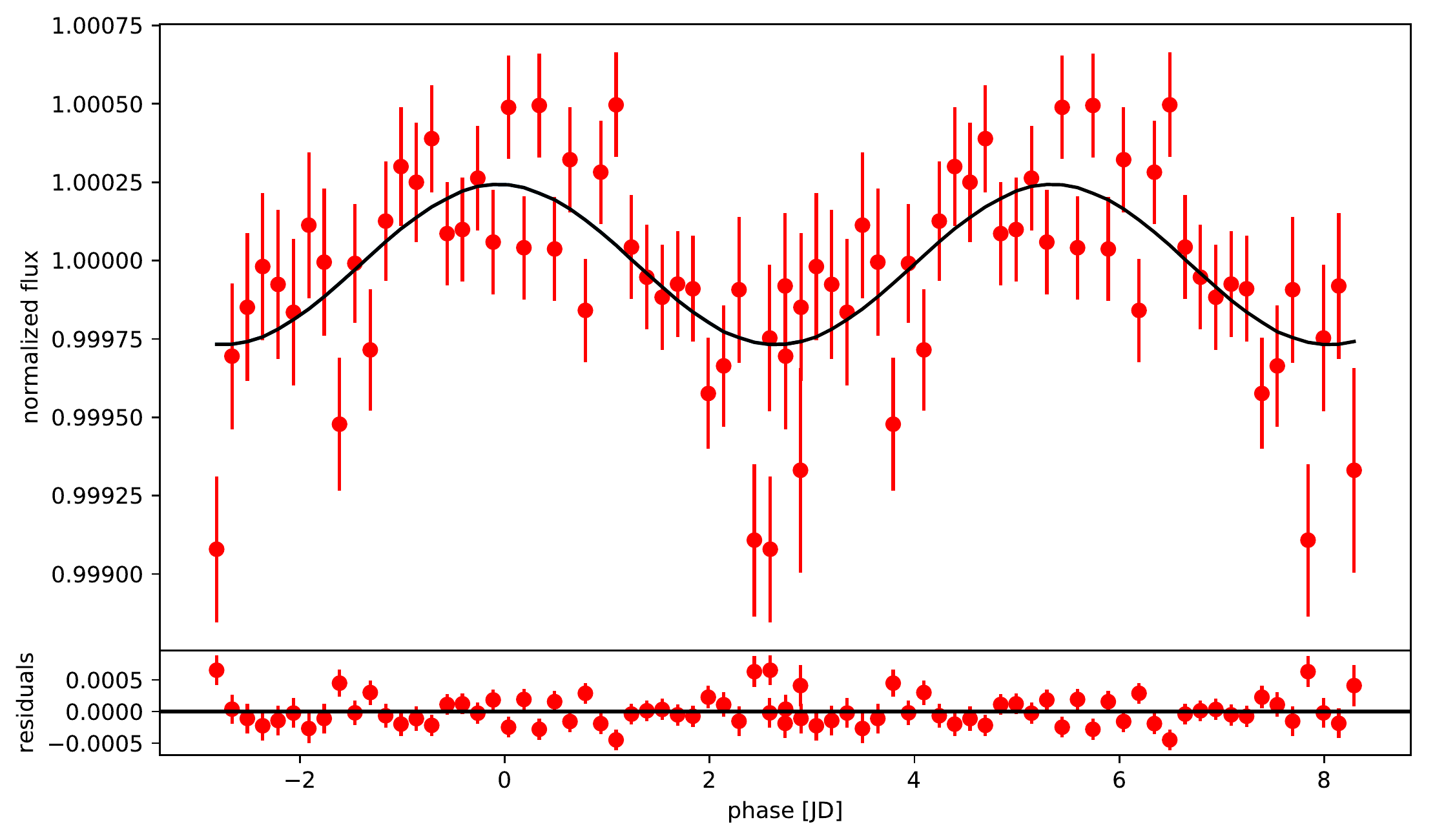}
        \caption{CD-24731}
    \end{subfigure}\hfill       
    \caption{Binned light curve of the newly confirmed sdB+WD systems with best model fit shown with the black line and the residuals in the lower panel (continued).}
    \label{ell3}
\end{figure*}
\FloatBarrier

\twocolumn

\end{appendix}
%
%

\end{document}